%% file: TOMS2017.tex
\documentclass[format=acmsmall, review=false, natbib, screen=true]{acmart}
\pdfoutput=1
\usepackage{booktabs} % For formal tables

\usepackage[ruled]{algorithm2e} % For algorithms

\SetAlFnt{\small}
\SetAlCapFnt{\small}
\SetAlCapNameFnt{\small}
\SetAlCapHSkip{0pt}
\IncMargin{-\parindent}

\usepackage{lipsum}  
\input{macros}

\setcopyright{rightsretained}
\newcommand{\myrightleftarrows}{\mathrel{\substack{\longleftarrow \\[-.6ex] \longrightarrow}}}

\begin{document}

% Title portion. Note the short title for running heads 
\title[Arrangements of cellular complexes]{{Regularized} arrangements of cellular complexes}  
\author{Alberto Paoluzzi}
%\orcid{1234-5678-9012-3456}
\affiliation{%
  {Roma Tre University,}
  {Rome,}
  {Italy}
}
\author{Vadim Shapiro}
\affiliation{%
  {University of Wisconsin-Madison \& ICSI,}
  {United States}
}
\author{Antonio DiCarlo}
\affiliation{%
  {CECAM-IT-SIMUL Node,}
  {Rome,}
  {Italy}
}

\begin{abstract}
In this paper we propose a novel algorithm to combine two or more cellular complexes, providing a minimal fragmentation of the cells of the resulting complex. We introduce here the idea of arrangement generated by a collection of cellular complexes, producing a cellular decomposition of the embedding space. The algorithm that executes this computation is called \emph{Merge} of complexes. The arrangements of line segments in 2D and polygons in 3D are special cases, as well as the combination of closed triangulated surfaces or meshed models.
This algorithm has several important applications, including Boolean  and other set operations over large geometric models, the extraction of solid models of biomedical structures at the cellular scale, the detailed geometric modeling of buildings, the combination of 3D meshes, and the repair of graphical models. The algorithm is efficiently implemented using  the Linear Algebraic Representation (LAR) of argument complexes, i.e., on sparse representation of binary characteristic matrices of $d$-cell bases, well-suited for implementation in last generation accelerators and GPGPU applications. 
\end{abstract}


\begin{CCSXML}
<ccs2012>
<concept>
<concept_id>10010147.10010371.10010396.10010401</concept_id>
<concept_desc>Computing methodologies~Volumetric models</concept_desc>
<concept_significance>500</concept_significance>
</concept>
<concept>
<concept_id>10010147.10010341.10010342.10010343</concept_id>
<concept_desc>Computing methodologies~Modeling methodologies</concept_desc>
<concept_significance>300</concept_significance>
</concept>
</ccs2012>





\thanks{This work is partially supported  from SOGEI S.p.A. ---  the ICT company of the Italian Ministry of Economy and Finance, by grant 2016-17, and by the  ERASMUS+ EU project medtrain3dmodsim. {V.S. is supported in part by National Science Foundation grant CMMI-1344205 and National Institute of Standards and Technology. }

  Authors addresses: A.~Paoluzzi, Department of Mathematics and Physics, Roma Tre University; 
  V.~Shapiro, University of Wisconsin at Madison, and International Computer Science Institute, Berkeley; A.~DiCarlo, CECAM-IT-SIMUL Node.
}

\renewcommand{\shortauthors}{A.~Paoluzzi et al.}

\nonstopmode

\maketitle

\input{s3p2017}

% Bibliography
\bibliographystyle{ACM-Reference-Format}
\bibliography{TOMS2017}
\end{document}

%% file: macros.tex
\def\E{\mathbb{E}}
\def\R{\mathbb{R}}
\def\Z{\mathbb{Z}}

\def\p#1{{\bf #1}}

\def\mat#1{{\left(\begin{array}{cccccccccccccccccccccccccccc}#1\end{array}\right)}}

%% file: s3p2017.tex
\section{Introduction}\label{introduction}

Given a finite collection $\mathcal{S}$ of cellular complexes in $\E^d$,
$d \in \{2,3\}$, the \emph{arrangement} $\mathcal{A}(\mathcal{S})$ is the
decomposition of $\E^d$ into connected cells of dimensions
$0, 1, \ldots, d$ induced by $\mathcal{S}$.

In this paper, we discuss the computation of the arrangement produced by
a given set of cellular complexes in either 2D or 3D. Our goal is to provide a
complete description of the plane or space decomposition induced by the
input, into cells of dimensions 0, 1, 2 or 3.

A planar collection $\mathcal{S}$ may include line segments, open or closed
polygonal lines, polygons, two-dimensional meshes, and discrete images
in 2D. A space collection may include 3D polygons, polygonal meshes, B-reps of solid models---either
manifold or non-manifold, three-di\-men\-sional CAE meshes, and volumetric
images in 3D.

The result of the computation discussed in this paper is the arrangement
$\mathcal{A}(\mathcal{S}) = X$, with $X := \bigcup_{k=0}^d$$X_k$, where $X_k$ is called the $k$-skeleton of the cellular complex $X$,  usually with
$d \in \{ 2,3 \}$, providing a cellular
decomposition of the space $\E^d$ where the {underlying space} (point-set) of $\mathcal{S}$ is embedded.

For example, you may consider a set {$\mathcal{L} = \{l_h\}$ of line
segments}, and the plane
arrangement generated by it (see~\cite{Chazelle:1993:CFA:165062.165181}), 
i.e.,~the 2-dimensional complex made by
open cells of dimension 0, 1 and 2. 
Analogously, you may consider a
set {$\mathcal{P} = \{p_k\}$ of planar polygons} in
$\E^3$. In this case the space arrangement
$\mathcal{A}(\mathcal{P})$ is a 3-dimensional complex $X$ made by closed 0-,
1-, 2- and 3-cells. A similar result $\mathcal{A}(\mathcal{S})$ is produced by any set $\mathcal{S}$ of 3D meshes and/or 2D meshes decomposing either open or closed surfaces, and with any kind of connected cells. {Finally,} {let us remember that, given a set $A$, the regularized set $A^*$ is the closure of the interior of $A$. Our generated arrangements are regularized by construction, since all cells not contained in an $d$-dimensional cell are removed. }

\subsection{Problem statement and results}

In general, we want to compute the (regularized) \emph{cellular $d$-complex} $X$ generated by a set of 
$(d-1)$-complexes of linear, bounded and connected cells {in $\E^d$}. 
This problem can be identified with the construction,
by induction, of a $d$-dimensional cellular complex, through the
filtration (or stratification) of its \emph{skeletons}
$X_0, X_1, \ldots, X_{d-1}, X_d$. In fact, the unknown $d$-cells of the
output arrangement are generated starting from $0$-cells, used as \emph{0-chains} to compute the coboundary operator $\delta_0$, used in turn to
compute the $1$-cells and the $\delta_1$ operator over 1-chains, and so on, by
iteratively growing in dimension.

The above introduces an important {feature} of the algorithm presented in this paper, since in computing the arrangement $\mathcal{A}(\mathcal{S})$ induced by $\mathcal{S}$, and hence the cellular complex $X := \mathcal{A}(\mathcal{S})$, we actually compute the whole \emph{chain complex} $C_\bullet$ generated by $X$. For example, in 3D we get to know all objects and arrows (morphisms) in the diagram below, and hence we get a complete knowledge of space subdivision topology.
Given the input collection $\mathcal{S}$, we generate the chain complex
\[
C_3 \mathrel{\substack{\delta_2\\\myrightleftarrows\\\partial_3}}  
C_2 \mathrel{\substack{\delta_1\\\myrightleftarrows\\\partial_2}}  
C_1 \mathrel{\substack{\delta_0\\\myrightleftarrows\\\partial_1}}  
C_0,
\]
where $C_d$ is a linear space of $d$-chains (subsets  of $d$-cells with algebraic structure), and where $\delta_{d-1} = \partial_d^\top$. 
\emph{Per se}, boundary operators $\partial_1,\dots,\partial_d$ belong in the chain complex, while coboundary operators $\delta_0,\dots,\delta_{d-1}$ belong in the \emph{dual} cochain complex. As a consequence, taking the coboundary of a chain only makes sense after chains and cochains have been identified, as explained in Section~\ref{sec:cells-chains}. 
Note that, in extracting the $d$-cells of the arrangement, we actually compute the sparse matrix of the operator $\partial_d$, and for this construction we need the $(d-1)$-cells and $\partial_{d-1}$, and so on backward, until $\partial_1$ is trivially constructed from the most elementary data (0- and 1-cells). 

{One component of our algorithmic workflow is the \emph{topological gift-wrapping}, reminiscent of the ``gift-wrapping" algorithm for computing convex hulls of 2D and 3D discrete sets of points~\cite{Jarvis:1973:ICH,Cormen:2009:IAT:1614191}. Actually, our topology-based algorithm broadly generalizes the former, been applicable also to \emph{non-convex}  contractible polyhedra of any dimension. }

The robustness of geometrical and topological computations is approached here by lowering the dimension of numerical computations whenever it is possible, and by solving independently the resulting set of subproblems.
{\emph{E.g.}}, the intersection of 3-cells is reduced to a set of intersections of {bounding} 2-cells, and each of those to {a set of pairwise intersections of} {bounding} line segments in 2D. The topology is finally reconstructed bottom-up, by successive identification of geometrical elements (reduction to quotient spaces) by nearest neighborhood queries on vertices, and syntactical identification of coincident cells in canonical form, i.e., as sorted lists of vertex indices.

The problem studied in this paper has a number of useful geometric applications, including the motion planning of robots, the
{variadic}\footnote{Which accepts a variable number of arguments.}
computation of Bool\-ean operations (union, intersection, difference,
symmetric difference), and the topology repair of graphical meshes, all starting from a set of cellular complexes embedded in the same Euclidean space. In particular, we are currently using chain complexes and boundary operators to extract the models of neurons and vessels from extreme-resolution 3D images of brain tissue, and to dramatically reduce the complexity of their representation, while preserving the homotopy type, in order to piecewise compute the connectome of brain structures~\cite{ClementiSSPP-CAD16,doi:10.1080/16864360.2016.1168216}.

\subsection{Previous work}

The construction of arrangements of lines, segments, planes and other geometrical objects can be found in~\cite{fhktww-a-07} together with a description of the CGAL software~\cite{Fabri:2000:DCC:358668.358687}, implementing 2D/3D arrangements with Nef polyhedra~\cite{Hachenberger:2007:BOS:1247750.1248141}. A wide analysis of papers and algorithms concerning the construction and counting of cells may be found in the dedicated survey chapter on Arrangements in the Handbook of Discrete and Computational Geometry~\cite{Goodman:2017:HDC:285869}. 
The arrangements of polytopes, hyperplanes and $d$-circles are discussed in~\cite{Ziegler:92}. 
The above references deal with space arrangements generated by analytical subspaces.

Some early papers were concerned with efficient representation of 3D cellular decompositions. 
In particular, ~\cite{Dobkin:1987:PMT:41958.41967} defined the polygon-edge data structure, to represent orientable and non-punctured 3D decompositions and their duals, manipulated by intricate operations with specialized Euler operators.
The much simpler and compact  SOT (Sorted Object Table) representation is proposed in~\cite{Gurung:2009:SCR:1629255.1629266} for object decompositions with tetrahedral meshes. 
{Discrete Exterior Calculus (DEC) with simplicial complexes was introduced by \cite{Hirani:2003:DEC:959640} and made popular by \cite{Desbrun:2006:DDF:1185657.1185665} and \cite{Elcott:2006:BYO:1185657.1185666}.}
More recently,  a systematic recipe   has been proposed  in~\cite{Zhou:2016:MAS:2897824.2925901} for constructing a family of exact constructive solid geometry operations starting from a collection of triangle meshes.
To the best of our knowledge, no literature exists for the computational problem introduced and discussed in this paper, written as a guide {to build} a dedicated software library.

{Most of the above algorithms and procedures are restricted  to  specific data structures optimized for representation of selected class of geometric objects under consideration.   In contrast, our formulation is cast in terms {of} (co)chain complexes and (co)boundary operators that may be implemented using variety of data structures.  Our reference implementation relies in Linear Algebraic Representation (LAR)~\cite{Dicarlo:2014:TNL:2543138.2543294}, that is particularly well suited for efficient implementation of topological operations in terms {of} matrix algebra over cellular complexes with cells of a general kind.   }

\subsection{Paper preview}

In Section~\ref{background} the main definitions concerning cell complexes and chain complexes are recalled, together with the characteristic features of the Linear Algebraic Representation (LAR) scheme.
In Section~\ref{sec:gentle-intro} a gentle introduction to our computational approach is given, supported also by a cartoon chronicle of a simple 3D example.
The main {contribution} of the paper, i.e., the novel algebraic algorithm for computing the merging of $d$-complexes, is {explained} in Section~\ref{merge-algorithm}. Section~\ref{sec:complexity} provides some pseudocode and discusses the complexity of the main computational steps.
{For clarity, Section~\ref{applications-and-examples} exemplifies the developed algorithms on  simple examples.}
The closing Section~\ref{conclusion} {summarizes} the work and highlights its salient features.
Some very simple examples of LAR input/output and topology computations are given in the {Appendix}.

\section{Background}\label{background}

For the sake of readability, let us introduce the meaning of some symbols: $\Lambda = \Lambda(X)$ {is a cellular decomposition of the topological space $X$, i.e., a quasi-disjoint union of cells}; $\Lambda_{p}$ is the set of $p$-cells; $C_p$ is a linear space of $p$-chains of cells over a {field of coefficients}; $X_p$ is a $p$-complex, i.e.~the $p$-skeleton of the  $d$-complex  $X_d := \mathcal{A}(\mathcal{S})$. Finally, $\mathcal{S}$ is a collection of cellular $d$-complexes and/or $(d-1)$-complexes embedded in $\E^d$ space. We use greek letter for \emph{cells} and latin letters for \emph{chains}, i.e., for signed combinations of cells.  With some abuse of language, cells in $\Lambda_p$ and singleton chains in $C_p$ are often identified.
Also, let us remind the reader that the \emph{characteristic function} $\chi_A: S\to\{0,1\}$ is a function defined on a set $S=\{s_j\}$, that indicates membership of an element $s_j$ in a subset $A\subseteq S$, having the value 1 for all elements of $A$ and the value 0 for all elements of $S$ not in $A$. 
We call \emph{characteristic matrix} $M$ of a collection of subsets $A_i\subseteq S$ ($i=1,\ldots,n$) the binary  matrix $M=(m_{ij})$, with $m_{ij} = \chi_{A_i}(s_j)$, that provides a basis for a linear space of \emph{$p$-chains}.

\subsection{Cellular complex and chain spaces}
\label{sec:cells-chains}

Let $X$ be a topological space, and
$\Lambda(X) = \bigcup_k\Lambda_k$ ($k \in {0, 1,\ldots,d}$) be a
{cellular decomposition} of $X$, with $\Lambda_k$ a set of {closed} and connected $k$-cells. 
A \emph{CW-structure} on the space $X$ is a filtration
$\emptyset = X_{-1} \subset X_0 \subset X_1 \subset \ldots \subset X = \bigcup_d X_d$,
such that, for each $k$, the \emph{skeleton} $X_k$ is homeomorphic
to a space obtained from $X_{k-1}$ by attachment of $k$-cells in
$\Lambda_k = \Lambda_k(X)$.

A \emph{CW-complex} is a space $X$ endowed with a CW-structure, and is also
called a \emph{cellular complex}. A cellular complex is \emph{finite}
when it contains a finite number of cells. A \emph{regularized}
$d$-complex is a complex where 
%every closed $d$-cell equates the closure ofits interior, and where 
every $k$-cell ($k < d$) is contained in the
boundary of a $d$-cell.
A $d$-complex is \emph{orientable} when its $d$-cells can be coherently oriented. Two $d$-cells are \emph{coherently oriented} when their common $(d-1)$-facets have opposite orientations. 

{In this paper, we restrict attention to CW-complexes, even though our merge operation may generate punctured $d$-cells, i.e., cells with holes. As a matter of fact, such spaces are handled by combining standard CW-complexes (i.e., with cells homeomorphic to balls) by the adjunction of $p$-cells ($0\leq p\leq d$) to the interior of $d$-cells. See, e.g., the Algorithm 2 of Section~\ref{sec:shells}.}

\subsubsection{{Chains}}
\label{sec:thechains}

Let $(G,+)$ be a nontrivial commutative group, {whose identity element will be denoted $0$}. A $p$-chain of $X$ with coefficients in $G$ is a mapping $c_p : X \to G$ such that, for each $\sigma \in X_p$, reversing a cell orientation changes the sign of the chain value:
$$
c_p(-\sigma) = -c_p(\sigma).
$$

Chain addition is defined by addition of chain values: if $c_{p_1}, c_{p_2}$ are $p$-chains, then $(c_{p_1} + c_{p_2})(\sigma) = c_{p_1}(\sigma) + c_{p_2}(\sigma)$, for each $\sigma \in X_p$. The resulting group is denoted $C_p(X;G)$. When clear from the context, the group $G$ is often left implied, writing $C_p(X)$.

Let $\sigma$ be an oriented cell in $X$ and $g \in G$. The \emph{elementary chain} whose value is $g$ on $\sigma$, $-g$ on $-\sigma$ and 0 on any other cell in $X$ is denoted $g\sigma$. Each chain can then be written in a unique way as a sum of elementary chains.
With abuse of notation, we do not distinguish between cells and \emph{singleton chains} (i.e., the elementary chains whose value is $1\,\sigma$ for some cell $\sigma$), used as elements of the standard bases of chain groups.

         Chains are often thought of as attaching orientation and multiplicity to cells: if coefficients are taken from the group {$G = (\{-1,0,1\}, +) \simeq (\Z_3,+)$}, then cells can only be discarded or selected, possibly inverting their orientation (see~\cite{ieee-tase}).
It is useful to select a conventional choice to orient the singleton chains (single cells) automatically. 0-cells are considered all positive. The $p$-cells, for $1\leq p\leq d-1$, can be given a \emph{coherent (internal) orientation} according to the orientation of the first $(p-1)$-cell in their \emph{canonical} (sorted on facet indices) representation. Finally, a $d$-cell may be oriented as the sign of its oriented volume.

\subsubsection{{Cochains}}
\label{sec:cochains}

          Cochains are dual to {chains}: a $p$-cochain attaches additively an element of the group $G$ to each $p$-chain. Singleton cochains, attaching the identity element 1 to {singleton chains}, form the standard bases of cochains groups. The groups of $p$-chains and those of $p$-cochains may be identified with each other in {infinitely different ways}. Different legitimate identifications, while affecting the \emph{metric} properties of the chain-cochain complex~\cite{DiCarlo:2009:DPU:1629255.1629273}, do not change the \emph{topology} of finite complexes. Since we shall only use the topological properties of finite chain-cochain complexes, we feel free to chose the simplest possible identification, obtained by identifying the elements of the standard chain bases with the corresponding elements of the standard cochain bases. In this paper, we take for granted that chains and cochains are identified in this trivial way.

\subsubsection{{Examples}}

{
For reader's convenience, we include here some very simple examples of cellular complexes, and some basic computations with boundary and coboundary operators. In Figure~\ref{fig:ex0-abc} we show the same space partition into cells of dimension 0 ($\nu_i$, $1\leq i\leq 6$), dimension 1 ($\eta_j$, $1\leq j\leq 8$), and dimension 2 ($\gamma_k$, $1\leq k\leq 3$), associated with different additive groups of coefficients.
}

\begin{figure}[htbp] %  figure placement: here, top, bottom, or page
   \centering
   \includegraphics[width=0.23\textwidth]{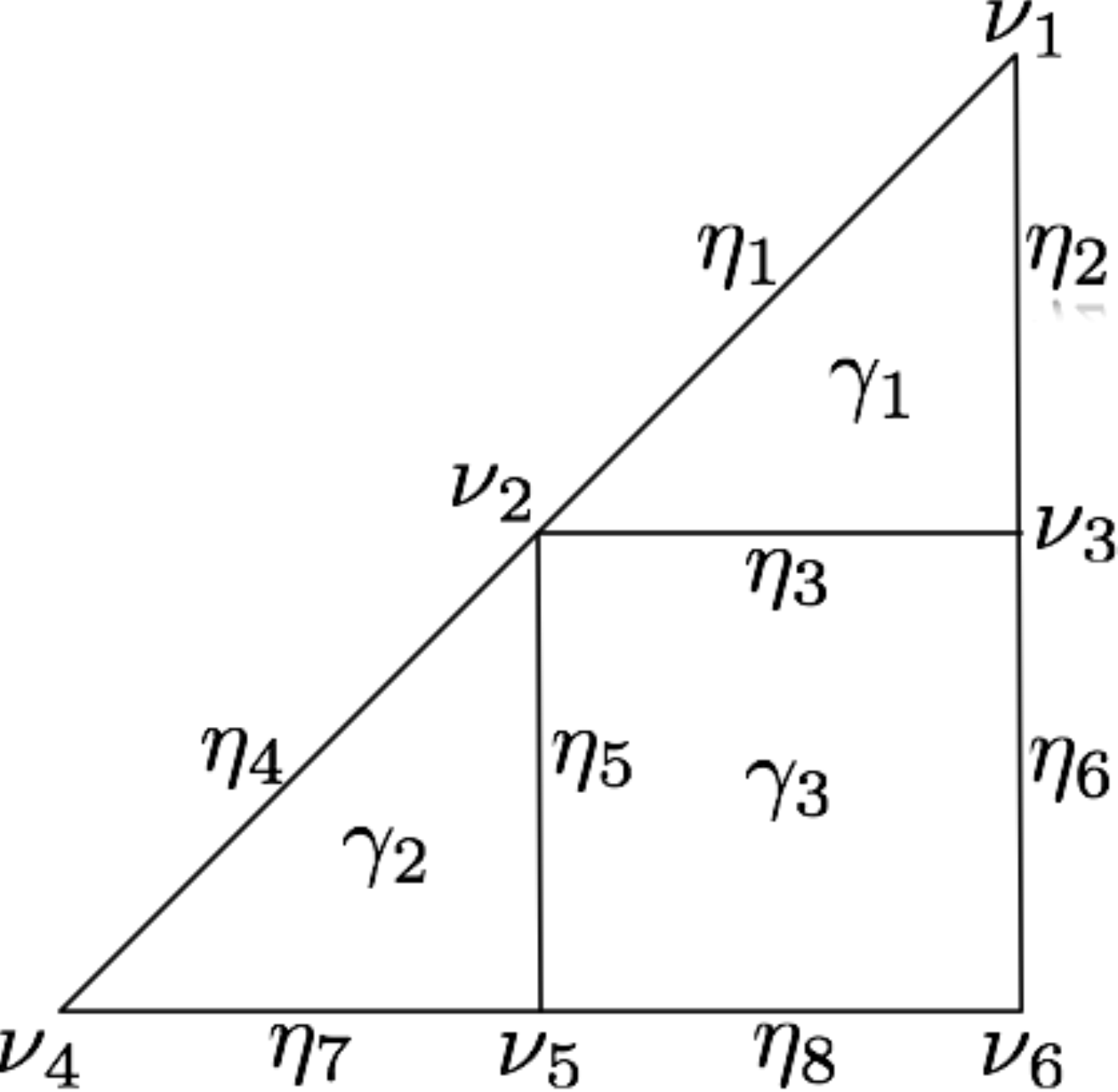}
   \hfill
   \includegraphics[width=0.23\textwidth]{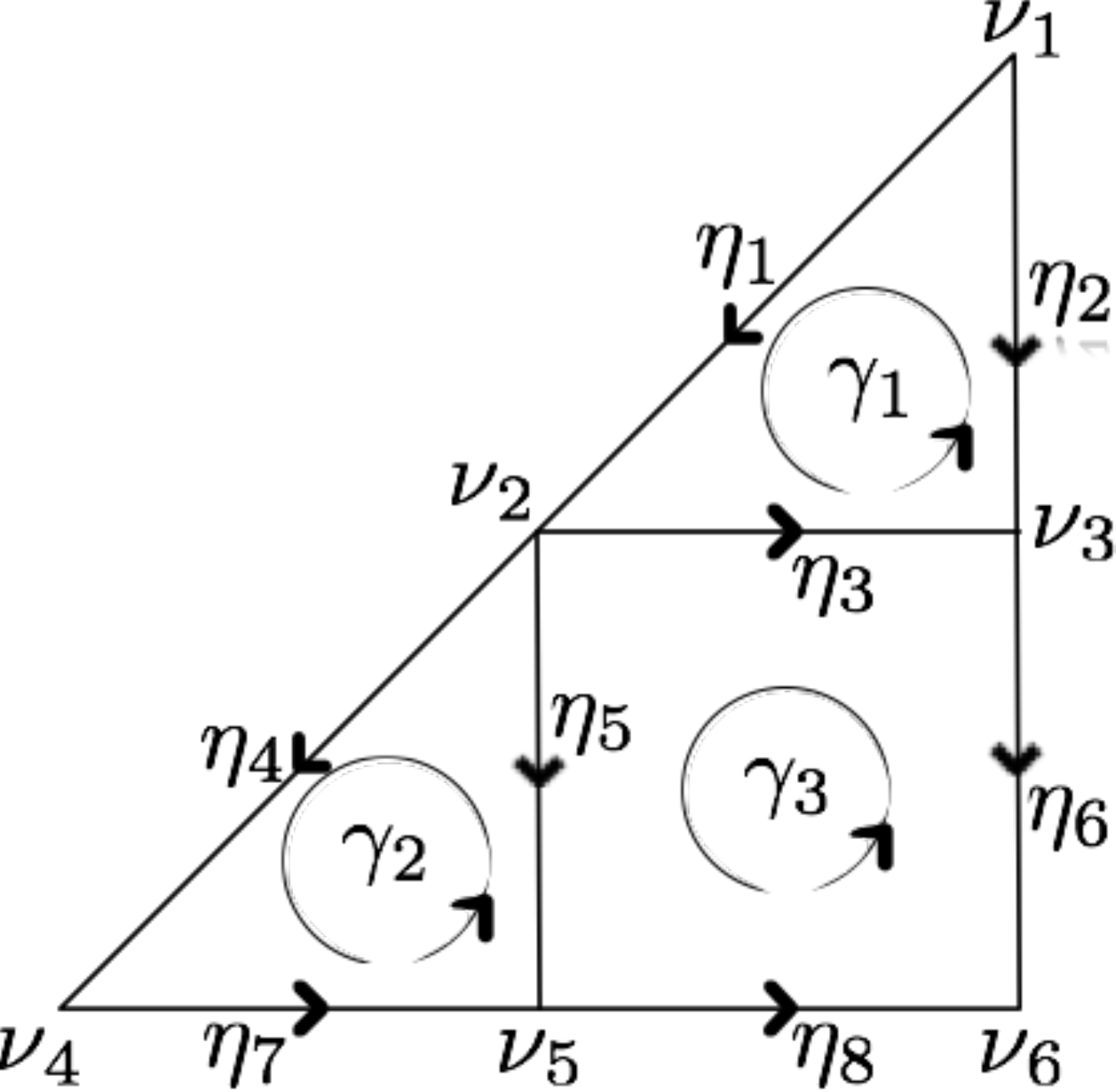}
   \hfill
   \includegraphics[width=0.23\textwidth]{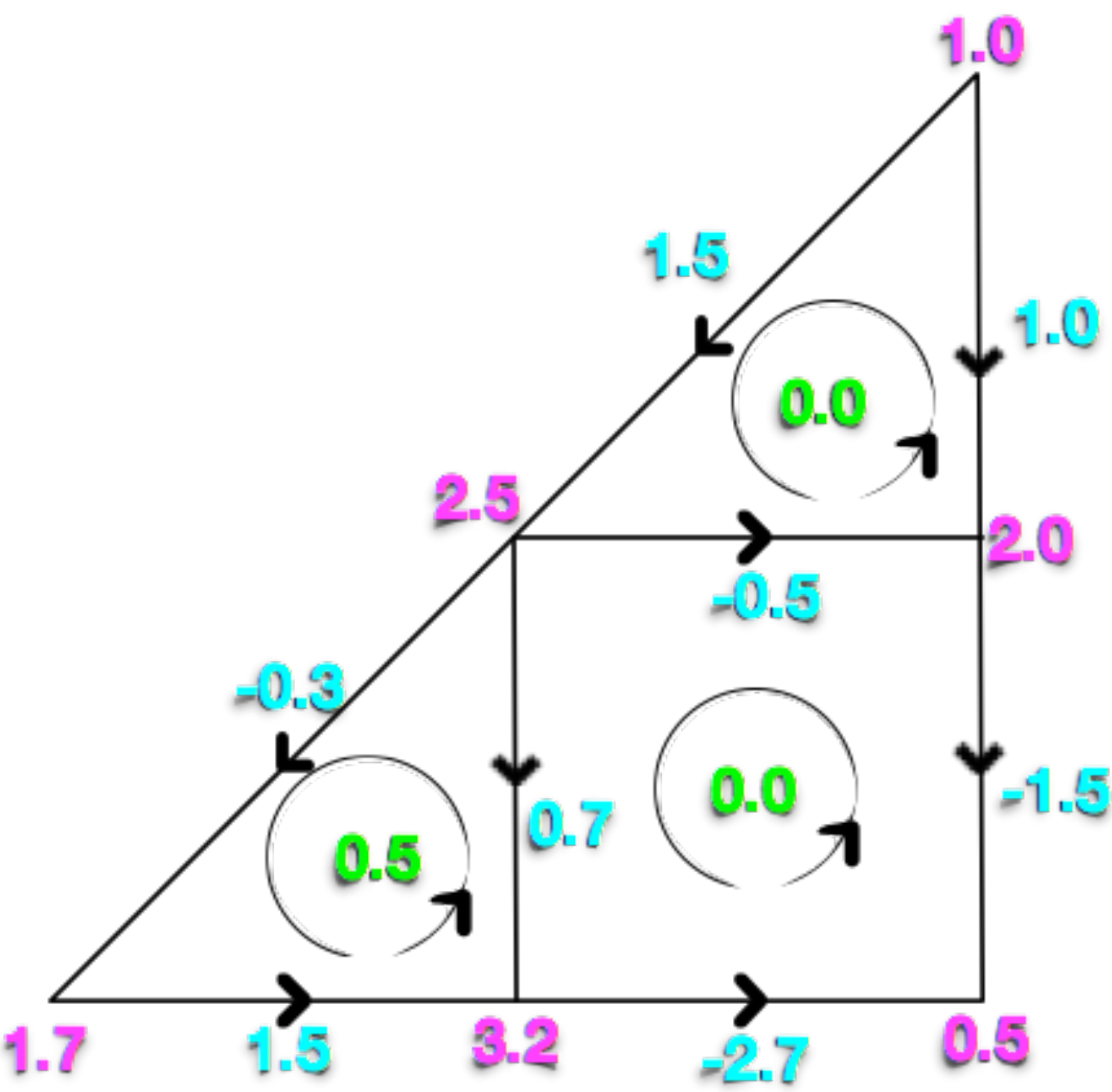}
   \hfill
   \includegraphics[width=0.23\textwidth]{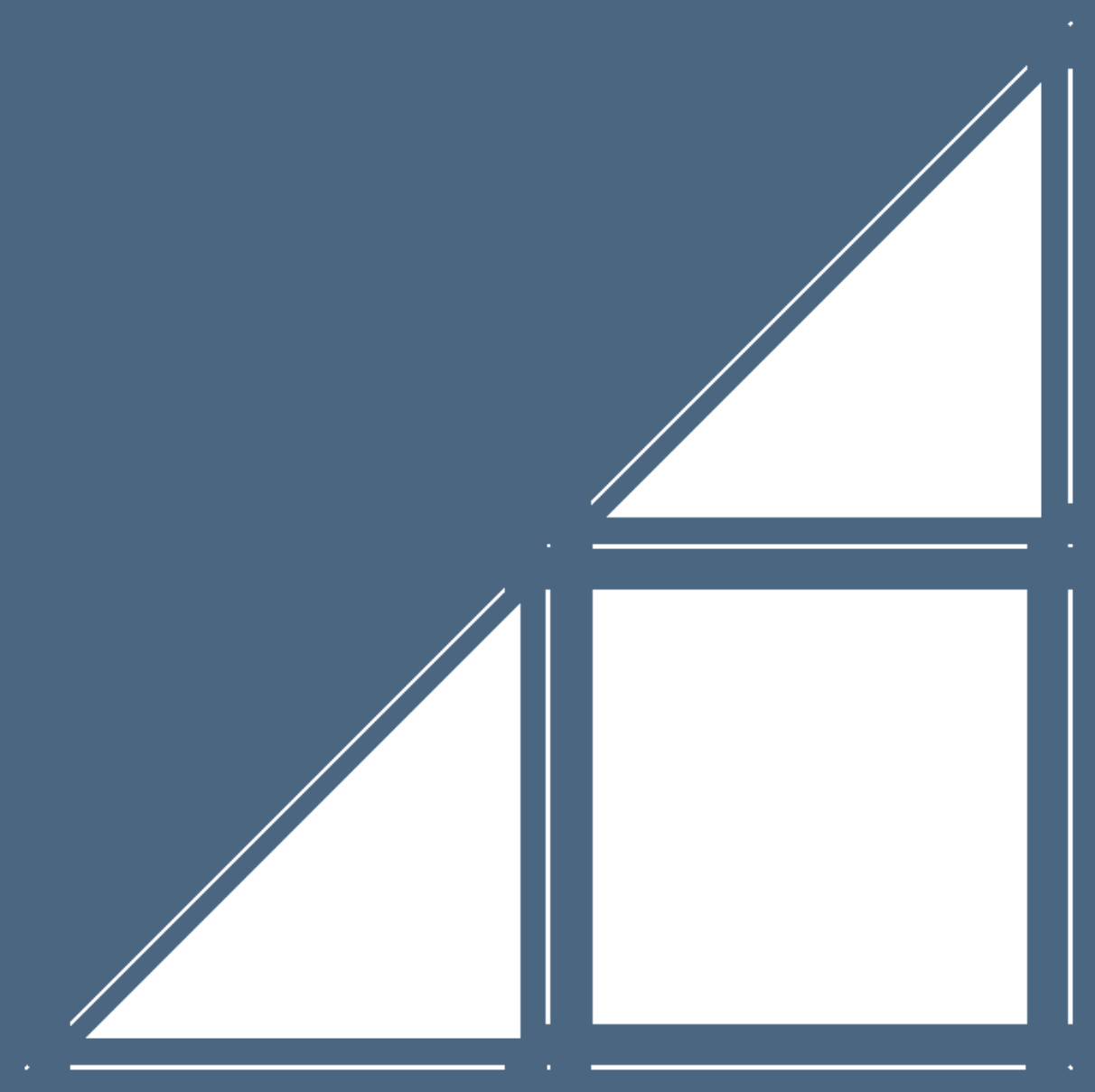}
   \caption{{
Cellular complexes with 0-cells {in} $\Lambda_0=\{\nu_1, \ldots, \nu_6\}$,  1-cells {in} $\Lambda_1=\{\eta_1, \ldots, \eta_8\}$, and 2-cells {in} $\Lambda_2=\{\gamma_1, \gamma_2, \gamma_3\}$: (a) non-oriented complex, with cell coefficients in $\Z_2=\{0,1\}$; (b) oriented complex, with cell coefficients in $G=\{-1,0,+1\}$; (c) oriented complex, with cell coefficients in $\R$, using different colors for the maps from $\Lambda_0$,  $\Lambda_1$, and  $\Lambda_2$ to $\R$. To interpret the real numbers here, see Example~\ref{examplee}; (d) the exploded 2-complex with $|\Lambda_0| + |\Lambda_1| + |\Lambda_2| = 6+8+3$ cells.}
}
   \label{fig:ex0-abc}
\end{figure}

{
\begin{example}[Chains]\label{examplea}
Unoriented chains take coefficients from $\Z_2=\{0,1\}$. E.g., {a} 0-chain $c\in C_0$ shown in Figure~\ref{fig:ex0-abc}a is given by $c = 1\nu_{1} + 1\nu_{2} + 1\nu_{3} + 1\nu_{5}$. Hence, the coefficients associated to all other cells are zero. The coordinate vector of $c$ with respect to the (ordered) basis $(\nu_1, \nu_2, \ldots, \nu_6)$ is hence $[1,1,1,0,1,0]^t$.
Analogously for the 1-chain $d\in C_1$ and the 2-chain $e\in C_2$, written by dropping the 1 coefficients, as $d = \eta_2 + \eta_3 + \eta_5$ and $e = \gamma_1 + \gamma_3$, with coordinate vectors $[0,1,1,0,1,0,0,0]^t$ and $[1,0,1]^t$, respectively.
\end{example}
}

{
\begin{example}[Orientation]\label{examplec}
In Figure~\ref{fig:ex0-abc}b it is shown an oriented version of the cellular complex $\Lambda = \Lambda_0 \cup \Lambda_1 \cup \Lambda_2$, where the 1-cells are oriented from the vertex with lesser index to the vertex with greater index, and where {all 2-cells are counterclockwise oriented}. Let us note that the orientation of every cell may be fixed arbitrarily, since can always be reversed by the associated coefficient, that is now taken from the set  $\{-1,0,+1\}$.  So, the oriented 1-chain having first vertex $\nu_1$ and last vertex $\nu_5$ is now given as $d' = \eta_2 -\eta_3 + \eta_5$, with coordinate vector $[0,1,-1,0,1,0,0,0]^t$
\end{example}
}

{
\begin{example}[Boundary]\label{exampled}
The boundary operators are maps $C_p\to C_{p-1}$, with $1\leq p\leq d$, hence for a 2-complex we have two operators, denoted as $\partial_2: C_2 \to C_1$ and $\partial_1: C_1 \to C_0$, respectively. 
Since they are linear maps between linear spaces may be represented by matrices of coefficients $[\partial_2]$ and $[\partial_1]$ from the corresponding groups. For the unsigned and the signed case (Figures~\ref{fig:ex0-abc}a and ~\ref{fig:ex0-abc}b) we have, respectively:
\begin{equation}
[\partial_2] = \mat{
1 & 0 & 0\\
1 & 0 & 0\\
1 & 0 & 1\\
0 & 1 & 0\\
0 & 1 & 1\\
0 & 0 & 1\\
0 & 1 & 0\\
0 & 0 & 1\\
}\ ,
\ \mbox{{and}}\quad
\arraycolsep=2.9pt%\def\arraystretch{2.2}
[\partial'_2] = \mat{
1 & 0 & 0\\
-1 & 0 & 0\\
1 & 0 & -1\\
0 & 1 & 0\\
0 & -1 & 1\\
0 & 0 & -1\\
0 & 1 & 0\\
0 & 0 & 1\\
}\ .
\label{eq:delta2}
\end{equation}
Analogously, for the unsigned and the signed matrices of the $\partial_1$ operator, we have: 
\begin{equation}
[\partial_1] = \mat{
1 & 1 & 0 & 0 & 0 & 0 & 0 & 0\\
1 & 0 & 1 & 1 & 1 & 0 & 0 & 0\\
0 & 1 & 1 & 0 & 0 & 1 & 0 & 0\\
0 & 0 & 0 & 1 & 0 & 0 & 1 & 0\\
0 & 0 & 0 & 0 & 1 & 0 & 1 & 1\\
0 & 0 & 0 & 0 & 0 & 1 & 0 & 1\\
}\ ,
\ \mbox{{and}}\quad
\arraycolsep=2.9pt%\def\arraystretch{2.2}
[\partial'_1] = \mat{
-1 & -1 & 0 & 0 & 0 & 0 & 0 & 0\\
1 & 0 & -1 & -1 & -1 & 0 & 0 & 0\\
0 & 1 & 1 & 0 & 0 & -1 & 0 & 0\\
0 & 0 & 0 & 1 & 0 & 0 & -1 & 0\\
0 & 0 & 0 & 0 & 1 & 0 & 1 & -1\\
0 & 0 & 0 & 0 & 0 & 1 & 0 & 1\\
}\ .
\label{eq:delta1}
\end{equation}
As a check, let compute the 0-boundary of the coordinate representations of the unsigned 1-chains $[d] = [0,1,1,0,1,0,0,0]^t$ and {of} the signed 1-chain $[d'] = [0,1,-1,0,1,0,0,0]^t$:
\[
[\partial_1][d]^t\ {mod\ 2} = [1,0,0,0,1,0]^t = \nu_1 + \nu_5 \in C_0,
\]
where the matrix product is computed $mod\ 2$, and where
\[
[\partial'_1][d']^t = [-1,0,0,0,1,0]^t = \nu_5 - \nu_1 \in C'_0.
\]
\end{example}
}

{
\begin{example}[Dual cochains]\label{exampleb}
The concept of cochain $\phi^p$ in a group $C^p$ of linear maps from chains $C_p$ to $\R$ allows for the association of numbers not only to single cells, as done by chains, but also to assemblies of cells. 
{A cochain is hence the association of  discretized subdomains of a cell complex with a global numeric quantity, usually resulting from a discrete integration over a chain. }
Each cochain $\phi^p \in C^p$ can be seen as a linear combination of the {
unit $p$-cochains $\phi_1^p,\ldots, \phi_k^p$}\footnote{{Coincident with $\eta^1_p,\ldots, \eta^k_p$ by identification of primal and dual bases.}} whose value is 1 on a unit $p$-chain and 0 on all others.
The evaluation of a real-valued cochain is denoted as a duality pairing, in order to stress its bilinear property:
\[
\phi^p ( c_p ) = \langle \phi^p, c_p \rangle .
\]
This mapping is orientation-dependent, and linear with respect to the assemblies of cells, modeled by chains~\cite{hinzl:thesis:2007}. Also, remember here that we identify chain and cochain spaces (see Section~\ref{sec:cochains}).
\end{example}
}

{
\begin{example}[Coboundary]\label{examplee}
The coboundary operator $\delta^p : C^p \to C^{p+1}$  acts on $p$-cochains as the dual of the boundary operator $\partial_{p+1}$ on ($p+1$)-chains. For all $\phi^p \in C^p$ and $c_{p+1} \in C_{p+1}$, it is:
\[
\langle \delta^p \phi^p, c_{p+1} \rangle = \langle \phi^p, \partial_{p+1} c_{p+1} \rangle.
\]
 Recalling that chain-cochain duality means integration, the reader will recognize this defining property as the combinatorial archetype of Stokes' theorem. 
 It is possible to see~\cite{DiCarlo:2009:DPU:1629255.1629273} that since we use dual bases, matrices representing dual operators are the transpose of each other: for all $p = 0,\ldots,d-1$, 
\[
  [\delta^p]^t = [\partial_{p+1}] .
\]
In Figure~\ref{fig:ex0-abc}c, coefficients from $\R$ are associated to elementary (co)chains, as resulting from the evaluation of cochain functions on elementary chains. 
When cochain coefficients are taken from $G=\{-1,0,+1\}$, we get, from Eqs.~(\ref{eq:delta2}) and~(\ref{eq:delta1}):
\[
[\delta^1] = [\partial'_2]^t\quad\mbox{and}\quad [\delta^0] = [\partial'_1]^t
\]
so that, with $f = [0,0,0,0,1,0,0,0]\in C^1\equiv C_1$, we get
\[
[\delta^1][f]^t = [0,-1,1]^t = \gamma_3-\gamma_2 \in C^2\equiv C_2,
\] 
as you can check by looking to Figure~\ref{fig:ex0-abc}b.
\end{example}
}

\subsection{Linear Algebraic Representation}

LAR (Linear Algebraic Representation)~\cite{Dicarlo:2014:TNL:2543138.2543294} is a 
representation scheme~\cite{Requicha:1980:RRS:356827.356833} for
geometric and solid modeling. The domain of the scheme is provided by
 \emph{cellular complexes}, while its codomain is
the set of \emph{sparse matrices}. 
The topology in LAR is given just by the
binary characteristic matrix $M_d$ of $d$-cells for polytopal complexes, or by the pair
$M_d, M_{d-1}$ for more general (non convex {and/or} non contractible) types of cells~\cite{Dicarlo:2014:TNL:2543138.2543294}. 

The LAR polyhedral domain coincides with complexes of
connected $d$-cells, including non-convex and multiply connected cells. 
Sparse matrices are 
stored in memory using either the CSR (Compressed
Sparse Row) or the CSC (Compressed Sparse Column) memory format~\cite{gemmexp}. 
Note that the sparsity of matrices representing a cellular complex grows quadratically with the number $n$ of cells, so that the sparse  matrix representation of a cellular complex is $O(n)$.

The very general shape allowed for cells makes the LAR scheme notably
appropriate for biomedical applications like the modeling of neuronal tissues~\cite{doi:10.1080/16864360.2016.1168216,ClementiSSPP-CAD16}
and the solid modeling of buildings and their components~\cite{visigrapp17:cvdlab}. E.g.,
the whole facade of a building can be described by a single 3-cell
in its solid model. Also, the algebraic foundation of LAR allows not
only for fast queries about incidence and adjacency of cells, but also
for extracting---via fast SpMV computational kernels---the boundary or coboundary 
 of \emph{any} 3D subset of the building model.

LAR provides a direct management of all subsets
of cells and their physical properties through the linear spaces of
\emph{chains} induced by the model partitioning, and their dual spaces
of \emph{cochains}. 
The linear operators of \emph{boundary} and
\emph{coboundary} between such linear spaces, suitably implemented by
sparse matrices, directly supply the discrete differential operators of \emph{gradient},
\emph{curl} and \emph{divergence}, while their combination gives the
\emph{Laplacian}~\cite{ieee-tase}. A word of warning is in order here: contrary to gradient, curl and divergence, the Laplacian operator substantially depends on metric properties, hence on the specific chain-cochain identification.

\subsection{LAR of a cellular complex}\label{representation}

{A} common representation of a $d$-complex in both commercial and academic systems is some---often very intricate---data structure  storing its $d$-boundary. For 3D meshes discretising the space of a simulation, or {when the representation scheme is some other cell decomposition}~\cite{Requicha:1980:RRS:356827.356833}, the object of the storage structure is the set of either $d$- or $(d-1)$-cells, or both. 
The representation of cells is normally supplemented by the storage of subsets of the incidence relations between topological elements, often paired with some ordering though linked lists of pointers.  

{The details of such data structures vary greatly depending on the type  of cells, dimension, or specific intended applications,  leading to many specialized and ad hoc computational procedures that must be redesigned for each new data structure.    For example,  boundary evaluation or boundary traversal algorithms depend on many specific assumptions in the data structure and do not generalize across dimensions.}

LAR is a more concise and simpler representation that supports the same queries and operators (incidence, boundary, coboundary, etc.)  without additional computational overhead.
{In Appendix~\ref{sec:lar-examples} we show some very simple examples of LAR definition of cellular complexes and computation of their properties. }
With the LAR scheme, only the characteristic functions of $d$-cells as vertex subsets are  necessary for representing polytopal complexes. Examples include simplicial, cubical, and Voronoi complexes. 
For more general cell complexes, including non-convex and non-contractible cells, useful in some applications, both the $d$ and $(d-1)$-skeletons are needed~\cite{Dicarlo:2014:TNL:2543138.2543294}. 
In all cases, boundary and coboundary operators and all topological queries are supported by computational kernels for sparse matrix multiplication implemented on GPUs.

It is worth noting that this {dimension-independent} representation supports a description of cellular and chain complexes, including topological operators, \emph{without special data structures}, but simply as either \emph{signed integers} or \emph{arrays of signed integers}, either dense or sparse depending on the size of the complex. 
The LAR scheme enjoys several useful properties, including simplicity, compactness, readability, and direct usability for calculus. The reader is referred to~\cite{Dicarlo:2014:TNL:2543138.2543294} for further discussion and details. 

\begin{example}[2D cellular complex] % example 0
{
The LAR of the simple example in Figure~\ref{fig:ex0-abc} is given below. Here, arrays \texttt{V}, \texttt{EV}, \texttt{FV} contain, respectively, the coordinates of vertices, the indices of cell vertices for edges, and the same  for faces.}
\begin{verbatim}
V = [[1,1],[0.5,0.5],[1,0.5],[0,0],[0.5,0],[1,0]]
EV = [[0,1],[0,2],[1,2], [1,3],[1,4],[2,5], [3,4],[4,5]]
FV = [[0,1,2],[1,3,4],[1,2,4,5]]
\end{verbatim}
{
It is worth noting that, by using the \emph{Merge} algorithm introduced in this paper, the \emph{{complete representation}} of geometry and topology can be reduced to }
\begin{verbatim}
V = [[1,1],[0.5,0.5],[1,0.5],[0,0],[0.5,0],[1,0]]
EV = [[0,1],[0,2],[1,2], [1,3],[1,4],[2,5], [3,4],[4,5]]
\end{verbatim}
{
since the 2-cells $\texttt{FV}$ may {be computed by} the plane arrangement $\mathcal{A}(\mathcal{E})$ induced by the set $\mathcal{E}$ of line segments codified here by $\texttt{V}$ and $\texttt{EV}$ arrays. See, e.g., the~Example~\ref{example3}.}
\end{example}

\section{A gentle introduction}
\label{sec:gentle-intro}

{To perform a Boolean set operation on two or more solids represented by their boundaries,} you need to compute how two (or more) cellular complexes (boundaries of solids) break each other into pieces, and select the 3D pieces that enter the Boolean result. {We deal with the first step of this process, which is commonly called a \emph{merge operation}. } This problem becomes much harder when using a three-dimensional mesh decomposing your solids, i.e., when using 3D decompositions (see, e.g., \cite{Dobkin:1987:PMT:41958.41967}). {Whereas these problems have treated independently in the past,  a unified approach is preferable in order to deal with increasingly complex and diverse models that may include variety of cellular models.   For example,} you may require to combine two or more 3D meshes with open or closed surfaces which partition the interesting parts of a model.

In a few words, you may often need to merge two or more cellular complexes, even of different dimensions, and compute the resulting space arrangement. In discrete geometry, an arrangement is the decomposition of the $d$-dimensional linear, affine, or projective space into connected open cells of lower dimensions, induced {by an intersection of a} finite collection of geometric objects.
This computation, using simple methods well founded on basic mathematical operations of algebraic topology, is the topic of this paper. {We} adopt here a dimension-independent notation, since concepts and methods are the same in 2D and in 3D, so that their implementation can be unified and made simpler, and even used for applications in higher dimensions.

The main idea of this paper is to reduce the intersections of higher dimensional discrete varieties to a number of (much) simpler intersections, between pairs of lines in the 2D plane. For this purpose, we (i) reduce the intersection of a $(d-1)$-cell $\sigma$ with all the possibly intersecting others to an independent set of intersections of their 1-faces with the $z=0$ plane, (ii) calculate the resulting (easily computed) arrangements in 2D, and finally (iii) come back to higher dimension, while at the same time identifying the possibly coincident instances of (back-transformed) faces of $\sigma$, that were independently generated from each other.

A cartoon chronicle of the computation of the {arrangement of $\E^3$} {(space decomposition including the unbounded exterior cell)} generated by two very simple cellular 2-complexes, i.e., by the boundaries of two translated and rotated unit cubes, is shown in Figures~\ref{fig:cartoon-1}, \ref{fig:cartoon-2}, \ref{fig:cartoon-3}, and~\ref{fig:cartoon-4}.

\begin{figure}[htbp] %  figure placement: here, top, bottom, or page
   \centering
\includegraphics[height=0.248\linewidth,width=0.248\linewidth]{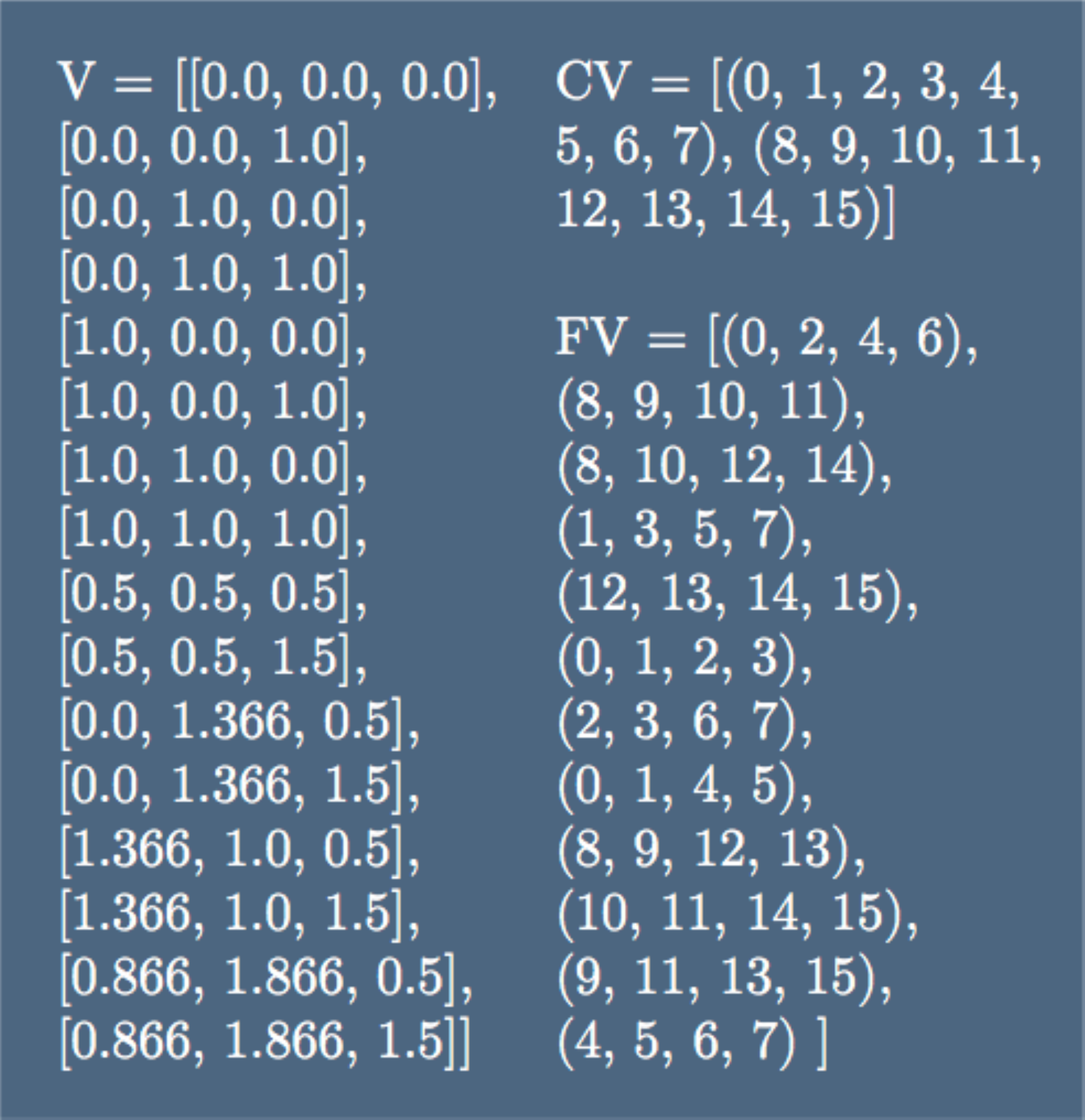}\hfill%
\includegraphics[height=0.248\linewidth,width=0.248\linewidth]{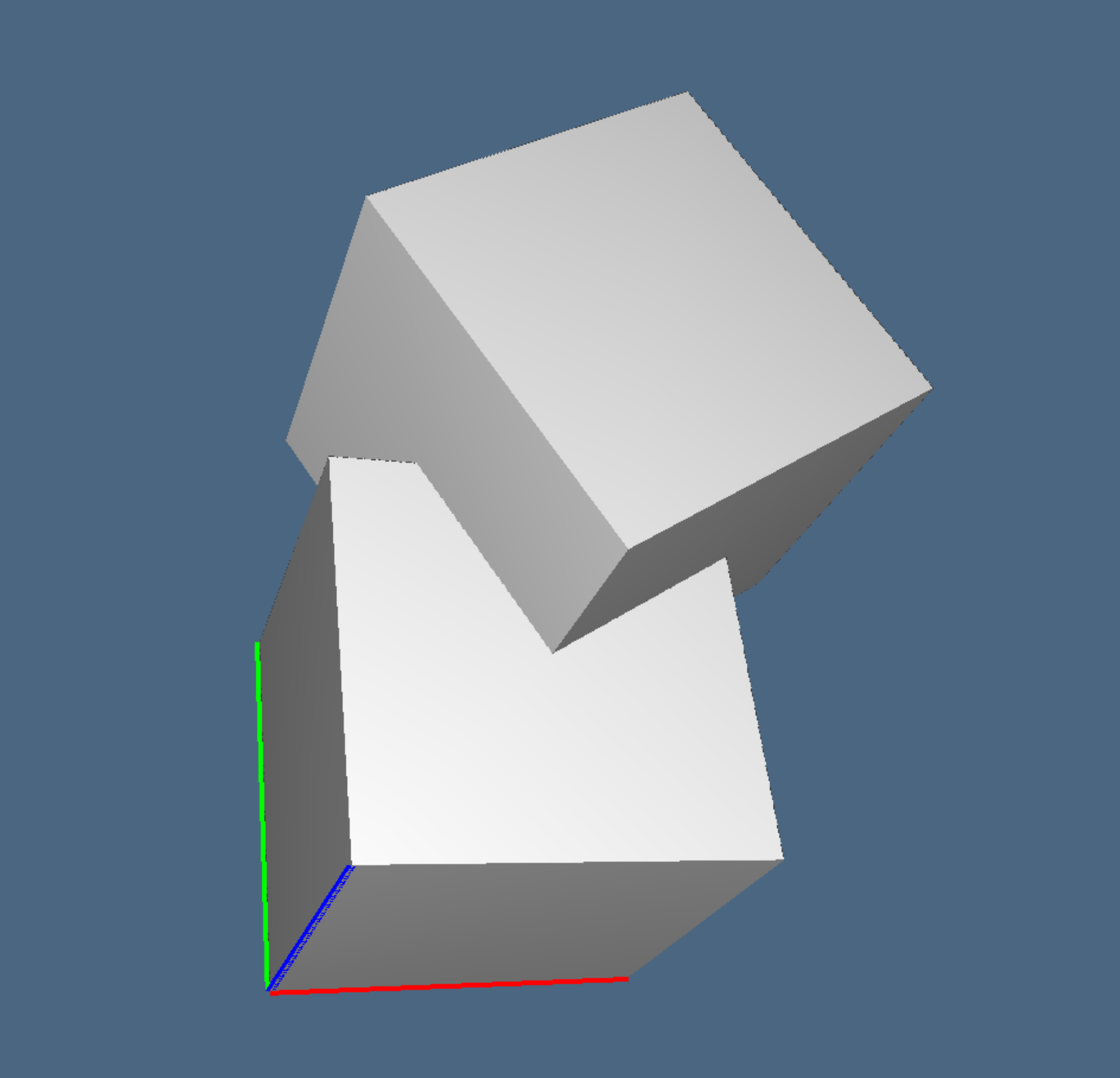}\hfill%
\includegraphics[height=0.248\linewidth,width=0.248\linewidth]{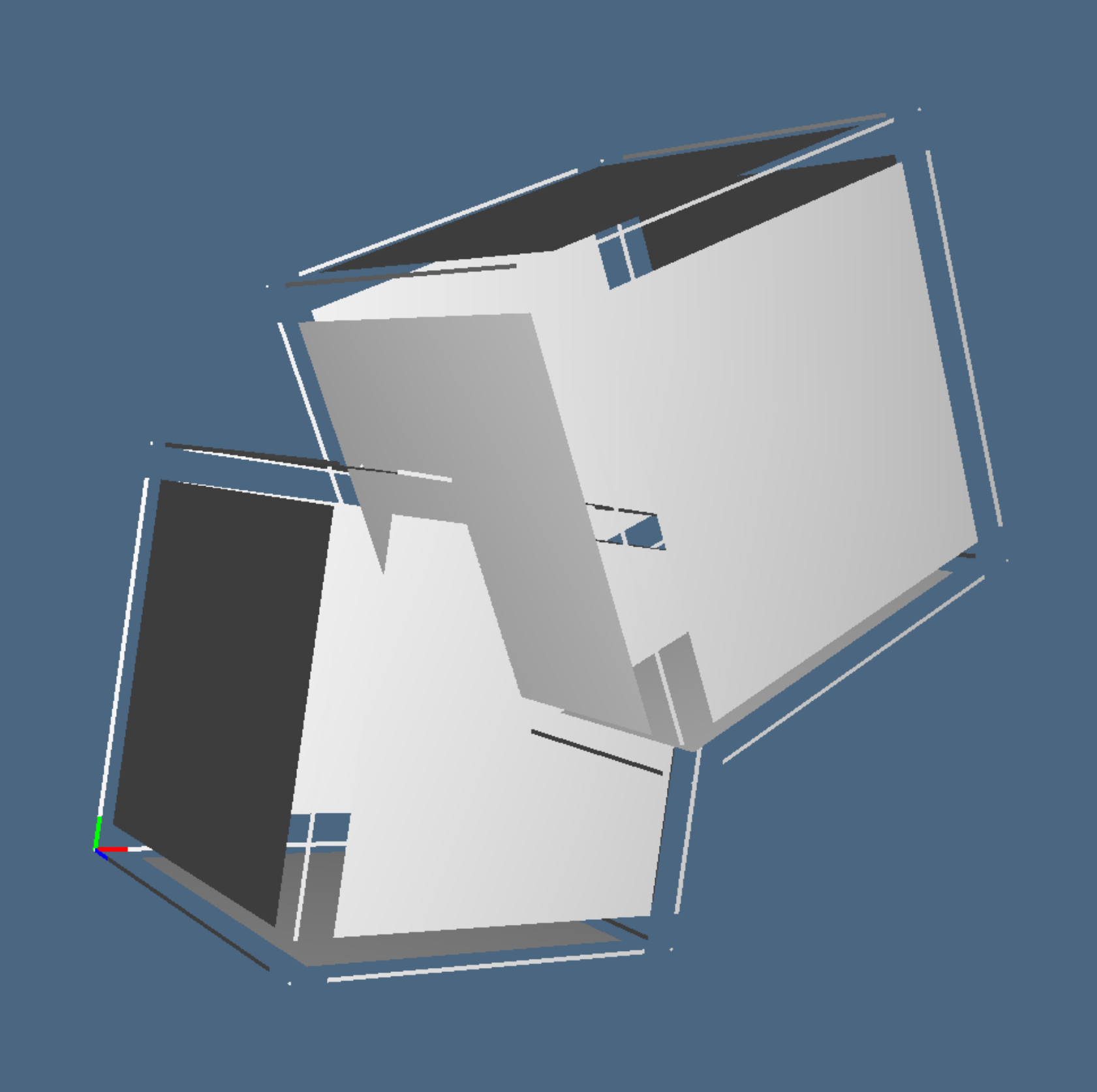}\hfill%
\includegraphics[height=0.248\linewidth,width=0.248\linewidth]{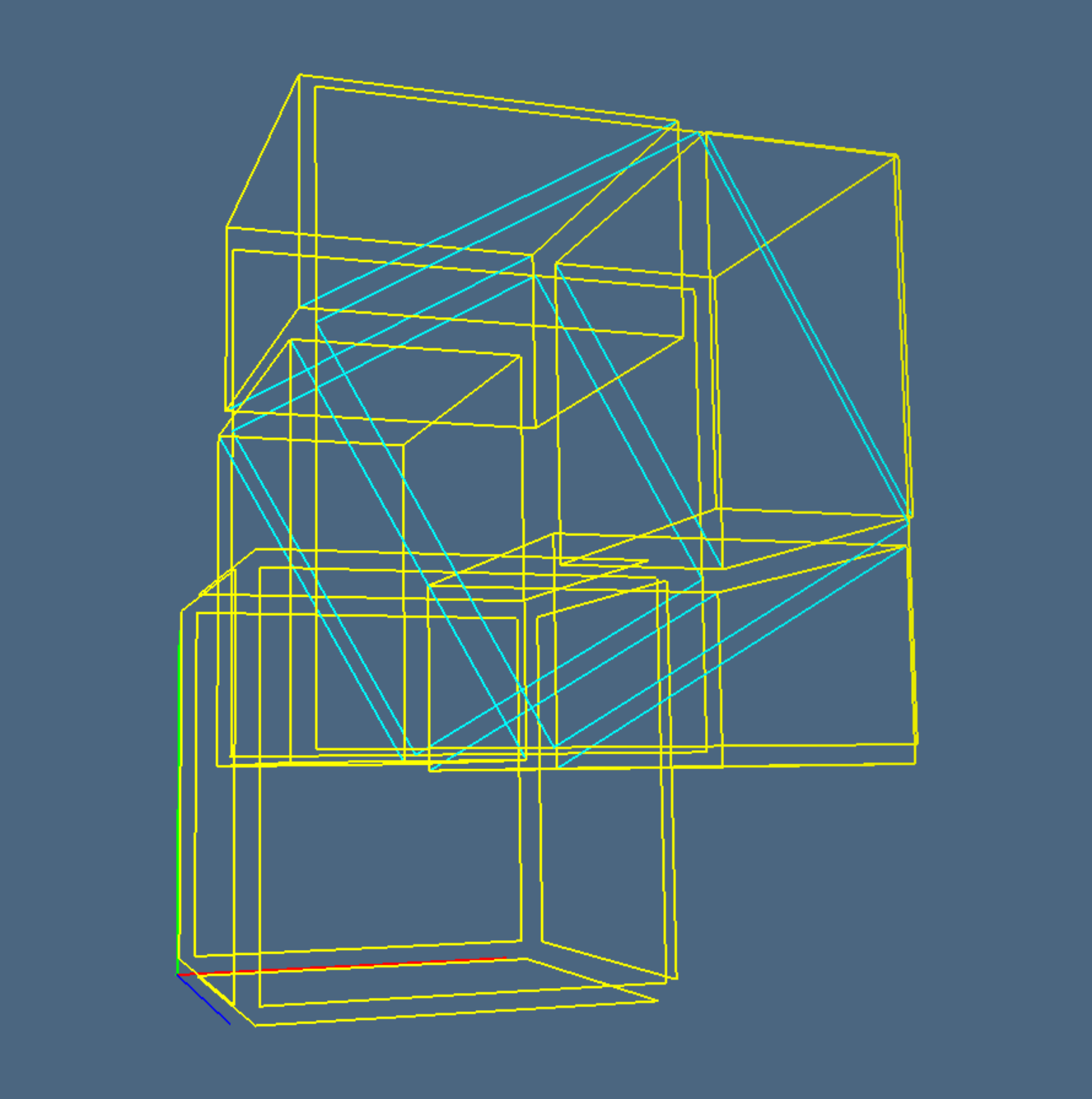}\hfill%

\vspace{-1.5mm}
{\footnotesize\hspace{.12\textwidth}(a)\hfill(b)\hfill(c)\hfill(d)\hspace{.12\textwidth}}
\vspace{-3.5mm}
   \caption{ Input of two 3D cellular complexes $\{X^1_3, X^2_3\}$ in $\E^3$: 
   (a) LAR input data: V := vertices; CV := 3-cells; FV := 2-cells; 
   (b) rotated and translated unit cubes {$\mathcal{S}_3 = \{\, X^h_3,\, h\in \{1,2\}\,\}$} in $\E^3$; 
   (c) the (exploded) set of 2-complexes $B = \bigcup_{h\in\{1,2\}} X^h_2$. Note that the 2-cells in 3-space have no orientation; 
   (d) image of the (exploded) spatial index over 2-cells, using {(yellow)} {3D containment boxes} \& three 1D interval trees {(not included in the image)}. \label{fig:cartoon-1} }

   \vspace{3mm}

\includegraphics[height=0.248\linewidth,width=0.248\linewidth]{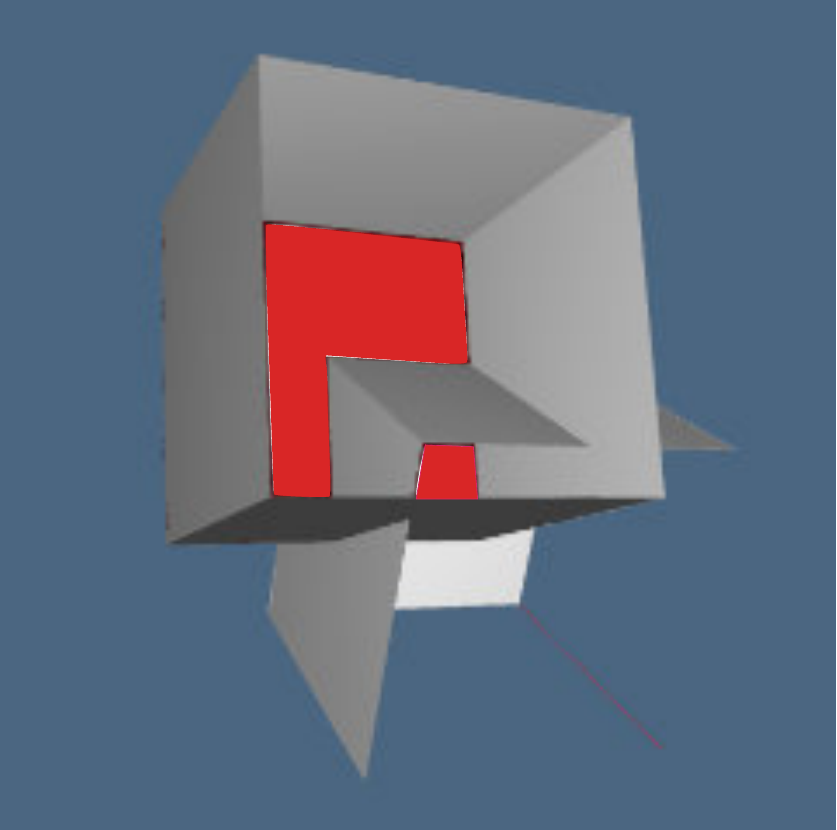}\hfill%
\includegraphics[height=0.248\linewidth,width=0.248\linewidth]{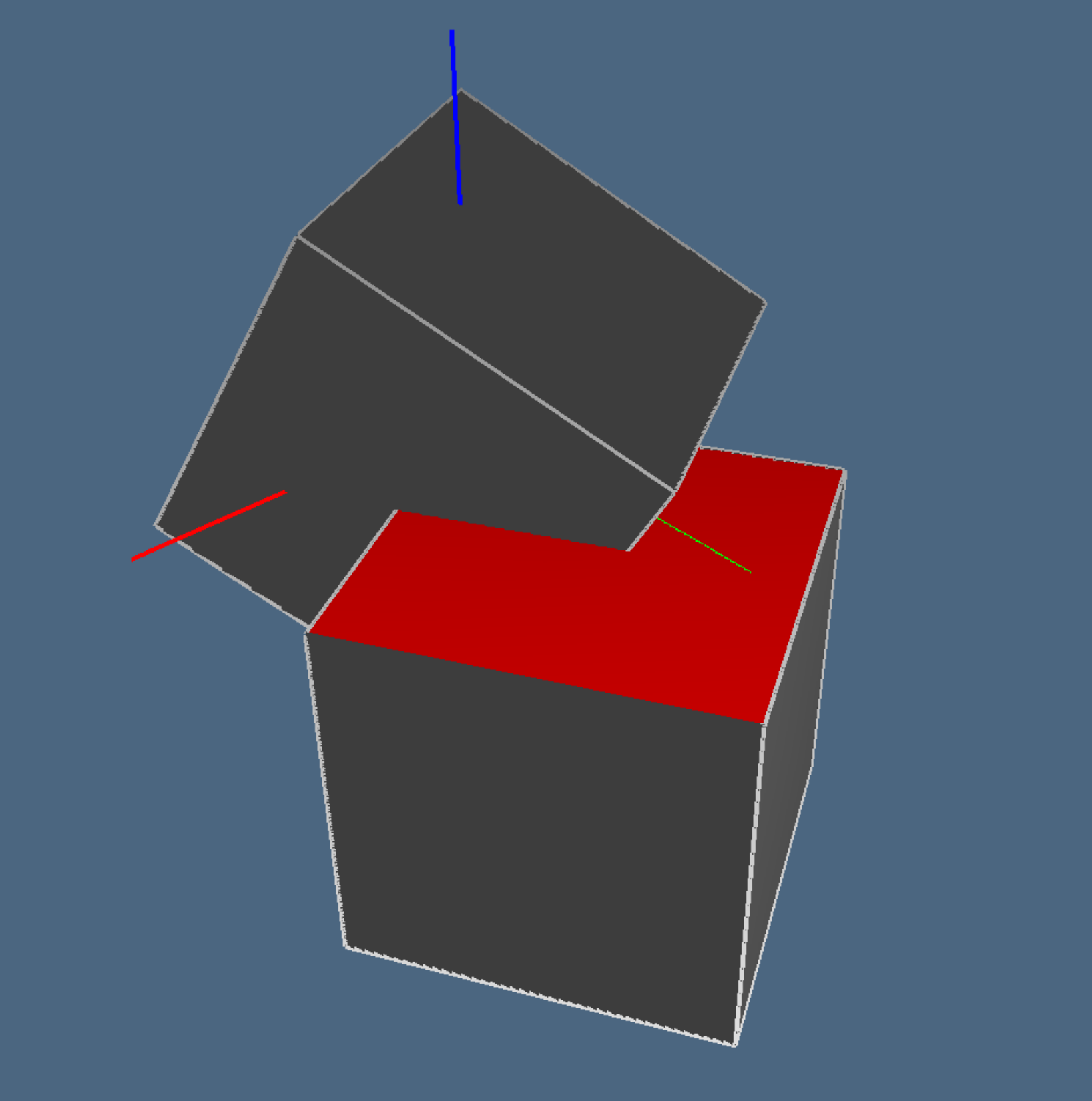}\hfill%
\includegraphics[height=0.248\linewidth,width=0.248\linewidth]{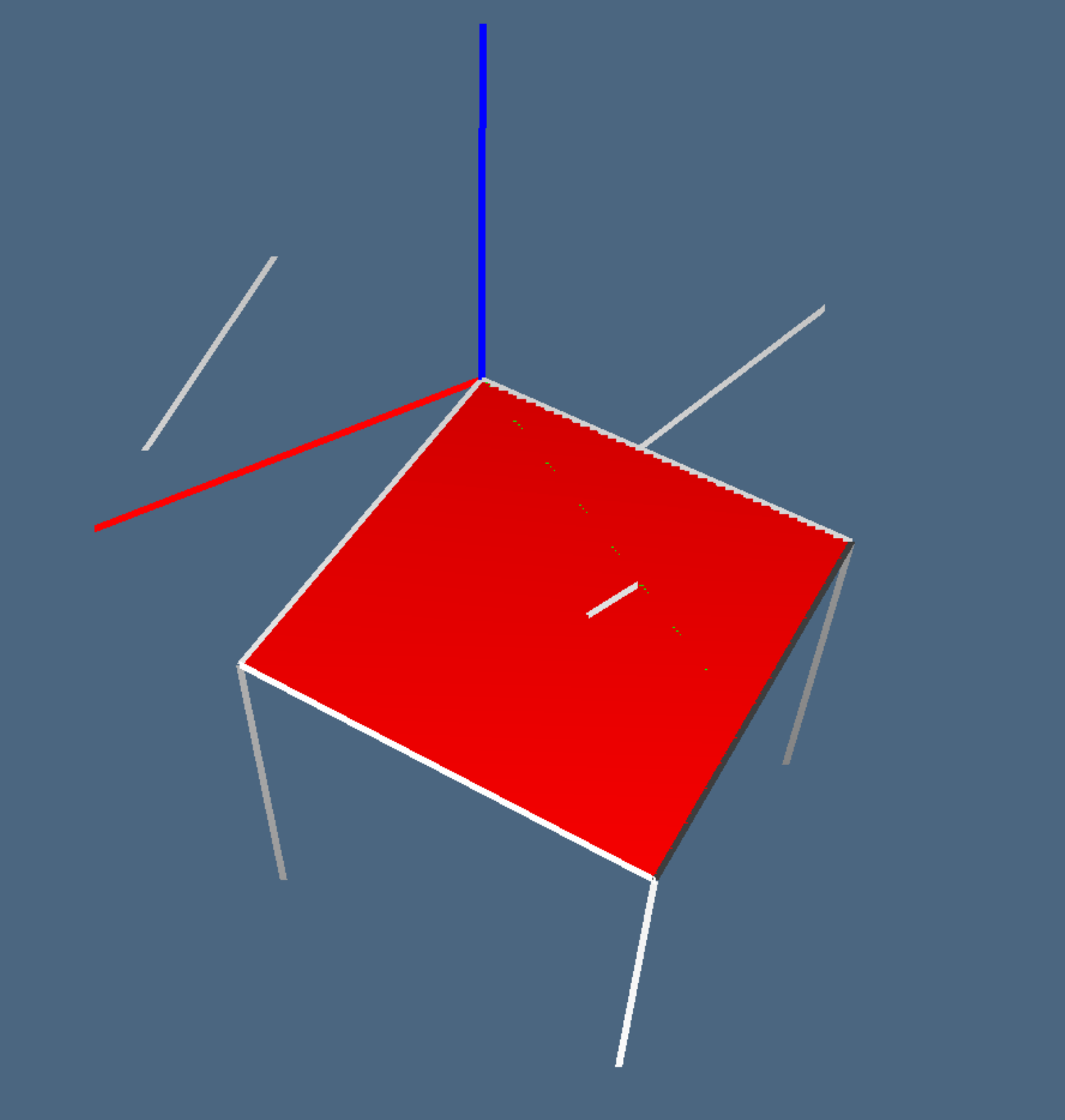}\hfill%
\includegraphics[height=0.248\linewidth,width=0.248\linewidth]{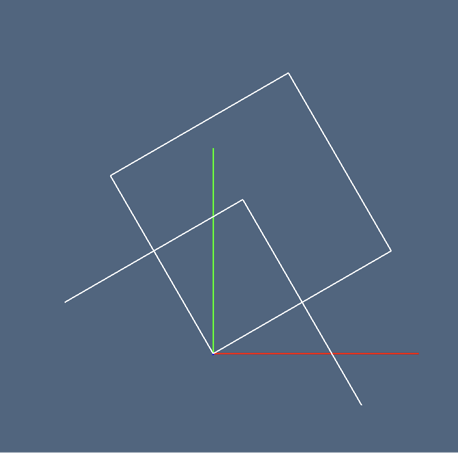}%

\vspace{-1.5mm}
{\footnotesize\hspace{.12\textwidth}(a)\hfill(b)\hfill(c)\hfill{(d)}\hspace{.12\textwidth}}
\vspace{-3.5mm}
   \caption{ Single facet 2D decomposition: 
   (a) the (red) reference facet $\sigma$ and the set $\mathcal{I}(\sigma)$ of possibly intersecting facets; 
   (b) $\Sigma = \{\sigma\}\cup\mathcal{I}(\sigma)$ after the transformation mapping $\sigma$ into $x_3=0$; 
   (c) $\sigma$ facet and 1-cells in $\Sigma$ intersecting the subspace $x_3 =0$; 
   (d) the six 1D generated intersections in $\E^2\times \E$ between $\mathcal{I}(\sigma)$ and $x_3=0$. The output is a $\mathcal{L}({\Sigma})$ collection of {six} line segments in the {$x_3=0$ subspace (the 2D plane)}
   \label{fig:cartoon-2} }

   \vspace{3mm}

\includegraphics[height=0.248\linewidth,width=0.248\linewidth]{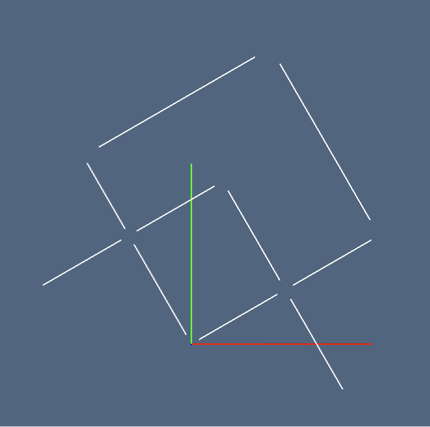}\hfill%
\includegraphics[height=0.248\linewidth,width=0.248\linewidth]{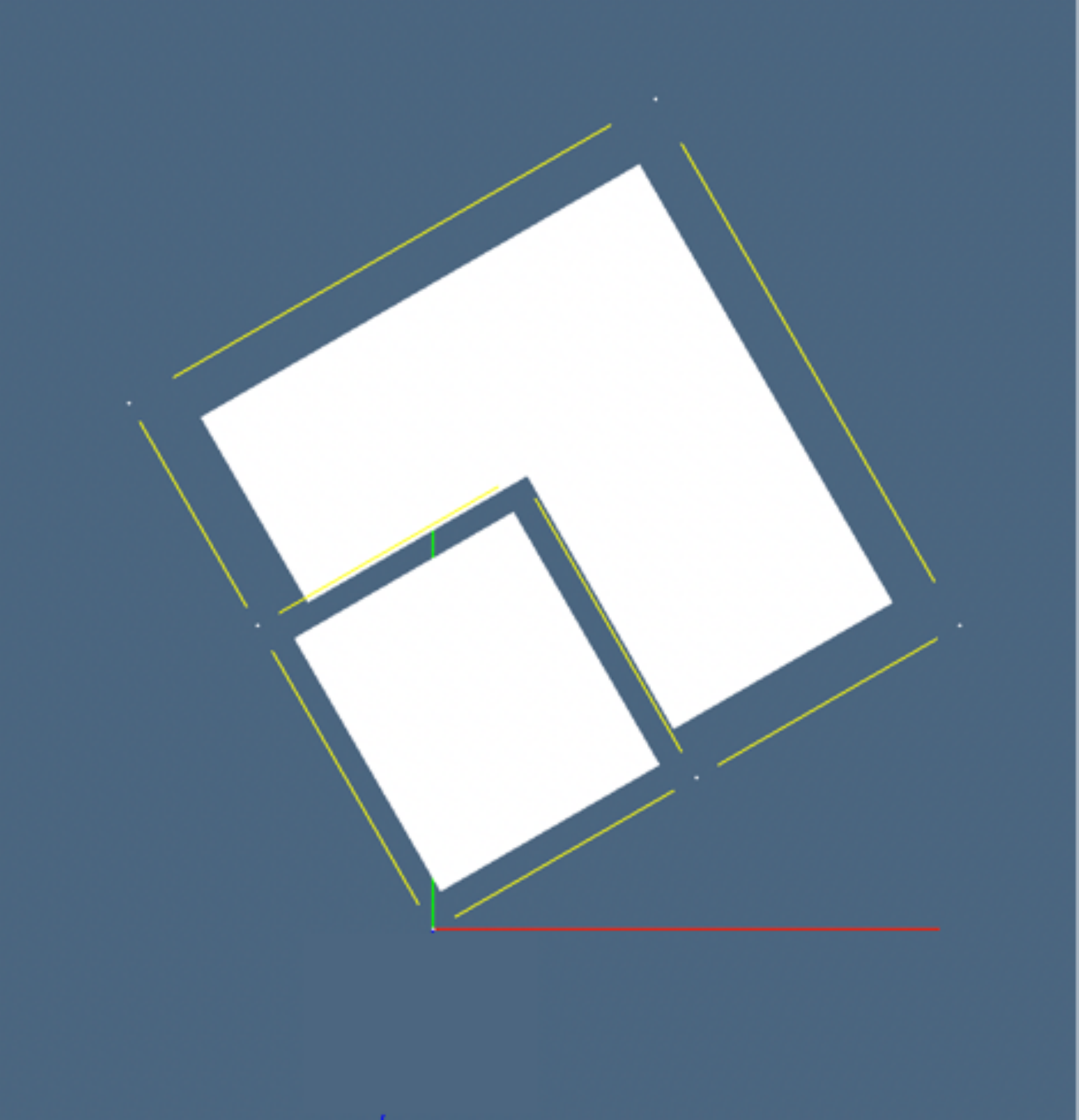}\hfill%
\includegraphics[height=0.248\linewidth,width=0.248\linewidth]{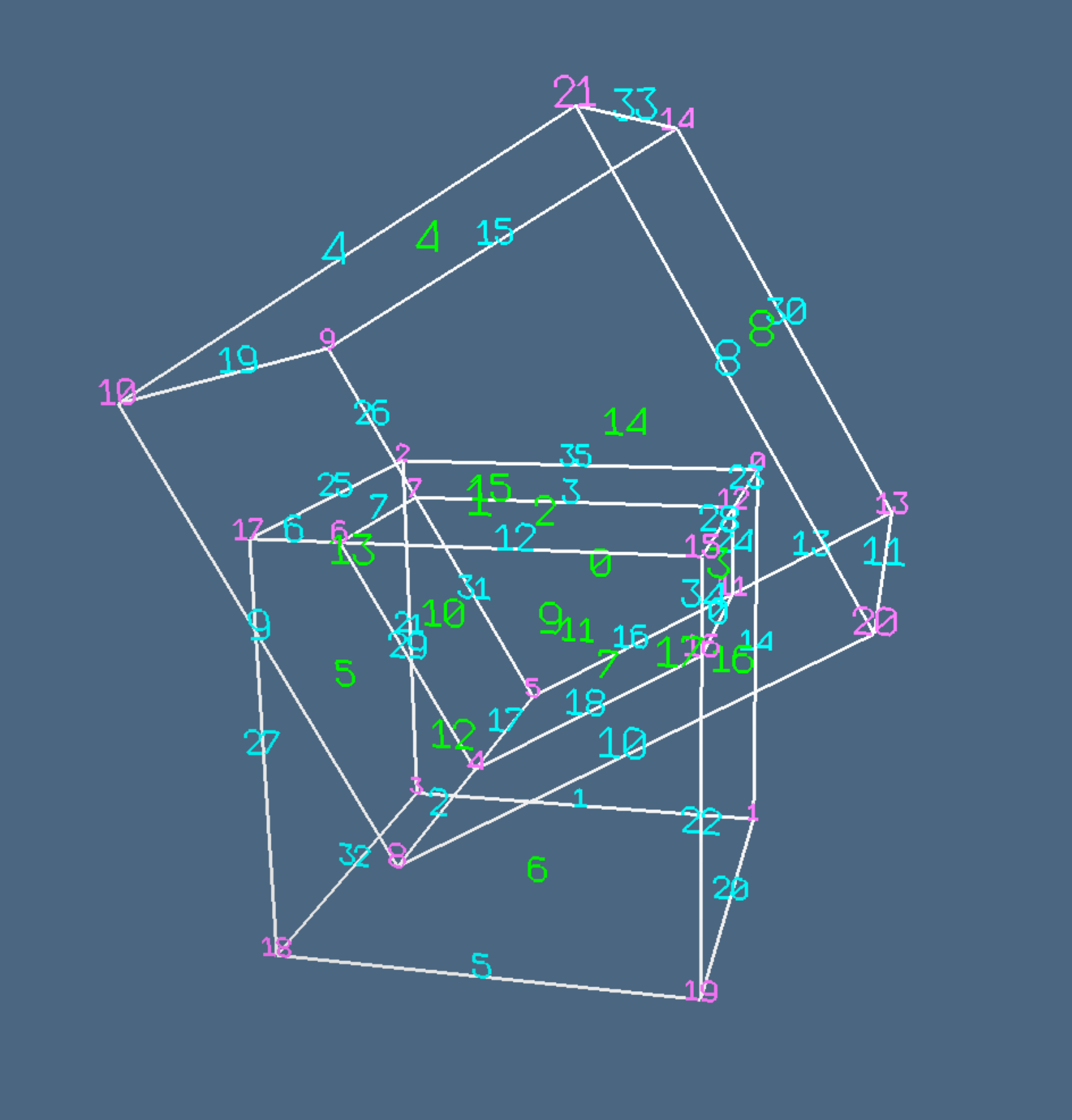}\hfill%
\includegraphics[height=0.248\linewidth,width=0.248\linewidth]{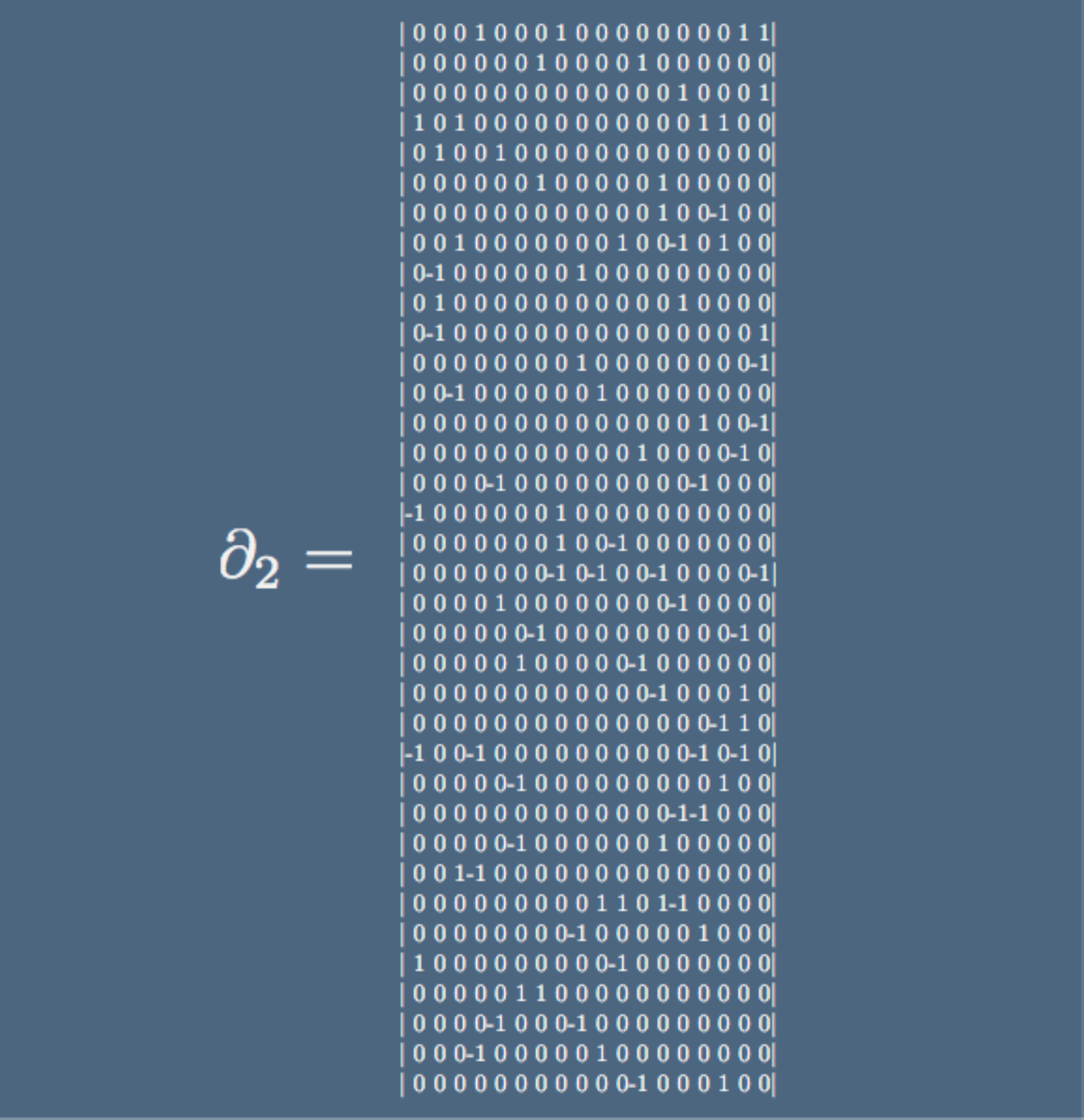}%

\vspace{-1.5mm}
{\footnotesize\hspace{.12\textwidth}{(a)}\hfill(b)\hfill(c)\hfill(d)\hspace{.12\textwidth}}
\vspace{-3.5mm}
   \caption{ 3D reconstruction: 
   (a) view (exploded) of 2D line segments produced by pair-wise intersection of elements in $\mathcal{L}({\Sigma})$; 
   (b) 2-complex $X_2({\Sigma})$ generated as $\mathcal{A}(X_1({\Sigma}))$, where { the $X_1$ skeleton is generated by computing} the maximal biconnected subgraph {induced by $\mathcal{L}({\Sigma})$}, with nodes at the edge intersections; 
   (c)~numbered 0-, 1- and 2-cells (with different colors) of the reconstructed $X_2(B)$ after identification of coincident cells generated by the embedding in $\E^3$ of all $X_2({\Sigma})$, for each $\sigma\in B$;
   (d) matrix $[\partial_2]$, assembling by column the signed 1-chain representation of 2-cells, {using ``quotient 0- and 1-cells''}. {The reader should remember that a quotient set is a set derived from another by an equivalence relation. In this case, two cells are equivalent if (and only if) the have the same support. The representatives of equivalence classes are computed by identification of coincident vertices (for 0-cells), and  by equality of canonical representations (for 1-cells).}
   Note that {2-cells are uniquely generated, whereas} 3-cells are still undetected. 
   \label{fig:cartoon-3} }
   
\end{figure}

\subsection{Input and setup}

First we need to assemble the input collection of data into a single representation, by properly concatenating the arrays of vertices and cells.
Figure~\ref{fig:cartoon-1}a shows the LAR input, made by the vertex array \texttt{V} and by the arrays \texttt{CV}, \texttt{FV} of indices of vertices of each 3-cell and 2-cell of the cubes. \texttt{CV}, \texttt{FV} stand for ``cells by vertices" and ``faces by vertices", respectively. Note that the input \emph{is not} a cellular complex, since faces intersect outside of their boundary.  The relative positions of cubes, and {both their (explosed) boundary 2-complex $X^h_2$ ($h\in \{1,2\}$)} are shown in Figures~\ref{fig:cartoon-1}b and~\ref{fig:cartoon-1}c.  
The bounding boxes of each $\sigma\in B$, with $B=\Lambda_2^1 \cup \Lambda_2^2$, and $\Lambda_2^h$ the sets of 2-cells, are displayed in yellow in Figure~\ref{fig:cartoon-1}d. Note that some 3D bounding boxes of 2-cells degenerate to rectangles, since their 2-faces are aligned with the coordinate planes, while others are not.

\subsection{Decomposition of input 2-cells  (facets: $2=d-1$)}
Figure~\ref{fig:cartoon-2} gives the sequence of computations on a generic 2-cell $\sigma$ (i.e.~a cell of dimension $d-1$) in the input set $B$. First the subset $\mathcal{I}(\sigma)$ of 2-cells of \emph{possible intersection} with $\sigma$ is computed, by intersecting the results of queries upon three (i.e.,~$d$) interval trees, based on sides of containment boxes of 2-cells. Then the set $\Sigma = \{\sigma\}\cup\mathcal{I}(\sigma)$ is transformed in such a way that the 2-cell $\sigma$ is mapped into the $z=0$ subspace, i.e.~to $x_3=0$.  
In this space $\E^2\times\E$ the 1-cells in $\Sigma$ are used to compute a \emph{set of line segments} in $\E^2$, generated by intersection of edges of each 2-cell in $\mathcal{I}(\sigma)$ with $z=0$, and by join (convex combination) of alternate pairs of intersection points of boundary edges along the intersection line of a 2-face with $z=0$. 

\subsection{Construction of $\partial_2$ boundary matrix}

The {planar processing of the 2-cell $\sigma$} continues in Figure~\ref{fig:cartoon-3}, where the  arrangement {$X_2(\Sigma) := \mathcal{A}(\Sigma)$} induced by $\Sigma$ on $\E^2$ is computed (see Figure~\ref{fig:cartoon-3}b). This computation by intersection of lines produces a linear graph, shown exploded in Figure~\ref{fig:cartoon-3}a. The dangling edges (and  dangling tree subgraphs) are removed, by computing the maximal 2-point-connected subgraphs~\cite{vialar2016handbook} via the~\cite{Hopcroft:1973:AEA:362248.362272} algorithm.  The resulting graph is actually a representation of the $\partial_1(\sigma)$ and $\delta_0(\sigma)$ operators of the chain 2-complex associated to a plane arrangement.

Finally, the oriented 2-cells of the partition $\mathcal{A}(\Sigma)$ are computed  as shown in Figure~\ref{fig:cartoon-3}b, so generating the $\partial_2(\sigma)$ and $\delta_1(\sigma)$ operators of the plane arragement.
The process is repeated for each $\sigma\in B$, each $X_2(\sigma) = \mathcal{A}(\Sigma)$ is mapped back in $\E^3$, and coincident 0- and 1-cells are identified numerically or {syntactically}, making use of their \emph{unique} canonical LAR representation. {The canonical representation of a cell is the tuple (array)} of sorted indices of cell vertices, after identification of  numerically nearby-coincident vertices using a $kd$-tree.  The resulting 2-complex $X_2(B)$, embedded in $\E^3$, is shown in Figure~\ref{fig:cartoon-3}c. Its 2-cells, written as 1-chains, i.e.~as linear combinations of signed 1-cells, are stored by column in the matrix of the operator $\partial_2: C_2\to C_1$, shown in Figure~\ref{fig:cartoon-3}d. Let us remind that a 1-cell $\tau$, is written\footnote{As a 0-chain of signed 0-cells in the matrix representation of $\partial_1: C_1\to C_0$.} by convention as $\nu_k - \nu_h$ {when $k>h$},  and is oriented from $\nu_h$ to $\nu_k$.  {The conventional rules used in this paper about sign and orientation of cells are summarized at the end of Section~\ref{sec:thechains}.}

\begin{figure}[htb]
\includegraphics[height=0.248\linewidth,width=0.248\linewidth]{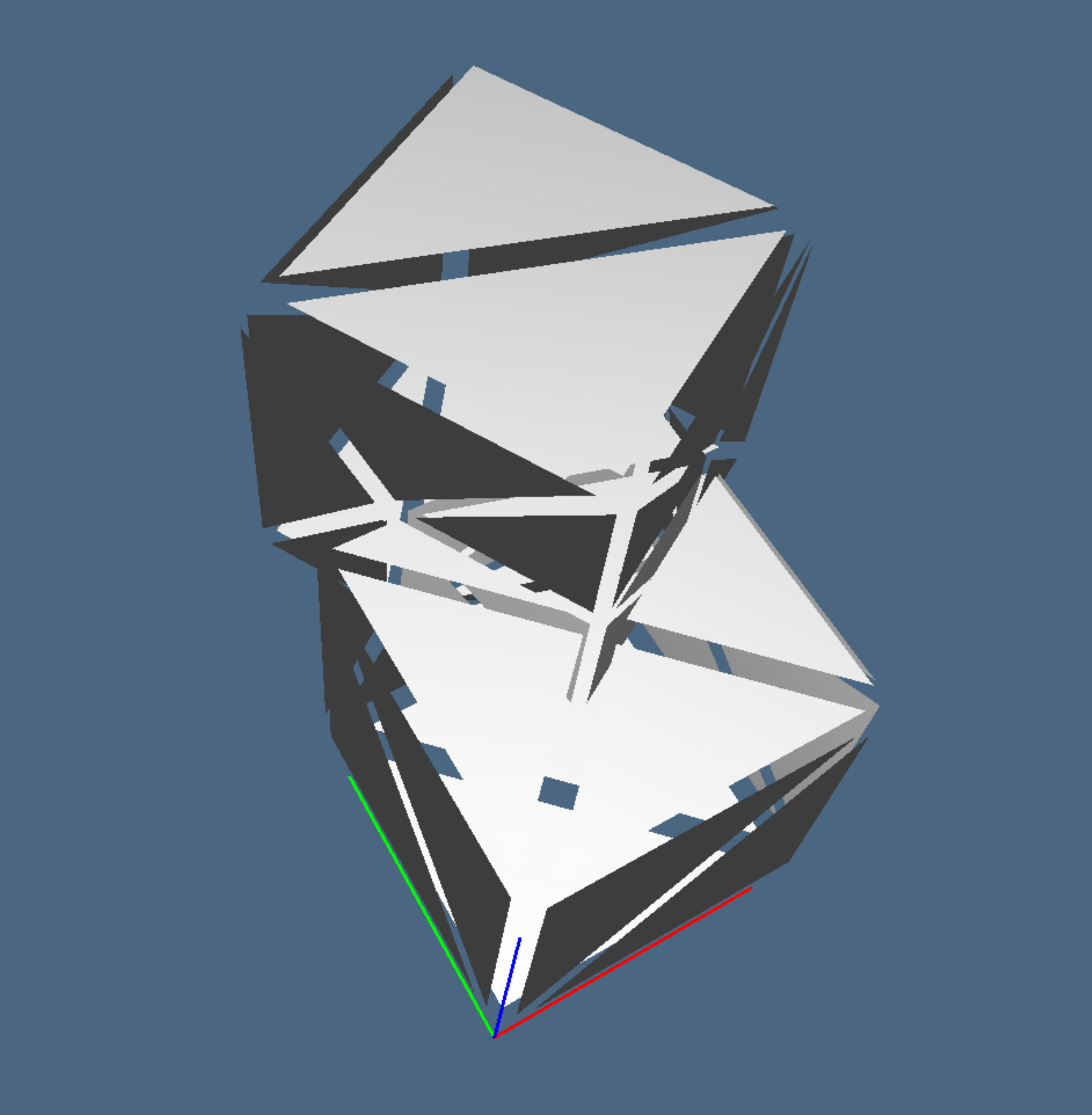}\hfill%
\includegraphics[height=0.248\linewidth,width=0.248\linewidth]{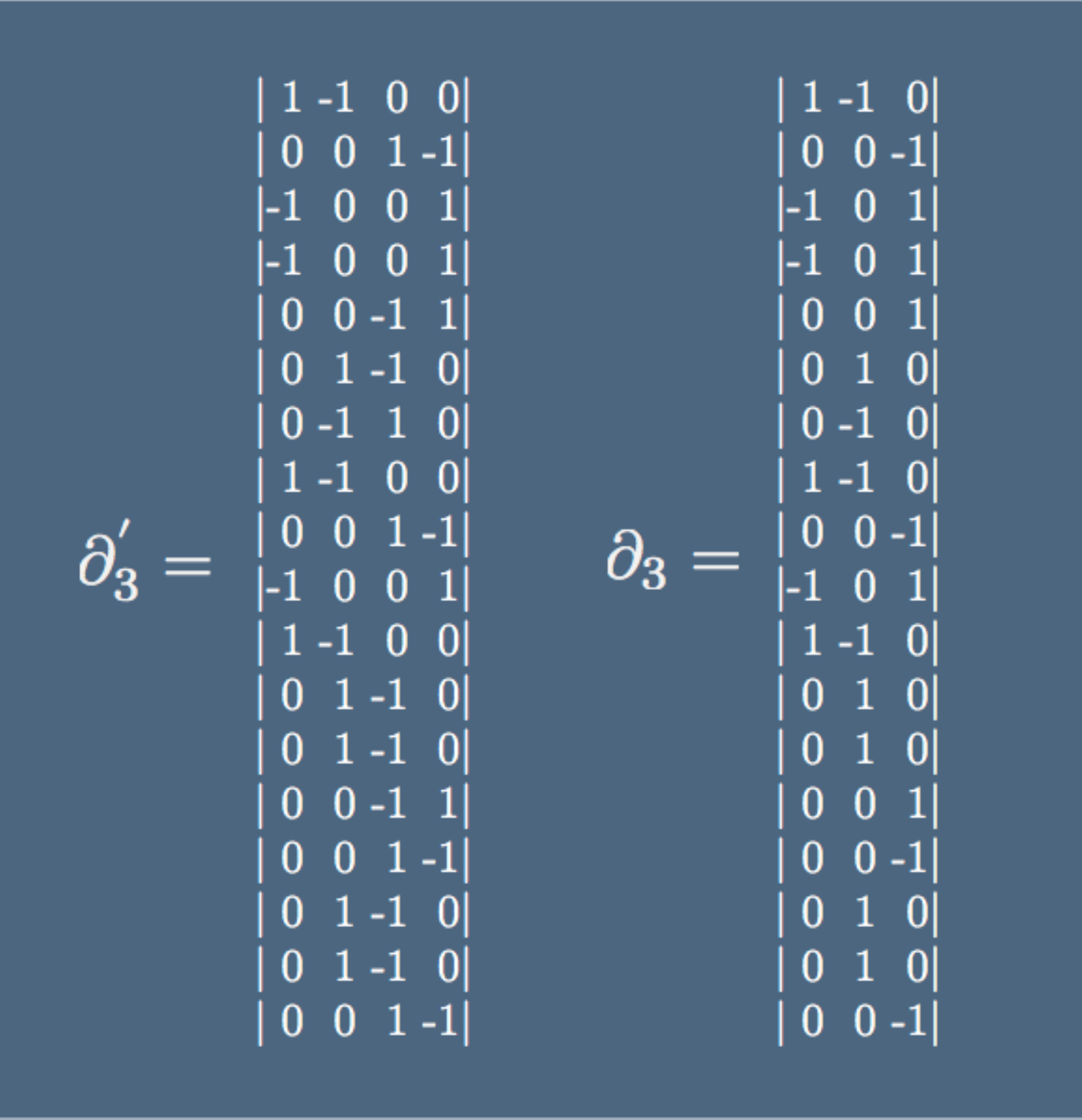}\hfill%
\includegraphics[height=0.248\linewidth,width=0.248\linewidth]{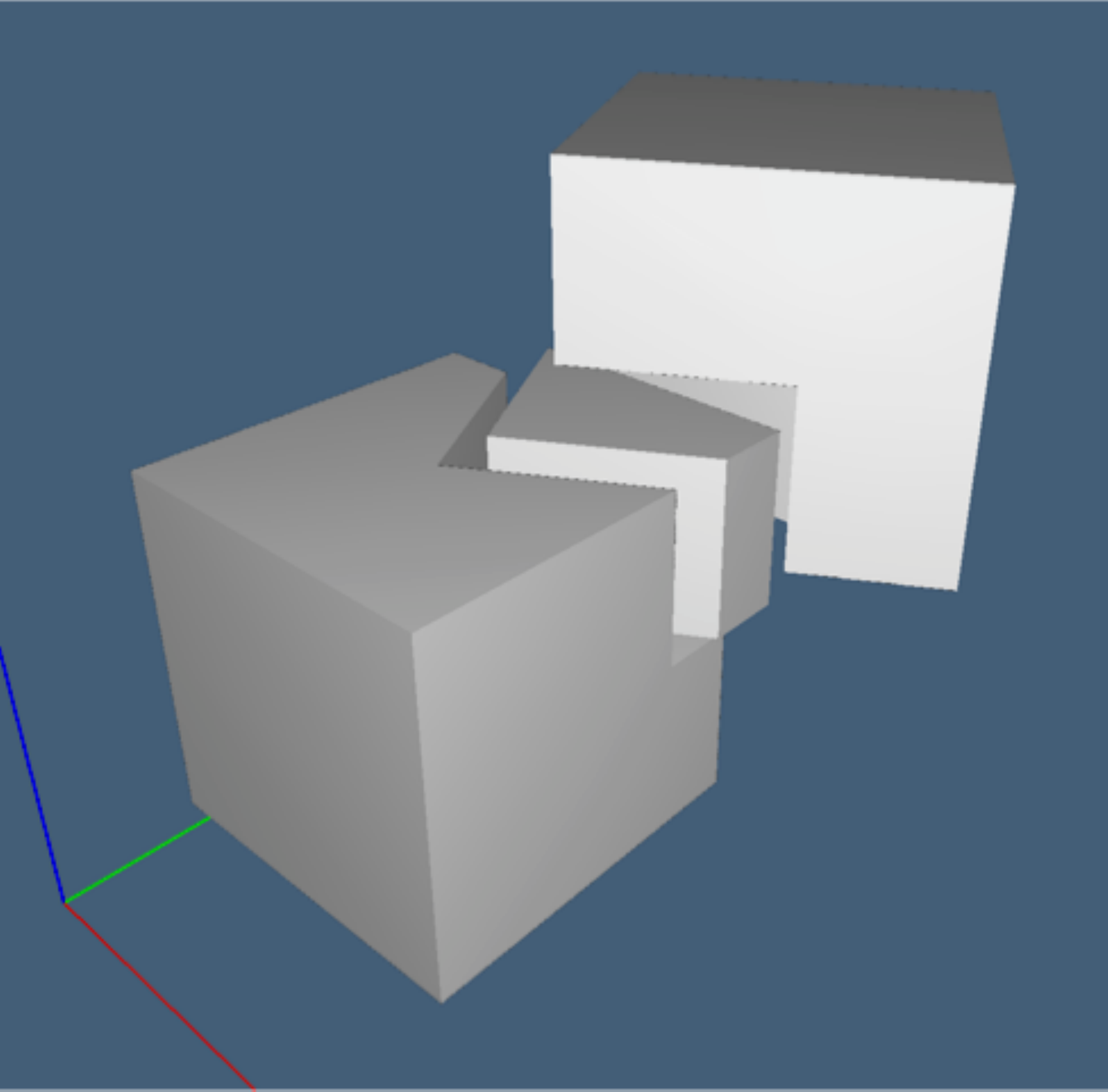}\hfill%
\includegraphics[height=0.248\linewidth,width=0.248\linewidth]{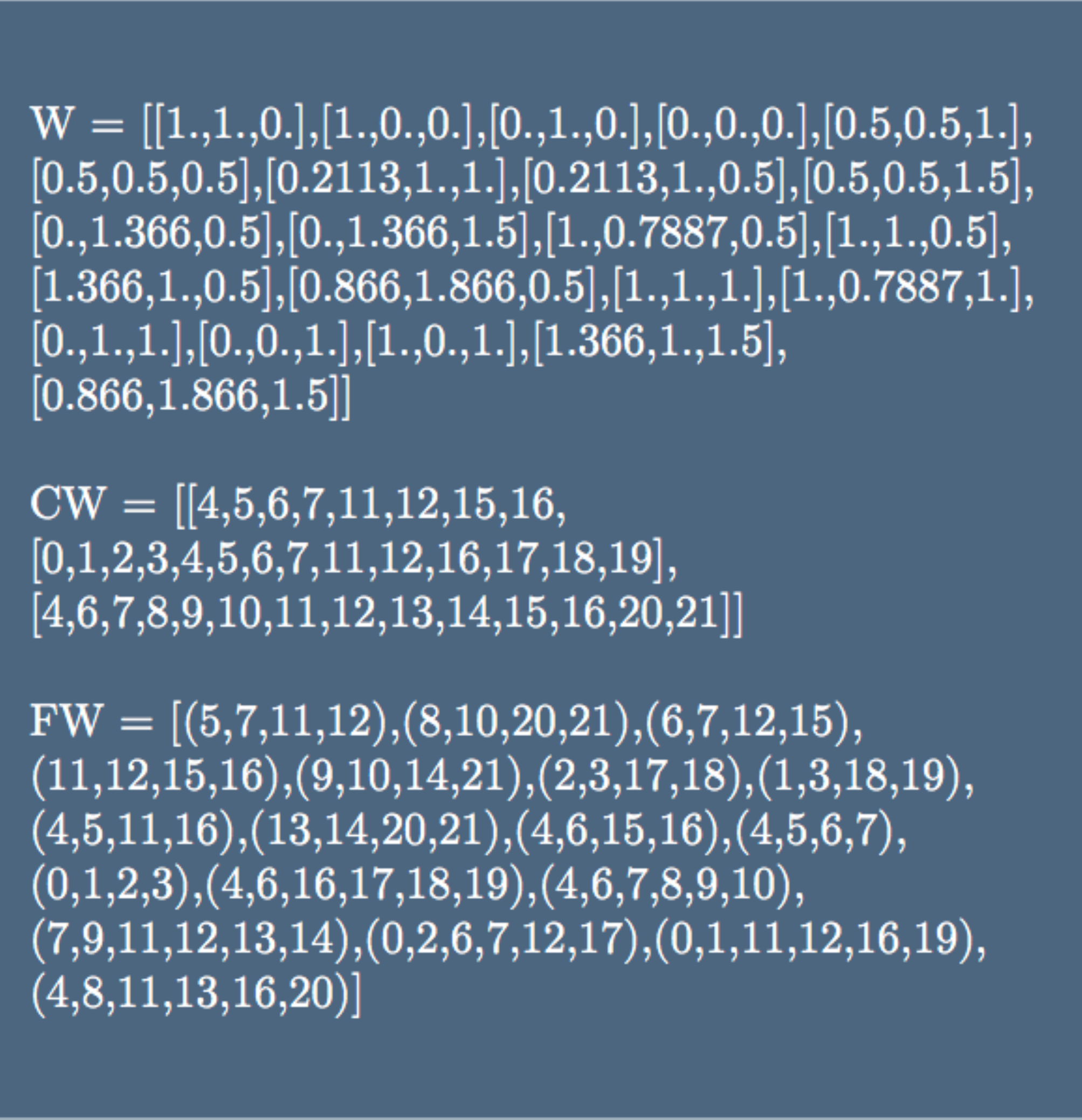}%

\vspace{-1.5mm}
{\footnotesize\hspace{.12\textwidth}(a)\hfill(b)\hfill(c)\hfill(d)\hspace{.12\textwidth}}
\vspace{-3.5mm}
   \caption{ Output: 
   (a) Triangulation of 2-cells, needed to compute the 1-cell coboundaries, sorted on angles; 
   (b)~$\partial^{+}_3: C_3\to C_2$, and boundary operator $\partial_3$ removed of exterior cell, providing a basis for chains of bounded cells; 
   (c)~3-cells (exploded) in $X_3 := \mathcal{A}(\mathcal{S})$; 
   (d) output LAR model of the cellular complex $X_3$. }
   \label{fig:cartoon-4}
\end{figure}

\subsection{$\partial_3$ boundary matrix and $\mathcal{A}(X_3)$ output}

Finally, the 3-cells of the 3-space partition induced by $B$ are computed, using the algorithm of Sections~\ref{sec:cycles} and~\ref{sec:gift-wrapping}, pseudo-coded in Section~\ref{sec:pseudocode},  whose results are shown in Figure~\ref{fig:cartoon-4}. 

{For each 1-cell in the $X_1$ skeleton we compute the cyclic order of 2-cells incident on it, in order to extract the 3-cells via our \emph{topological gift wrapping} algorithm (see Figure~\ref{fig:3D})}
\footnote{{In our current implementation, a CDT triangulation~\cite{Shewchuk2008} of all 2-cells is used to compute correctly, also for non-convex cells, the cyclic orders of 2-cells about 1-cells. This triangulation, while  not mandatory, is the quickest robust method we found to implement exact cyclic permutations of non-convex 2-cells around 1-cells. Alternatively, neighborhood classification via offsetting and point-in-set methods could be used.}}. Permutations $\mathit{Next}: C_2\to C_2$ and $\mathit{Prev} := \mathit{Next}^{-1}$ are used to compute the signed 2-cycle representation of basis 3-cells. Such closed (without boundary) 2-cycles are detected and stored  by column in the $\partial_3^+$ matrix (see Figure~\ref{fig:cartoon-4}b). 
This one is rewritten as the $\partial_3$ operator by removing the column of the exterior {unbounded} cell.
The {bounded} 3-cells of the complex $X_3 := \mathcal{A}(B)$ are displayed in {Figures~\ref{fig:cartoon-4}c and~\ref{fig:cartoon-4}d.}

The LAR representation of the output complex $X_3$ is finally given in Figure~\ref{fig:cartoon-4}d, using a user-readable version of CSR (Compressed Sparse Row) format for binary matrices, where the non-zeros of binary data do not need storage.  This representation is useful for input/output of LAR models, as the reader may see in Appendix~\ref{sec:lar-examples},  and for their long-term storage in document format. 

{It is easy to see that, for  B-reps of 2D manifolds without boundary}, the space required by LAR is {$|FV| = 2|E|$}~\cite{woo:85}, i.e.,~$1/4$ of the well-known \emph{winged-edge} representation~\cite{Baumgart:1972:WEP:891970}. {$|FV|$ is the cardinality of the incidence relation between faces and vertices. In LAR, it is given by the sum of lengths of elements of the array \texttt{FV}. As an example, consider the B-rep of a 3-cube: each face contains 4 vertices ($6\times 4 = 24 = 2\times 12$), where 12 is the number of edges. }
Note that the actually computed $\partial^+_3$ matrix contains one more column (the \emph{exterior} 3-cell) which is a linear combination of the other columns. {Hence,} in order to get a basis for the linear space $C_3$, and {the matrix representation of $\partial_3$ with the correct number of independent elements}, this column has to be located (see Section~\ref{sec:extrema}) and removed, as shown in Figure~\ref{fig:cartoon-4}b.  

The time complexity of sequential $[\partial_3]$ construction from $[\partial_2]$ is $O(n m\log m)$, where $n,m$ are the numbers of 3- and 2-cells,  in the worst case of unbounded complexity of $3$-cells (see Section~\ref{sec:complexity1}), and $O(n k\log k)$ when the number of 2-cells on the frontier of 3-cells is bounded by a (small) integer $k$.

\section{Merge algorithm}\label{merge-algorithm}

The computation of the arrangement of $\E^d$ 
generated by a collection of $d$-complexes, here called
\emph{merge of complexes}, is discussed in this section using the
mathematical language of chain complexes, i.e., the linear spaces of
chains (of cells) and the linear (co)boundary maps between such spaces.

It is worth noting that the support space (point-set union) of the $(d-1)$-skeleton of the output complex equals the union of the corresponding input skeletons. {By contrast, approaches based on spatial indices, like~\cite{Ayala:1985:ORM:3973.3975,Thibault:1987:SOP:37402.37421}, lead to much larger fragmentation of the underlying space and correspondingly larger size of the computed cellular arrangement.} The reader may easily appreciate how much this differs from other dimension-independent approaches based on spatial indices, like $2^d$-trees or BSP-trees. 

\subsection{Dimension-independent workflow}

Let $d$ be the dimension of the embedding space $\E^d$, and
\[
\mathcal{S}_d = \{\, X^h,\, h\in H\,\} 
\]
the input collection of $d$-complexes, and denote with $X^h_{d-1}\subset X_d$ the $(d-1)$-th
skeleton of the $h$-th $d$-complex. We proceed to compute the arrangement of $\E^d$ generated by $\mathcal{S}_d$, or, more precisely, the regularized $d$-complex $X=\mathcal{A}(\mathcal{S}_{d-1})$.
{Note that, in a full-fledged dimension-independent implementation, the algorithm would iterate on dimensions, starting from dimension 2. Consequently, no recursion is needed, and this allows for an efficient parallel implementation. We leave this matter to a further paper.}
For the sake of concreteness, the reader may here assume $d=3$.

\subsubsection{Assemble the $(d-1)$-skeletons}

We start by considering the set $B = \bigcup_{h\in H} X^h_{d-1}$, combinatorial union of
$(d-1)$-cells in $\mathcal{S}_d$, where $H$ is
a set indexing the input complexes.

\subsubsection{Compute the spatial index $\mathcal{I}: B\to 2^B$}
\label{sec:sigma}

An \emph{interval tree} is a  data structure to hold intervals~\cite{Preparata:1985:CGI:4333}, that allows to efficiently find all intervals that overlap with any given interval or point. It is mainly used for windowing queries.
We compute a spatial index $\mathcal{I}: B\to 2^B$, mapping each
$(d-1)$-cell $\sigma$ to the subset of $(d-1)$-cells $\tau$ such
that $\mathit{box}(\sigma)\cap \mathit{box}(\tau) \not= \emptyset$, where $\mathit{box}(\gamma)$ is the \emph{containment \mbox{box}} of $\gamma\in B$, i.e.~the minimal $d$-parallelepiped parallel to the Cartesian frame, and such that $\gamma\subseteq \mathit{box}(\gamma)$. The $\mathcal{I}$ map is computed by set intersection of the outputs of $d$ elementary queries on 1D coordinate interval trees.

\subsubsection{Compute the facet arrangments in $\E^{d-1}$}
\label{sec:facet-arrangment}

We compute the set $\mathcal{X}_{d-1}$ of \emph{facet arrangements} $\mathcal{A}(X_{d-1}(\sigma))$ induced in $\E^{d-1}$ by $\sigma$ elements of B, fragmented by all the incident cells $\mathcal{I}(\sigma)$:
$$
\mathcal{X}_{d-1} = \{\, \mathcal{A}(\sigma \cup \mathcal{I}(\sigma)),\, \sigma\in B \,\}.
$$ 
To compute the arrangements, for each $\sigma \in B$, a submanifold map\footnote{Submanifold map is a function that maps some submanifold to a coordinate subspace of a chart.} $M: \E^d \to \E^d$ is easily determined, mapping $\sigma$ into the subspace with implicit equation $x_d=0$.

Accordingly, for each $\sigma \in B$:
\begin{enumerate}
\item
  compute an affine transformation $M$ that maps $\sigma$ into the
  subspace $x_d=0$,
\item
  consider the set $\mathcal{S}_{d-1} = M(\Sigma)$, with
  $\Sigma = \{ \sigma \} \cup \mathcal{I}( \sigma )$,
\item
  compute the hyperplane arrangement
  $\mathcal{A}(\mathcal{S}_{d-1}) = X_{d-1}^\sigma$,
\item
  transform back the resulting $(d-1)$-complex to $\E^d$, i.e.~compute
  $M^{-1}(X_{d-1}^\sigma)$.
\end{enumerate}

All such $(d-1)$-complexes embedded in $\E^d$ are accumulated in $
\mathcal{X}_{d-1} = \{  M^{-1}( X_{d-1}^\sigma ), \sigma\in B \},
$ represented as the proper LAR structure.

\subsubsection{Assemble the output $(d-1)$-skeleton}

A $kd$-tree, or \emph{$k$-dimensional tree}, is a binary search tree used to organize a discrete set of $k$-dimensional points.  $Kd$-trees are very useful for range and nearest neighbour searches~\cite{Bentley:1975:MBS:361002.361007}.
Here, a $kd$-tree is computed over the vertices of the collection of
complexes $\mathcal{X}_{d-1}$, in order to
  collapse into a single point all the vertices sharing the same
  $\epsilon$-neighborhood, with small $\epsilon\in \R$. 
As a consequence of this reduction of the vertex set, we
\begin{itemize}
\item
  rewrite the representation of each $(d-1)$-cell using the new vertex
  indices;
\item
  sort the vertex indices of each $(d-1)$-cell to identify and remove
  the duplicate $(d-1)$-cells.
\end{itemize}

The resulting $(d-1)$-complex in $\E^d$ is the $(d-1)$-skeleton
$X_{d-1}$ of the output complex $X = \mathcal{A}(\mathcal{S}_d)$ generated by the
input collection $\mathcal{S}_d$. The LAR representation is also reduced in \emph{canonical form}, where the vertex indices are sorted in each cell.

\subsubsection{Extraction of d-cells from $(d-1)$-skeleton}
\label{sect:alg-1}

At the beginning of this stage of the algorithmic reconstruction of the space arrangement induced by a collection $\mathcal{S}$ of cellular complexes, we have a \emph{complete}~\cite{Requicha:1980:RRS:356827.356833} representation in $\E^d$ of the skeleton $X_{d-1}$  of $\mathcal{A}(\mathcal{S})$.

Then, a coordinate representation of the matrix of the coboundary
$\delta_{d-1}: C_{d-1}(X) \to C_d(X)$ is incrementally constructed, by
computing one by one the $d$-cells in $X$, i.e, the rows in $\delta_{d-1}$. This procedure will complete our reconstruction of $\mathcal{A}(\mathcal{S})$. It can be shown that this computation can be executed in parallel, with a number of independent tasks equal to the number of $d$-cells in $X$.

Let us consider the boundary map $\partial_d: C_d \to C_{d-1}$, with
$\partial_d=\delta_{d-1}^\top$. Each column in $\partial_d$ is the
representation of one element of the $C_d$ basis
(i.e., of a $d$-cell in $X_d$), in the basis of $C_{d-1}$, i.e., as a linear combination of the
cells in $X_{d-1}$.
The construction of $d$-cells is carried out by using two tools:

\begin{itemize}
\item
  the property that every $(d-1)$-cell in a $d$-complex is shared
  \emph{at most} by \emph{two} d-cells~\cite{hatcher:2002}. Note that if the complex includes also the \emph{exterior} (unbounded) \emph{cell}, then  such  incidence relation between cells in $X_{d-1}$ and in $X_d$ holds exactly;
\item
  the computation of the cyclic subgroups of permutations  of $(d-1)$-cells about their
  common $(d-2)$-cell, which implies the existence of a $\mathit{next}$ and
  $\mathit{prev} = \mathit{next}^{-1}$ bijections on such subsets.
\end{itemize}

Cyclic subgroups $\delta c$, with elementary (singleton) $c\in C_{d-2}$, have the following properties:

\begin{enumerate}
\item
  are in one-to-one correspondence with cells in $X_{d-2}$,
\item
  $next(-c) = next^{-1}(c)$.
\end{enumerate}

Since we mostly deal with \emph{oriented} chain complexes, we normally want to build a
\emph{coherently oriented} cellular complex $X$ from the arrangement
$\mathcal{A}(\mathcal{S}_d)$, enforcing the condition that every facet shared by two
$d$-cells in $X_d$ should have opposite orientations on them. This re-orientation of $d$-cells can be executed in a successive final stage, in linear time  with respect to the number of cells and the size of $X$. As a matter of fact, the construction of $d$-cells is done by computing the columns of $[\partial_d]$ matrix (see Section~\ref{sec:pseudocode} and Algorithm~\ref{alg:one}). Hence the reorientation of a $d$-cell just requires to invert the sign of non-zero coefficients of its (sparse) column, or (symbolically) to multiply the column times a $-1$ coefficient.

\subsubsection{$d$-cells are {bounded by} minimal $(d-1)$-cycles}
\label{sec:cycles}

A chain $c \in C(X)$ is said to be a \emph{cycle} if $\partial c=\emptyset$.
A formal algorithm for extraction of $d$-cells of $X$ starting from
$(d-1)$-cells in $X_{d-1}$ is given in Section~\ref{sec:pseudocode}. In particular, we generate
the linear representation of a basis $d$-cell (more exactly: of a singleton
$d$-chain, element of a $C_d$ basis) as \emph{minimal} combination of coherently oriented 
$(d-1)$-chains.

We say that a basis of $d$-chains is \emph{minimal} (or canonical) when the sum of cell numbers of its cycles, as combinations of facets with non-zero coefficients, equals the double of the number of facets. In other terms, if we denote the cardinality of a $p$-basis as $\# X_p$, we have, for a canonical $d$-basis and the {matrix $[\partial_d^+]=(a_{ij})$} with $a_{ij}\in\{-1,0,1\}$
\[
\sum_{i=1}^{\#X_{d-1}} \sum_{j=1}^{\#X_{d}} |a_{ij}|= 2\,\#X_{d-1}.
\]
{In other terms, every facet ($d-1$-cell) is used exactly twice in constructing the rows of $[\partial_d^+]$.}
This property, well known in solid modeling, is normally represented~\cite{woo:85} as an identity between the cardinalities of incidence relations between vertices, edges and faces in 3D boundary representations and/or graphs drawn on a 2-manifold. 

\subsubsection{Topological gift wrapping}
\label{sec:gift-wrapping}

{The algorithm given in this sections reminds of the "gift-wrapping" algorithm for computing convex hulls~\cite{Jarvis:1973:ICH,Cormen:2009:IAT:1614191}. Actually, the topology-based algorithm given here is a broad generalization, since it applies also to \emph{non-convex} polyhedra contractible to a point.   }
{In particular, we start a $(d-1)$-cycle $c$ from a single $(d-1)$-cell and update it iteratively by summing it a suitable $(d-1)$-chain until $c$ becomes closed. See Examples~\ref{example1}, \ref{example2} and Section~\ref{sec:pseudocode}.}
The minimality of each basis $d$-cell is guaranteed in Algorithm~\ref{alg:one} by the repeated
application of the map \[
\mathit{corolla}: C_{d-1}\to C_{d-1}, 
\] that takes as input a $(d-1)$-chain $c$,
starting from an initial ``seed'' $(d-1)$-cell, and returns a subset of ``petals'' from a ``corolla'' (see Figures~\ref{fig:3D}c, \ref{fig:3D}e, \ref{fig:3D}f) of
boundary cells at each repeated application, until $\partial c = \emptyset$. In formulas, we may write
\[
\mathit{corolla}(c) = c + \mbox{next}(\,\partial c\,) = c + \sum_{\tau\in \partial c} \mbox{next}\big((\delta\tau) \cap c\big).
\]
In words, a basis element of $C_d$, represented as a minimal $(d-1)$-cycle in $C_{d-1}$, is generated from $X_{d-1}$ by: (i)~choosing a
cell $\sigma\in X_{d-1}$ and setting
$c := 1\,\sigma$, then (ii)~by repeating $c := c + \mathit{corolla}(c)$, 
until (iii)~$\partial c = \emptyset$. See Figure~\ref{fig:3D} and Algorithm~\ref{alg:one}.

\begin{figure}[htbp] %  figure placement: here, top, bottom, or page
   \centering
\includegraphics[width=0.142\textwidth]{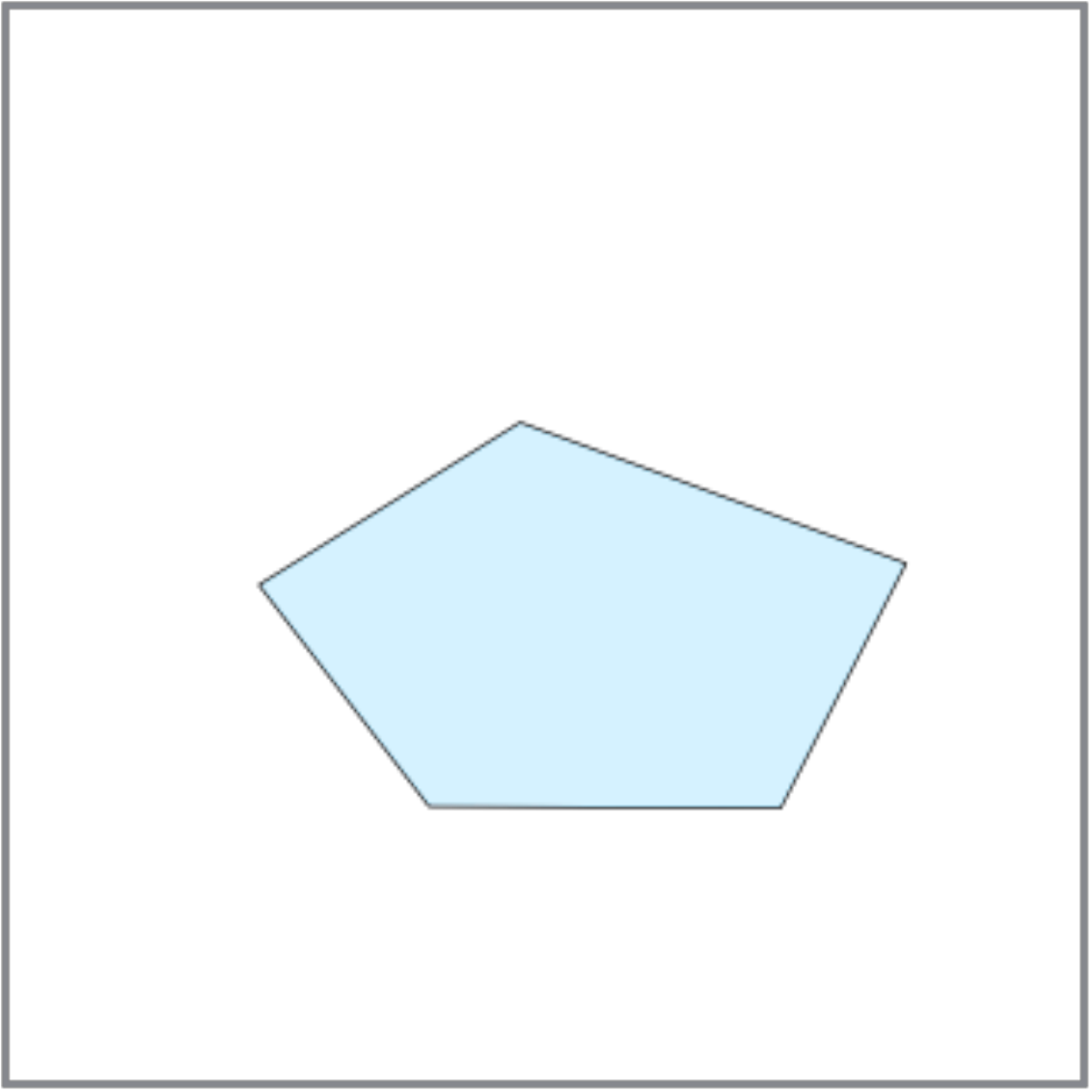}\includegraphics[width=0.142\textwidth]{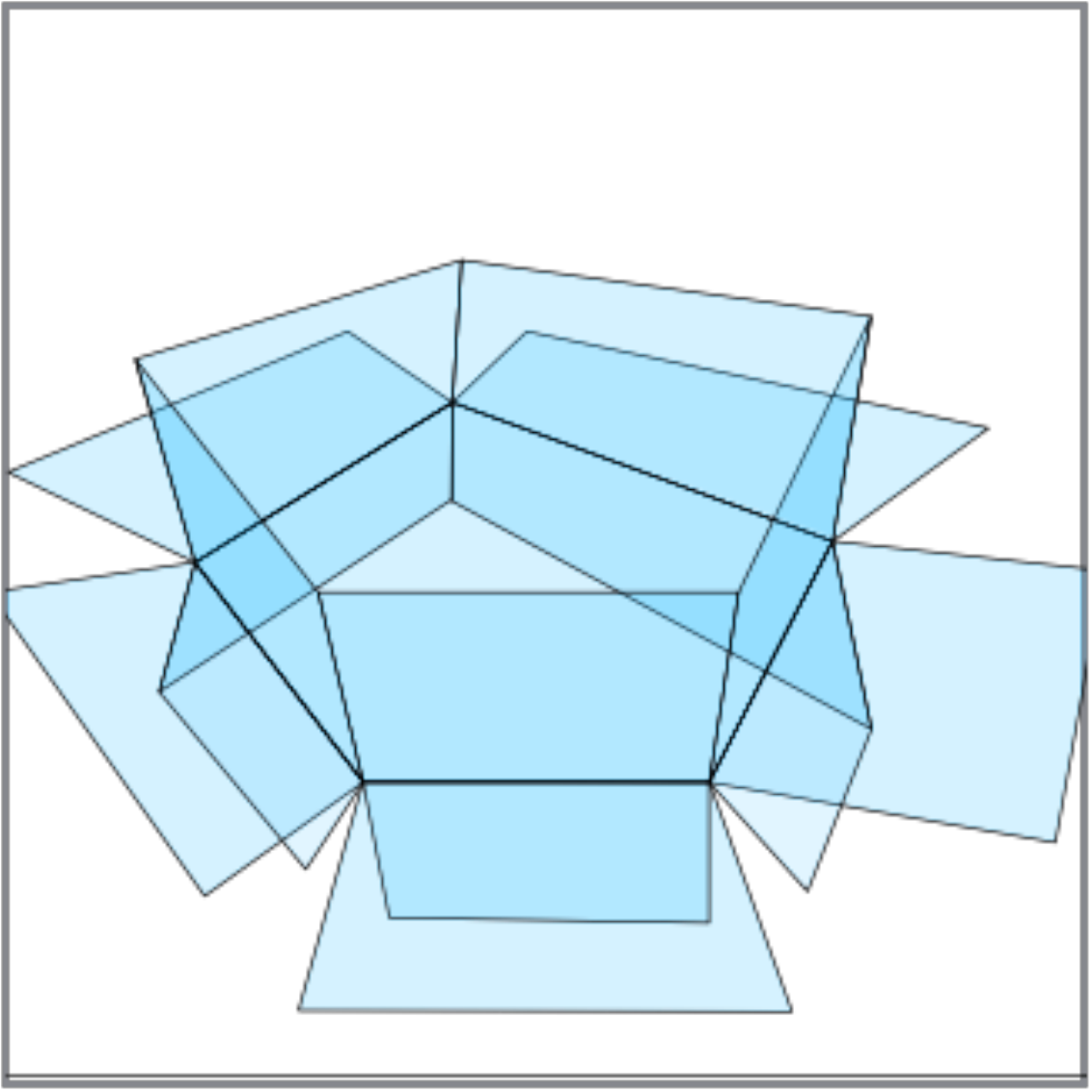}\includegraphics[width=0.142\textwidth]{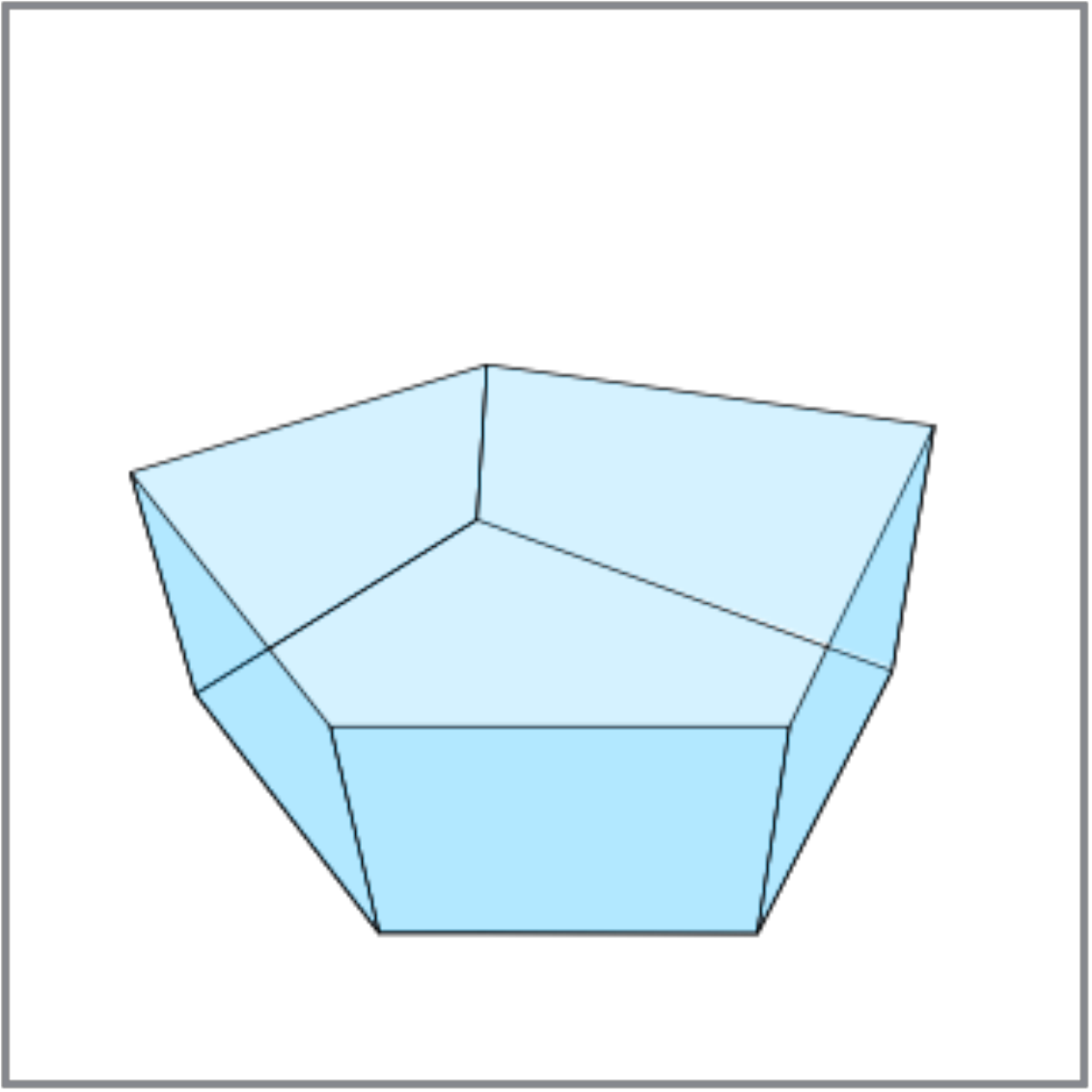}\includegraphics[width=0.142\textwidth]{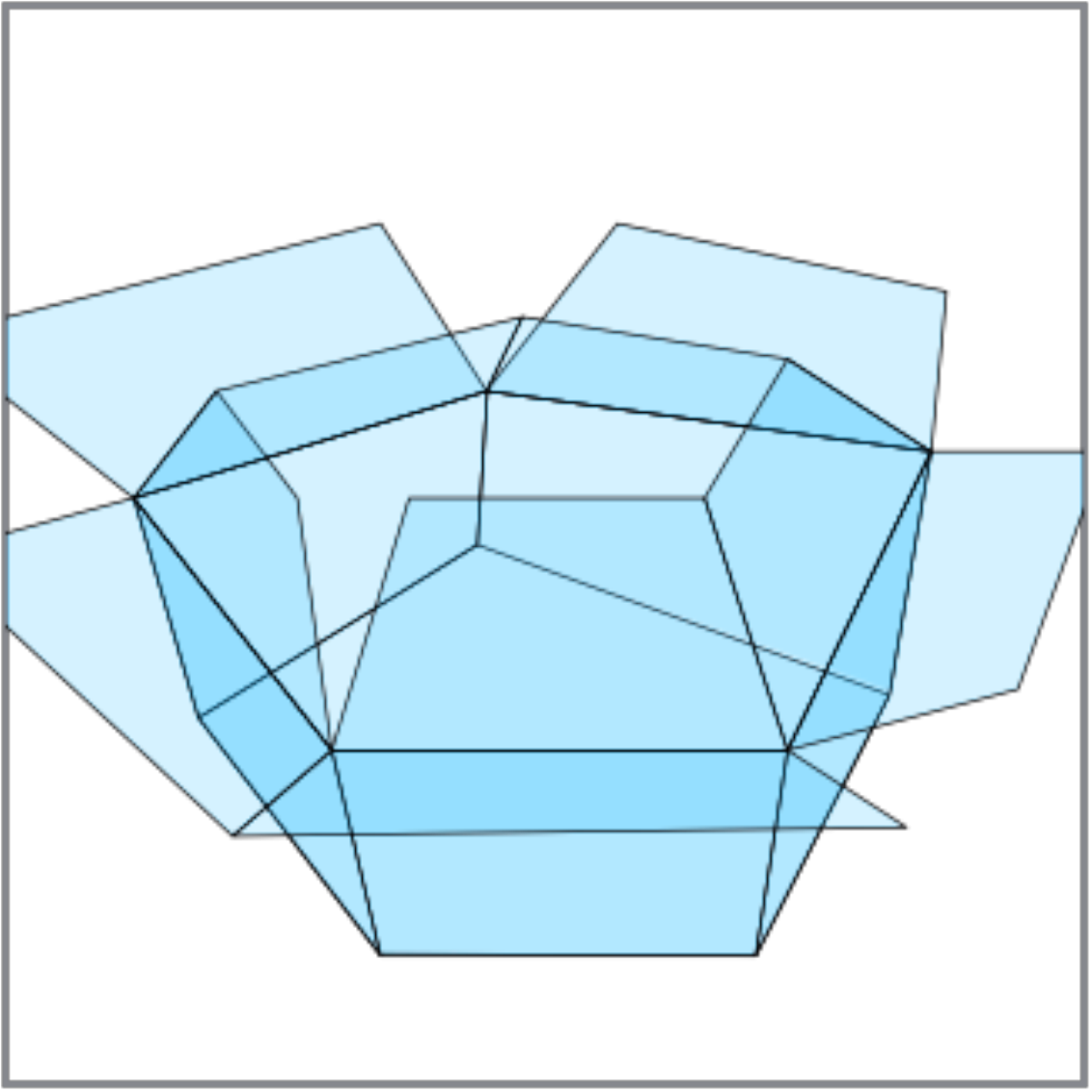}\includegraphics[width=0.142\textwidth]{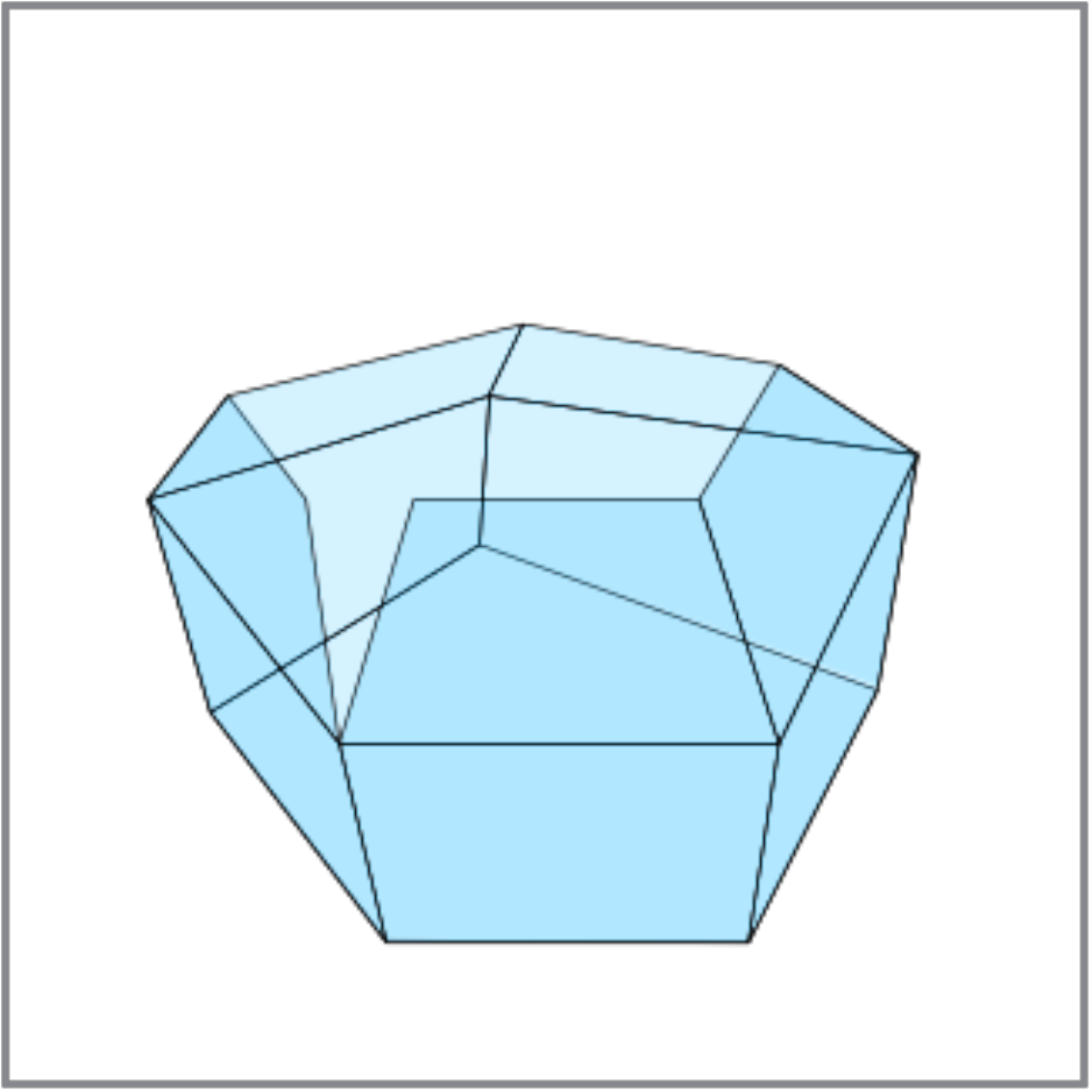}\includegraphics[width=0.142\textwidth]{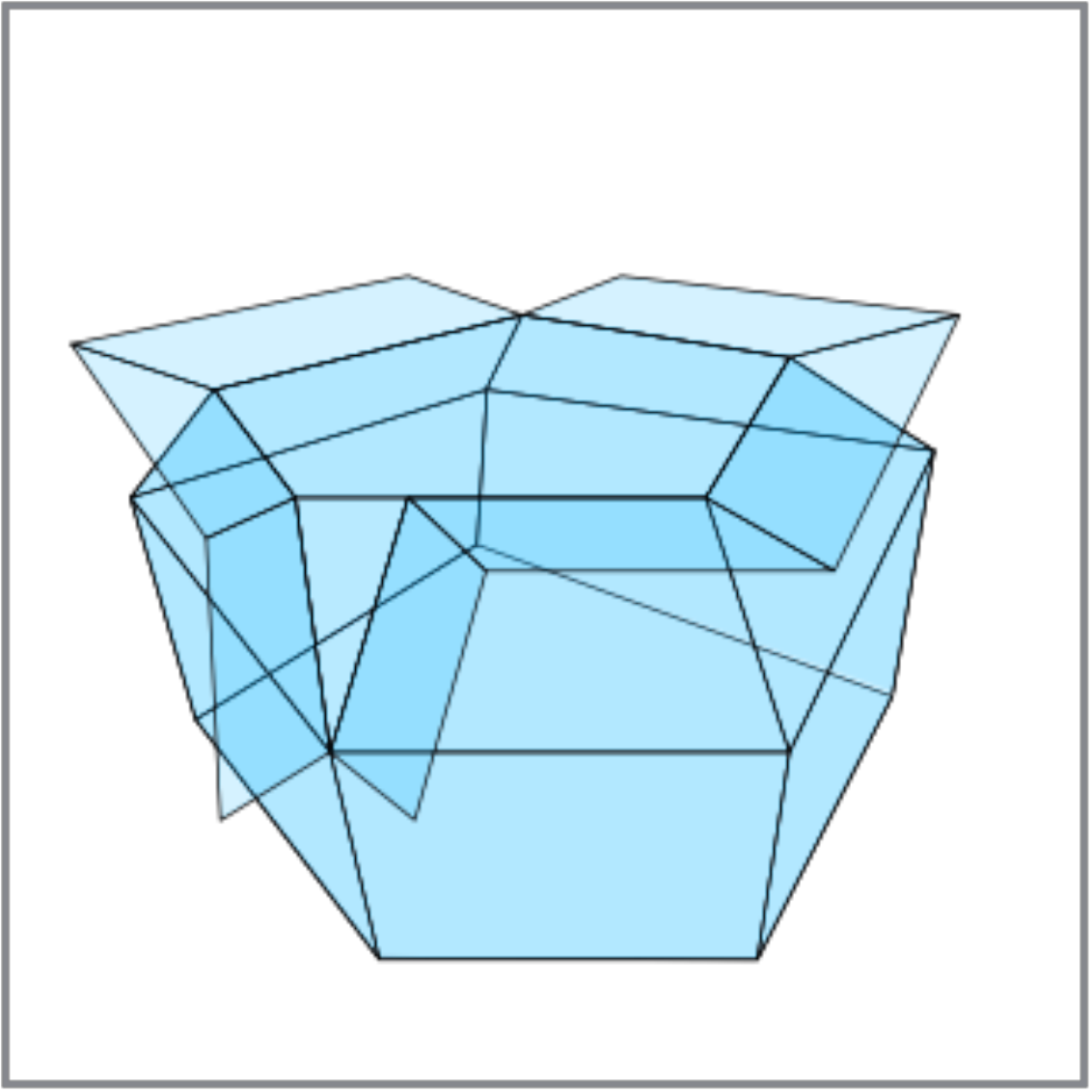}\includegraphics[width=0.142\textwidth]{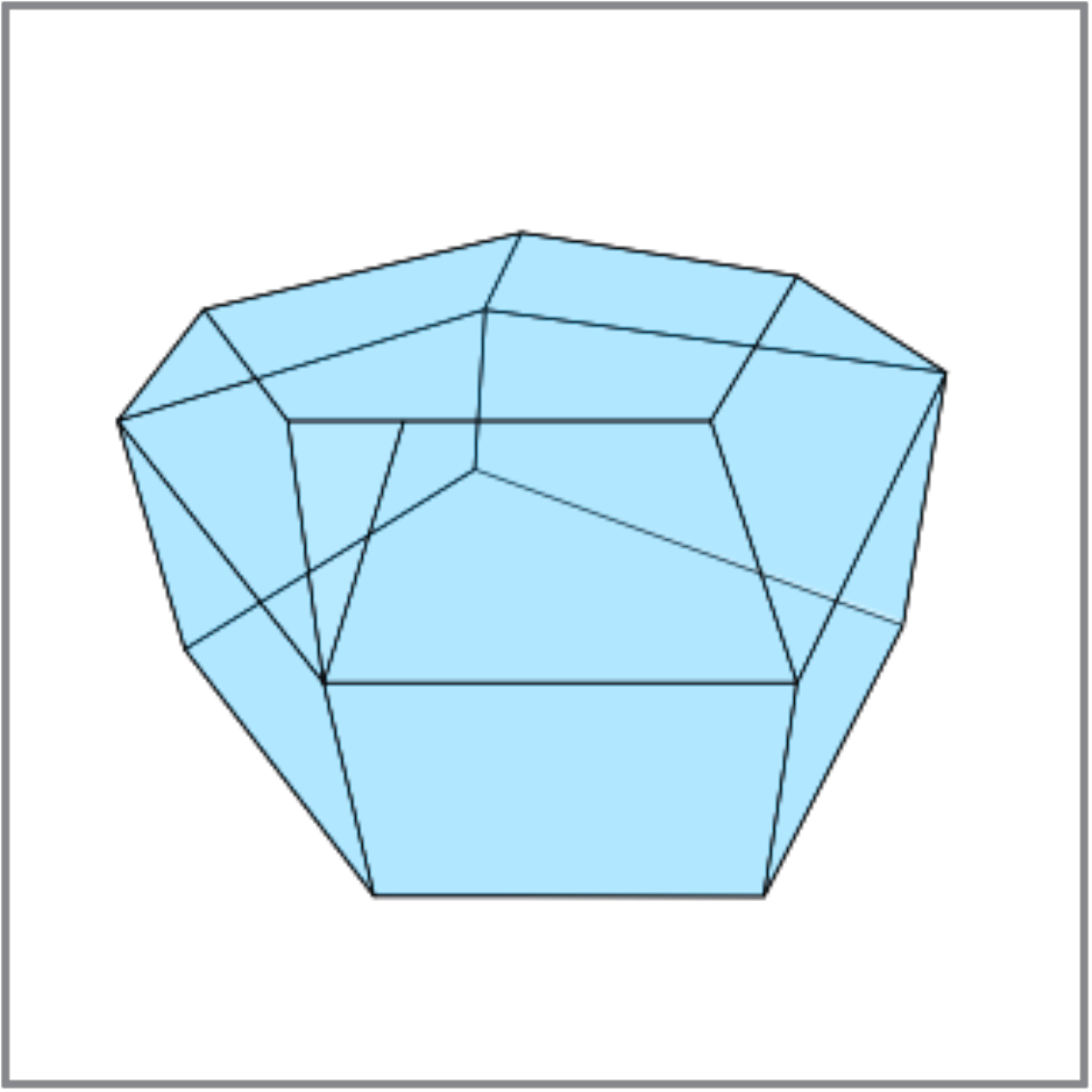}

{\footnotesize\hspace{.06\textwidth}(a)\hfill(b)\hfill(c)\hfill(d)\hfill(e)\hfill(f)\hfill(g)\hspace{.06\textwidth}}
\vspace{-2mm}   
\caption{Extraction of a minimal 2-cycle from $\mathcal{A}(X_2)$: (a) 0-th value for $c\in C_2$; (b) cyclic subgroups on  $\delta\partial c$; (c)~1-st value of $c$; (d) cyclic subgroups on $\delta\partial c$; (e) 2-nd value of $c$; (f) cyclic subgroups on $\delta\partial c$; (g) 3-rd value of $c$, such that $\partial c=0$, hence stop.}
   \label{fig:3D}
\end{figure}

\subsubsection{Cyclic subgroups of the $\delta\partial c$ chain}
\label{sec:subgroups}

We remark that in computing the representation $c\in C_{d-1}$ of a $C_d$ basis element, we iterate over ``corollas" --- $(d-1)$-chains --- of coherently oriented $X_{d-1}$ cells extracted from the coboundary $\delta\partial c\in C_{d-1}$ of a cycle $\partial c\in C_{d-2}$, with $n=\#\delta\partial c$ and $m=\#\partial c$. In the following we   {iterate over subsets of ``petals"} associated to each element of $\partial c$.

Let us consider the symmetric group $S_n$ of permutations of elements of $\delta\partial c$. By linearity of $\delta$, the group elements can be partitioned into $m$ cyclic subgroups ${R^k}\subset X_{d-1}$, one-to-one mapped to each element $\tau\in\partial c$, with $\sum_{k=1}^m \#{R^k} = \#\delta\partial c = n$. {In the concrete case with $d=2$ of Example~\ref{example1}, such subgroups are represented in Figures~\ref{fig:step-by-step}b and~\ref{fig:step-by-step}d  by arrowed circles}. 

It is easy to prove that each $R_k$ contains one and only one $c$ component, computable as $\sigma=R_k\cap c$, so that the generic \emph{corolla} application returns $m$ cells, each one generated either by \emph{next}$(\sigma)(R_k)$ or by \emph{prev}$(\sigma)(R_k)$, depending on the orientation of each ``hinge" cell $\tau\in C_{d-2}$ with respect to the orientation of $\partial c$, codified by either a $+1$ or a $-1$ coefficient in $\partial c$. 

The computation of the minimal boundary of a $d$-cell is shown for $d=2$ in Example~\ref{example1} and Figures~\ref{fig:step-by-step} and~\ref{fig:2D-full}. {The above description} also holds  for higher values of $d$, and in particular in 3-space, where the cyclic subgroups are around edges---see Figures~\ref{fig:3D}b, \ref{fig:3D}d, \ref{fig:3D}f.

\subsubsection{Linear independence of the extracted {cycles}}
\label{sec:extrema}

First remind that, with abuse of language, we often use \emph{cell} as synonym of \emph{singleton chain}, {element of a basis for a chain space.}

{Therefore, when constructing the matrix of a linear operator between linear spaces, say $\partial_d: C_d\to C_{d-1}$, by building it column-wise, we actually construct an element of the basis of the first space, represented as a linear combination of basis elements of the second space. In this sense we ``build" or ``extract" one $d$-cell as a ($d-1$)-cycle. The extraction of the \texttt{LAR} representation of a $d$-cell is lately completed, by union of the  corresponding rows of the characteristic matrix $M_{d-1}$, finally getting the list of vertices of the "built" or "extracted" $d$-cell. }

{Now, the $(d-1)$-cycles  generated in Section~\ref{sec:cycles} to construct the basis $d$-cells}   \emph{are not linearly independent}, since one of them, and in particular the external unbounded cycle, is computable as the sum of all the other ones~\cite{mayeda:72}. To remove the external cycle from the basis requires a dedicated test. We may find in linear time the vertices having an extremal (max or min) value of one coordinate, and select for each one the incident subset of $(d-1)$-cycles, i.e., the incident subset of $[\partial_d^+]$ columns. Their intersection will necessary contain only the exterior cell. 
In the worst case---the wildest non convex cells, having one vertex in all extrema positions---it may be necessary to compute the absolute value of the signed volume of cells~\cite{Cattani:1990:BIO:87203.87215,Bernardini:1991:IPO:115604.115609}. The cell greatest in volume will be removed from the basis.

\subsubsection{Poset of isolated boundary cycles}

When the boundary of the chain sum of all basis $d$-cells {(called \emph{total $d$-chain} in what follows)} is an unconnected set of $(d-1)$-cycles, these $n$  \emph{isolated}  boundary components have to be compared with each other, to determine the possible relative containment and consequently their orientation. To    this purpose an efficient point classification algorithm is {used}, in order to compute the poset (partially ordered set) {induced by the} containment relation of isolated components. 

The $n\times n$ binary and antisymmetric matrix $M=(m_{ij})$ of the containment relation between the external cycles of the $n$ maximal 2-point-connected subgraphs (see Section~\ref{sec:jeh}) of each planar subproblem is constructed, computing each element of the matrix with a single point-cycle containment test, because the two corresponding cycles (the one associated to row $i$ of $M$, from which   the vertex point is extracted, and the one associated to column $j$) certainly do not intersect. 

Then the attribute of each cycle $c_j$ as either external or internal (and hence its relative orientation) is given by the parity of $\sum_{p=1}^n m_{pj}$.
Further details on this point and pseudocode are given in Section~\ref{sec:shells} and Algorithm~\ref{alg:four}.

\subsection{{Base case}: arrangement of lines in $\E^2$}
\label{sec:basic-step}

The actual {base} case in $\E^2$, starting from a collection $\mathcal{L}$ of 1D complexes, i.e., from a collection of line segments, is discussed in depth in this section. Example~\ref{example3} and Figure~\ref{fig:2D-complex} illustrate this point.
{A similar procedure, using the bipartite graph of incidence between vertices and cells, is used in higher dimensions to regularize the result, i.e., to remove the higher dimensional dangling parts.}

\subsubsection{Segment {subdivision}: linear graph}

Each input line segment in $\mathcal{L}$ is {subdivided} against all the intersecting segments, detected using a spatial indexing based on interval trees (Figure~\ref{fig:2D-complex}b). Each generated sub-segment produces a pair of \emph{unique} vertices. Quasi-coincident vertices---i.e.~very close numerically--- are finally identified, and a single 1-complex, stored as vertices and edges in a complex $X_1$, is created. {Remember that a cellular 1-complex is just a graph, and} the generation of {line sub-segments (graph edges)} is embarrassingly parallel.

\subsubsection{Maximal 2-point-connected subgraphs}
\label{sec:jeh}

The $X_1$ complex is reduced to the union of its maximal 2-point-connected subgraphs, discovered by using the~\cite{Hopcroft:1973:AEA:362248.362272} algorithm, in order to remove all dangling edges and tree subgraphs. As a consequence, $\mathcal{A}(X_1)$ generates a partition of $\E^2$ with open, connected, and regularized, but still undetected, 2-cells (Figure~\ref{fig:2D-complex}c) to be computed in the next step.

\subsubsection{Topological extraction of $X_2$}
\label{sec:topo}
{The topological gift-wrapping algorithm in 2D is used for} computation of the 1-chain representation of cells in $X_2$, providing also the signed $\partial_2$ and $\delta_1$ operators, is discussed in Section~\ref{sec:cycles}, and shown in Figures~\ref{fig:step-by-step} and~\ref{fig:2D-full}. In Figures~\ref{fig:2D-complex}d and~\ref{fig:2D-complex}e we show the extraction of 2-cell elements in $X_2$, starting from a random set of line segments.

\subsection{Comments}

Our approach is \emph{embarrassingly parallel}, since no effort
is spent to separate the problem into a number of independent 
tasks: if $B_2$ is the collection of $2$-cells in $\mathcal{S}$, with $n_2=\#B_2$, we have
$n_2$ independent tasks for the computation of planar
arrangements $X_2^\sigma$, $\sigma \in B_2$.

For example, given two complexes in $\E^3$, intersect all their
polygonal facets with each possibly intersecting facet, to produce the 0-, 1-,  and 2-cells
of their planar arrangements independently from each other. Then, order the symbolic representations of cells both internally and externally, to discover quotient sets and to compute $X_2$ in $\E^3$. Finally, extract a basis of 3-cells for the linear space $C_3$ and compute the coboundary operator $\delta_2$. 

All geometric computations
on each target facet are essentially two-dimensional, since the focus,
after a proper affine map, is on partitioning the $z=0$ plane against the
affine hulls of incident polygons, or, more precisely, against the line segments intersecting $z=0$,
so always reducing the problem to the arrangement of line segments (see Figure~\ref{fig:2D-complex}).

If computing would entail only $k$-planes, e.g.~lines in
2D and planes in 3D, then all cells in their arrangement would be connected
and {convex~\cite{Ziegler:92}, leading} to straightforward geometric and topological
computations. Conversely, in our case, because the arrangements are generated by finite geometric objects, the
resulting cellular complex may contain general non convex cells, possibly multiply-connected, and needs
 more complicated algorithms for inclusion of hole vertices in LAR cells, that we remind are just sorted sets of vertex indices. 
{A CDT (Constrained Delaunay Triangulation) of every 2-cell is performed in our implementation,} both for correct computation of the permutation subgroups~\ref{sec:subgroups}, and for visualization of non convex cells.

\section{Pseudocodes and complexity}
\label{sec:complexity}

Here we provide a slightly simplified pseudocode of some algorithms discussed in the previous sections,
and discuss their worst case complexity.  The used pseudocode style {is a blend of} Python and Julia. 
Of course, the \emph{accumulated assignment} statement $A\ $+=$\ B$ stands for $A = A + B$, where the meaning of ``$+$" symbol depends on the contest, e.g. may stand either  for sum (of chains), or for union (of sets), or for concatenation (of matrix columns). Analogously, $A\ $-=$\ B$ stands for $A = A - B$.

\subsection{Signed boundary $\mathbf{\partial_{d} : C_{d} \to C_{{d}-1}}$ computation}
\label{sec:pseudocode}

This section deals with the pseudo-coded Algorithm~\ref{alg:one}, which describes the generation of the signed matrix $[\partial_d]$ of the boundary operator that maps $C_d$ into $C_{d-1}$, whose construction was discussed in Sections~\ref{sect:alg-1}--\ref{sec:subgroups}. 
Algorithm~\ref{alg:one} is written in a dimension-independent way, and works for both $d=2$ and  $d=3$. The reader may find beneficial to mentally map $(d, d-1, d-2)$ onto $(2,1,0)$ or onto $(3,2,1)$, according to Figures~\ref{fig:step-by-step} or~\ref{fig:3D}, respectively. 

Some preliminary words about pseudocode notations: we use greek letters for the \emph{cells} of a space partition, and latin letters for  \emph{chains} of cells, all actually coded in LAR as either signed integers or arrays of signed integers.  $[\partial_d]$ or $[c_d]$ stand for general matrices or column matrices, whereas $\partial_d[h,k]$ or $c_d[\sigma]$ stand for their indexed elements. Also, $\{|c_d|\}$ stands for the set of \emph{unsigned} (nonzero) indices of the (sparse) array $[c_d]$.  It may be also useful to recall that column $\partial_d[\cdot,\sigma]$  of the operator matrix is the chain representation of the $d$-cell $\sigma$ of  $C_d$ basis, written by using the $C_{d-1}$ basis, i.e.~ as linear combination of $(d-1)$-cells of the space partition. 

Note the precondition, warning that the algorithm can only be used to compute the $\partial_d$ matrix for a cell decomposition of a $d$-space. In fact, only in this case the $(d-1)$-cells are shared by exactly \emph{two} $d$-cells, including the exterior cell. The termination predicate is a consequence of this property: the algorithm terminates when all incidence numbers in the $marks$ array equal 2, so that their sum is exactly $2n$, where $n$ is the number of $(d-1)$-cells. Of course, the actual implementation in scientific languages like Python or Julia uses sparse arrays and coordinates in $\{-1,0,1\}$ to achieve an actual efficient execution in storage space and computation time.

% Algorithm
\begin{algorithm}[h]
\SetAlgoNoLine
\DontPrintSemicolon
\tcc{{\rm\textbf{Pre-condition:} $d$ equal to space dimension, s.t.~$(d-1)$-cells are shared by \emph{two} $d$-cells}}
\tcc{{\rm ~}}
\KwIn{$[\partial_{d-1}]\qquad~$ \color{blue}{\# Compressed Sparse Column (CSC) signed matrix} $(a_{ij})$, where $a_{ij}\in\{ -1,0,1 \}$}
\KwOut{$[\partial_d^+]\qquad~$ \color{blue}{\# CSC signed matrix}}
$[\partial_d^+] = []\ $; $m,n = [\partial_{d-1}].shape\ $; $marks = Zeros(n)\ $ $\qquad~$ {\color{blue}{\# initializations}}\; 
\While{\emph{Sum}$(marks) < 2n$}{
	$\sigma = Choose(marks)$ $\qquad$ {\color{blue}{\# select the $(d-1)$-cell seed of the column extraction\;}}
	\lIf{$marks[\sigma] == 0$}{$[c_{d-1}] = [\sigma] $}
	\lElseIf{$marks[\sigma] == 1$}{$ [c_{d-1}] = [-\sigma]$}
	$[c_{d-2}] = [\partial_{d-1}]\,[c_{d-1}]$ $\qquad$ {\color{blue}{\# compute boundary $c_{d-2}$ of seed cell\;}}
	\While($\qquad$ {\color{blue}{\# loop until boundary becomes empty}}) {$[c_{d-2}] \not= []$}{
		$corolla = []$\;
		\For($\qquad$ {\color{blue}{\# for each ``hinge'' $\tau$ cell}}){$\tau \in c_{d-2}$}{
			$[b_{d-1}] = [\tau]^t [\partial_{d-1}]$  $\qquad$ {\color{blue}{\# compute the $\tau$ coboundary}}\;
			$pivot = \{|b_{d-1}|\} \cap \{|c_{d-1}|\}$  $\qquad$ {\color{blue}{\# compute the $\tau$ support}}\;
			\lIf($\qquad$ {\color{blue}{\# compute the new adj cell}}){$\tau > 0$}{$adj = Next(pivot, Ord(b_{d-1}))$}
			\lElseIf{$\tau < 0$}{$adj = Prev(pivot, Ord(b_{d-1}))$}
			\lIf($\qquad$ {\color{blue}{\# orient adj}}){$\partial_{d-1}[\tau,adj] \not= \partial_{d-1}[\tau,pivot]$}{$corolla[adj] = c_{d-1}[pivot]$}
			\lElse{$corolla[adj] = -(c_{d-1}[pivot])$}
		}
		$[c_{d-1}]\, $+$= corolla$  $\qquad$ {\color{blue}{\# insert $corolla$ cells in current $c_{d-1}$\;}}
		$[c_{d-2}] = [\partial_{d-1}]\,[c_{d-1}]$  $\qquad$ {\color{blue}{\# compute again the boundary of $c_{d-1}$\;}}
	}
	\lFor($\qquad$ {\color{blue}{\# update the counters of used cells}}) {$\sigma\in c_{d-1}$}{$marks[\sigma]\ $+=$\ 1$}
	[$\partial_d^+]\ $+$= [c_{d-1}]$  $\qquad$ \color{blue}{\# append a new column to $[\partial_d^+]$ }
}
\Return{$[\partial_d^+]$}
\caption{ \emph{Computation of signed $[\partial_d^+]$ matrix} }
\label{alg:one}
\end{algorithm}

\subsubsection{Complexity of 3-cells extraction}
\label{sec:complexity1}

In three dimensions, Algorithm~\ref{alg:one} constructs one basis 3-cell at a time (as a 2-cycle, i.e., as a closed 2-chain), building the corresponding column of the matrix $[\partial_3^+]$, including one more boundary column for each connected component of the output complex, as detailed in Section~\ref{sec:shells}. The following Algorithm~\ref{alg:four} operates on $[\partial_3^+]$, used to actually generate the operator matrix $[\partial_3]$.

The complexity of each 3-cell is measured by a set of triples (\texttt{COO} representation of sparse matrices~\cite{gemmexp}), associated with a cycle of 2-cells, where each 2-cell is always shared by two 3-cells. Hence the total number of triples, i.e.~the space complexity of the \texttt{COO} representation of $[\partial_3^+]$, is exactly $2n$, where $n$ is the number of 2-cells in the $X_2$ skeleton. 

The construction of a single 3-cell requires the search of the adjacent $\mathit{adj}$ 2-cell for each $\mathit{pivot}$ 2-cell in the boundary shell. The search for $\mathit{next}$ or $\mathit{prev}$ 2-cell as $\mathit{adj}$ requires the circular sorting of each permutation subgroup of 2-cells incident to each 1-cell on each boundary of an incomplete 2-cycle. Consequently, we need a total number of sorts of small sets (normally bounded by a small integer---4 for cubical 3-complexes---hence $O(1)$ timewise) that is bounded by the number of 1-cells. 

The subsets to be sorted are encoded in the rows of the incidence relation between 1-cells and 2-cells, i.e., by the $i,j$ indices of  non-zero elements of $[\partial_2]$. The computation of (unsigned) $[\partial_2]$ is in turn performed through $SpMSpM$ multiplication of two sparse matrices (see~\cite{Dicarlo:2014:TNL:2543138.2543294}), and hence in time linear in the complexity of the output, i.e., in the number of non-zero elements of the   $[\partial_2]$ matrix.  Summing up, if $n$  is the number of $d$-cells and $m$ is the number of $(d-1)$-cells, the time complexity of this algorithm is $O(n m\log m)$ in the worst case of unbounded complexity of $d$-cells, and roughly $O(n k\log k)$ if their complexity is bounded by $k$ faces.

\subsection{Managements of contents of non-intersecting shells}
\label{sec:shells}

The cellular 3-complexes considered in this paper may contain cells with a number of holes, and/or inclusions of smaller cells and sub-complexes. 
In the general case, deep hierarchies of inclusion are possible.
A containment relation $R$ between isolated boundary cycles, called \emph{shells} in the following, must be handled. For this purpose we detect the shells, then discover the whole inclusion graph between shells, and compute its \emph{transitive reduction}, i.e.~the smallest relation having the transitive closure of $R$ as its transitive closure~\cite{doi:10.1137/0201008}. 

An example of the transitive reduction tree of $R$ is given in Figure~\ref{fig:shells}b for a 2D case.
When $d=3$, first we compute the set of disjoint parts of the $X_2$ complex in $\E^3$, given by the connected subgraphs, called \emph{components}, of the \texttt{FV} relation. For each connected component of $X_2$, we  extract the $[\partial_3^+]$ matrix  {and remove  from it the column that corresponds to the boundary of the component, i.e., to its exterior unbounded cell}. Note that each such ``isolated" component defines a connected boundary shell. We have to detect their relative containment, and possibly remove some cycles from the boundary operator matrix, so implicitly moving the cycle to the (unconnected) boundary of the \emph{exterior} cell.

\subsubsection{Preview of algorithm}

We need to consider two main concepts here: (i) the {maximal} connected components of $X_{d-1}$, producing {disconnected} \emph{$d$-components} of the output complex $X_d$; (ii) the possible inclusions of components within single cells of the output $d$-complex.  We list in the following the main stages of  the algorithm. Our goal is the computation of both the $X_d$ skeleton, and the $\partial_d$ operator. In other words, we ``extract'' the $d$-cells of a cellular complex from the knowledge of its $X_{d-1}$ skeleton:

\begin{enumerate}

\item
First, we have to split the (isolated) connected components of the input $X_{d-1}$ skeleton, using its LAR representation; 

\item
For each (connected) component $p$ of $(d-1)$-skeleton ($1\leq p\leq h$) compute the matrix $[\partial_d^+]^p$ and its boundary cycle, and remove it  from the matrix, moving into a \emph{shell array} $D$;

\item
compute the containment relation graph between shells in $D$, and the tree $T$ of its transitive reduction $R$.

\item 
for each arc $(c^i , c^j )\in T$ with even distance from the root,   look for the $d$-cell $\rho$ of {the component with bounding shell} $c^j$ {at minimal distance from the shell $c^i$}, and 
remove the $\rho$'s {bounding cycle} from the component matrix $[\partial_d]^j$, i.e.~ declare $\rho$ is empty.

\item produce the final $[\partial_d]$ matrix, by concatenation of component matrices $[\partial_d]^p$, ($1\leq p\leq h$), and the LAR representation of $X_d$.

\end{enumerate}

\begin{figure}[htbp] %  figure placement: here, top, bottom, or page
   \centering
   \includegraphics[width=0.25\textwidth]{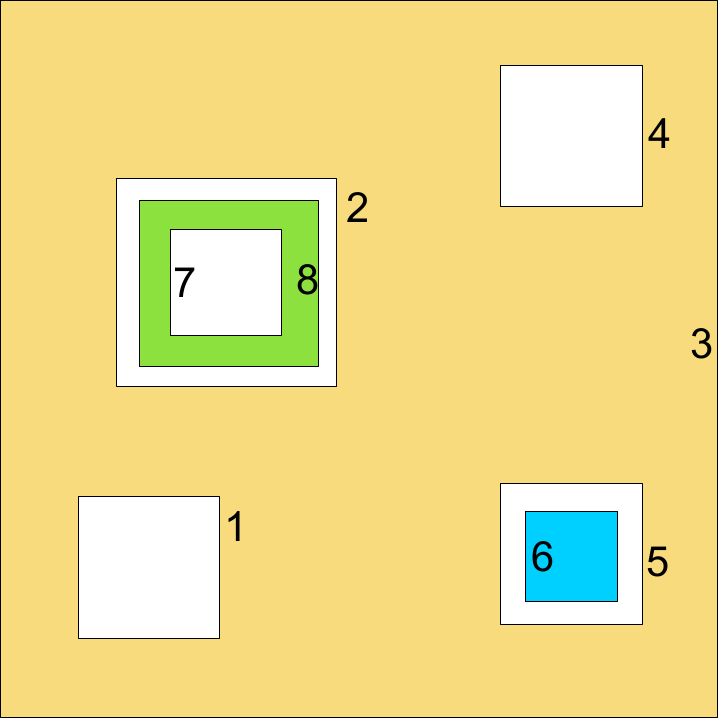} 
   \hspace{2mm}
   \includegraphics[width=0.25\textwidth]{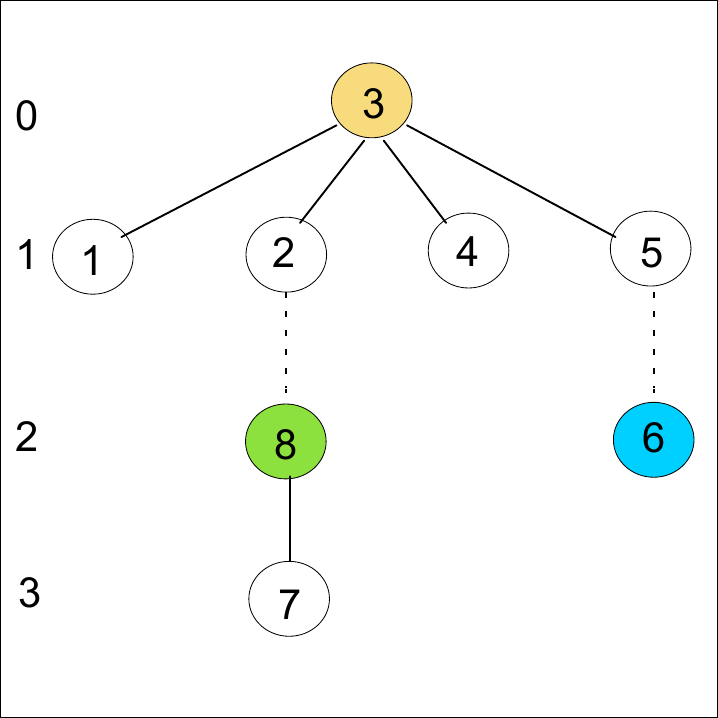} 
   \hspace{2mm}
\begin{minipage}[c]{0.3\textwidth}
	\vspace{-3.2cm}
	\small{$R = \mat{
	{\arraycolsep=1.9pt%\def\arraystretch{2.2}
	\begin{array}
	{cccccccc}
		- & 0 & 1 & 0 & 0 & 0 & 0 & 0  \\
		0 & - & 1 & 0 & 0 & 0 & 0 & 0  \\
		0 & 0 & - & 0 & 0 & 0 & 0 & 0  \\
		0 & 0 & 1 & - & 0 & 0 & 0 & 0  \\
		0 & 0 & 1 & 0 & - & 0 & 0 & 0  \\
		0 & 0 & 0 & 0 & 1 & - & 0 & 0  \\
		0 & 0 & 0 & 0 & 0 & 0 & - & 1  \\
		0 & 1 & 0 & 0 & 0 & 0 & 0 & -  \\	 
	\end{array}}}$}
\end{minipage}

\vspace{-1.5mm}
{\footnotesize\hspace{.2\textwidth}(a)\hfill(b)\hfill(c)\hspace{.2\textwidth}}
\vspace{-2.5mm}
   \caption{Non intersecting cycles within a 2D cellular complex with three connected components and only three cells, denoted by the image colors: (a) cellular complex; (b) graph of the \emph{reduced} containment relation $R$ between shells, with dashed arcs of even depth index; (c) matrix of transitively reduced R. Note that the ones equal the number of edges in the graph. }
   \label{fig:shells}
\end{figure}

\subsubsection{Pseudocode discussion}
For the sake of clarity, this subsection discusses the $d=3$ case.

Consider the bipartite graph {$G = (N, A)$, with $N= \Lambda_2 \cup \Lambda_0$, and $A \subseteq \Lambda_2 \times \Lambda_0$,} associated with the sparse matrix encoding the \texttt{FV} relation. $G$ has one node for each facet (2-cell), one node for each vertex (0-cell), and one arc for each incident pair. Therefore, the arcs in $A$ are in one-to-one correspondence with the nonzero elements of the \texttt{FV} matrix. By computing the {maximal} 2-connected components of $G$, we subdivide the $X_2$ skeleton into $h$ connected components: $\mathcal{X}_2 = \{ {X}_2^p \}$, {$1\leq p\leq h$.}

For each component ${X}_2^p$, repeat the following actions. First assemble the $[\partial_2]^p$ sparse matrix, and compute the corresponding $[\partial_3^+]^p$ generated by Algorithm~\ref{alg:one}. Then split it into the boundary operator $\partial_3^p: C_3^p\to C_2$ and  the column matrix $c^p = \partial_3^+[\sigma^p] \in C_2$ of the exterior cell $\sigma^p\in\Lambda_3$.
The set $D = \{ c^p \}$ of $h$ disjoint 2-cycles, is the initialization of the set of $X_d$ \emph{shells}. Some further (empty) shells of $X_d$ can be discovered later, resulting from mutual containment of $D$ elements. 

%Then we add to the set $D$ the empty shells already present in the input collection of complexes, when already included within some non-contractible cell. They are  called here \emph{input holes}. We discover the (unmodified) input holes by direct inspection of matrices $[\partial_3^p]$, since holes are contained there as \emph{pairs of columns with zero sum}. In fact each input hole, if non intersected by other data, returns unmodified, and produces \emph{two columns} in the component matrix $[\partial_3^p]$ it belongs to. Each corresponding pair of columns has all non-zero elements in the same rows, but with opposite signs (orientations).
%The discovering inspection is done using a sort of scan-line algorithm working on the rows of each $[\partial_3^p]$ matrix, that recognizes the emerging pairs of opposite columns, and stores each pair (candidate hole) until its contents eventually diverge. The  algorithm continues by moving from a row to the  next one, until the end of the matrix is reached and the set, possibly empty, of inner holes is returned. Each pair of columns is so removed from the component matrix $[\partial_3^p]$, and its 2-cycle is added to the set $D$ of shells.

% Algorithm
\begin{algorithm}[h]
\SetAlgoNoLine
\DontPrintSemicolon
\KwIn{\texttt{LAR}${}_{d-1}$, $[\partial_{d-1}]\qquad\qquad~$ \color{blue}{\# for $d=3$: \texttt{FV}, $\partial_2$}}
\KwOut{\texttt{LAR}${}_{d}$, $[\partial_d]\qquad\qquad~$ \color{blue}{\# for $d=3$: \texttt{CV}, $\partial_3$}}
$N = \Lambda_2 \cup \Lambda_0$; $\quad A \subseteq \Lambda_2\times\Lambda_0$; $\quad G = (N,A)$ $\qquad\qquad~$ {\color{blue}{\# initializations}}\; 
$\mathcal{G} = \{ G^p\ |\ 1\leq p\leq h \} \leftarrow \mathit{ConnectedComponents}(G)$  $\qquad\qquad~$ {\color{blue}{\# partition of $G$ into $h$ connected components}}\; 
$\mathcal{X}_{d-1} = \{ (X_{d-1}^p, \partial_{d-1}^p)\ |\ 1\leq p\leq h \} \leftarrow \mathit{Rearrange}(\mathcal{G})$  {$\qquad\qquad~$ {\color{blue}{\# partition of $X_{d-1}$ into $h$ connected components}}}\;
$D = []$ $\qquad\qquad~$ {\color{blue}{\# initialize the sparse array of \emph{shells}}}\; 
\For({$\qquad${\color{blue} \# for each connected component of $(d-1)$-skeleton}}){$p \in \{ 1,\ldots, h \}$}{
	$[\partial_d^+]^p = \mathit{Algorithm\_1}([\partial_{d-1}]^p)$ $\qquad$ {\color{blue}{\# compute the minimal $d$-cycles of a component of complex}}\;
	$(c^p, \partial_d^p) = \mathit{Split}([\partial_{d}^+]^p)$ $\qquad$ {\color{blue}{\# split the component into the exterior $(d-1)$-cycle and the boundary $\partial_d^p$}}\;
	$D\ $+=$\ [c]^p$ $\qquad\qquad~$ {\color{blue}{\# append the boundary shell to the shell array}}\; 
}
\For({$\qquad${\color{blue} \# for each shell pair $(c^i,c^j)\in R$,\ }}){$i,j \in \{ 1,\ldots, h \},i<j$}{
	$(R[i,j],R[j,i])::\mathit{Bool}\times\mathit{Bool} \leftarrow \mathit{PointSet}(\nu\in c^i,c^j)$ {$\qquad${\color{blue} \# containment test of $\nu$ in $c^j$ }}\;
}
$T = \{ (i,j) \} \leftarrow \mathit{Tree(TransitiveReduction}(R\,))$ {$\qquad${\color{blue} \# set of arcs of reduced containment tree of shells }}\;
\If({$\qquad${\color{blue} \# if the containment tree of shells is not empty}}){$T \not=\emptyset$}{
	\For({$\qquad${\color{blue} \# for each shell pair $(c^i,c^j)$ such that $dist(c^j)\%2\ $!=$\ 0$}}){$(i,j) \in T$}{
		$\rho = \mathit{FindContainerCell}(\nu,c^j,\texttt{LAR}_{d-1})$ {$\qquad${\color{blue} \# look for a $d$-cell $\rho$ such that $\nu\in  |c^i| \subseteq |\rho|\subseteq |c^j|$ }}\;
		$[\partial_d]^j\ $-=$\ \partial_d^j[\rho]$ {$\qquad${\color{blue} \# remove $\rho$ from $\partial_d^j$ }}\;
	}
	}
$\partial_d = [\partial_d^1 \cdots \partial_d^p \cdots \partial_d^h]$ {$\qquad${\color{blue} \# return the aggregate $\partial_d$ operator }} \;
$\texttt{LAR}_{d} = [\cup_k \texttt{LAR}_{d-1}(c^k = \partial_d[\cdot,k]),\ \mathit{for\ } k\in\mathit{Range}(Cols(\partial_d))]\qquad$ {\color{blue}{\# for $d=3$: \texttt{LAR}${}_{d} = \texttt{CV}$}}\;
\Return \texttt{LAR}${}_{d}$, $[\partial_d]$

\caption{ \emph{Non-intersecting shells}}
\label{alg:four}
\end{algorithm}

{Then we compute} the transitive reduction $R$ of the point-containment relation between shells in $D$, by computing for $i,j\in \mathit{range}(h), i<j$, the containment test $\mathit{pointSet}(\nu,c^j)$ between any vertex $\nu \in c^i$ and the cycle $c^j$, with $(c^i,c^j)\in R$.
If the edge set of the graph of $R$ is empty, then no disjoint component of $X_3$ is contained inside another one, and both $X_3$ and $\partial_3$ may by assembled by disjoint union of cells of $X_3^p$ and columns of $[\partial_3]^p$, respectively, for $1\leq p\leq h$.

If conversely the above is not true, then for each arc $(i,j)$ in the tree of $R$, with odd distance of $i$ node from the root, it is necessary to discover which cell of the container component $X^j$ actually contains the contained component $X^i$, i.e., its shell $c^i$. More complex intersecting situations are not possible {by construction}, since we know that the components are disjoint. Therefore, in case of containment, one component is necessarily contained in some \emph{empty cell} of the other.

To identify the 3-cell $\rho$ that is contained within a 2-cycle $c^j\in D$ and in turn contains a 0-cell $\nu\in c^i\in D$, and hence the whole $c^i$, is not difficult.
{It is achieved by shooting a ray} (halfline) in any direction from $\nu$ (e.g., in positive $x_1$ direction) against the 2-cells in $X_2^j$, i.e., the appropriate subset of rows of the sparse matrix \texttt{FV}, {and ordering parametrically} the resulting intersection points, and hence the 2-cells of $X_2^j$ intersecting the ray. The closest 2-cell will be shared by two 3-cells: the 3-cell $\rho\in X_3^j$ that does contain $\nu$, and the one that does not. 

Of course, $\rho$ is the \emph{empty cell} of $X_3^j$ that contains the \emph{whole} $X_3^i$. 
As a consequence, the column $\partial_3^j[\rho]$ has to be {removed} from $[\partial_3^j]$. We can check that (implicitly) the 2-cycle $\partial_3\rho$ is {automagically} added, with reversed coefficients, to the boundary of $X_3^j$, i.e., to the boundary cycle $[\partial_3]^j[c]^j \in C_2$ of the solid component $X_3^j$. Note that the resulting  boundary is {\emph{non} connected}.

When the above cancellation of empty cells has been performed for all  arcs of the relation tree, the new simplified matrices $[\partial_3]^p$ ($1\leq p\leq h$) can be assembled into the final $\partial_3$ operator matrix, whose column 2-cycles, suitably transformed into the union of corresponding LAR representations (subsets of vertices) of  2-cells, will provide the LAR representation of 3-cells in $X_3$.

\begin{example}\label{example4}
Consider the example in Figure~\ref{fig:shells}, where $d=2$.
In this case, by analysing the connection of $X_{d-1}=X_1$, {we obtain 8 connected subgraphs}. 
If we had directly applied Algorithm~\ref{alg:one} to the whole data set $X_1$, i.e.~to the whole matrix $[\partial_1]$, then we would have generated a matrix $[\partial_2^+]$ with dimension $e\times 16$, where $e$ is the number of edges and 8 is the number of shells. In fact, in this case every edge belongs to two {\emph{minimal} adjacent 1-cycles (shells from connected components of $X_1$ graph and their exterior cycles). }

According to the discussion above, about preliminary extraction of connected $d$-components, and to Algorithm~\ref{alg:four}, we actually extract from $X_1$ 8 shells, then 
we consider the $8\times 8$ matrix of the transitively reduced containment relation $R$. This one is antisymmetric, and we do not care about the reflexive part (see Figure~\ref{fig:shells}c).
{The final boundary matrix $[\partial_2]$ gets size $e\times 3$,} after the restructuring induced by the containment relation $R$ of shells, and considering also the parity of nodes of the tree of $R$, that determines the alternation between full and empty spaces.

{The resulting boundary matrix corresponds to} a 2-complex with \emph{three} 2-cells, denoted as $\sigma_2^1$ (yellow), $\sigma_2^2$ (green), and $\sigma_2^3$ (blue) in Figures~\ref{fig:shells}a and~\ref{fig:shells}b. Let us note that we have reduced the set of 8 isolated shells to three non-reducible 2-cells (in this example coincident one-to-one with isolated 2-components of the input complex).

\end{example}

\subsubsection{Complexity of shell management}
\label{sec:inclusion}

The computation of the connected components of a graph $G$ can be performed in linear time~\cite{Hopcroft:1973:AEA:362248.362272}. The recognition of the $h$ shells requires the computation of $[\partial_{d}^+]^p$ ($1\leq p\leq h$) and the extraction of the boundary of each connected component $X_{d}^p$ . To compute the reduced relation $R$ we execute $h^2/2$ point-cycle containment tests, linear in the size of a cycle, so spending a time $O(h^2\, k)$, where $h$ is the  umber of shells, and $k$ is the average size of cycles. Actually, the point-cycle containment test can be easily computed in parallel, with a minimal transmission overhead of the arguments.
The restructuring of boundary submatrices has the same cost of the read/rewrite of columns of a sparse matrix, depending on the number of non-zeros of $[\partial_3]$, and hence is $O(k\, \#\Lambda_d)$, i.e., linear with the product of the number of $d$-cells  $\#\Lambda_d$ and their average size $k$ as chains of $(d-1)$-cells. The parameter $k$ is a small constant for simplicial and cubical complexes.

%input holes can be done, as seen before, by scanning the rows of $[\partial_3^p]$ matrices, and hence in a time given by the product of number of rows, equal to the number of faces (2-cells), times the complexity of comparison of non-zero elements on the row with the current store of candidate pairs. This complexity equals that of merging such elements in a sorted and linked structure and is hence $O(k)$ where k is the number of non-zero elements in the row.
%The worst case is when all columns in each $[\partial_3^p]$ correspond to input holes. Therefore the worst-case complexity of shell recognition in 3D is $O(nm)$, where $n$ is the number of faces and $m$ is the number of shells in the component of highest genus.

% Algorithm
\begin{algorithm}[h]
\SetAlgoNoLine
\DontPrintSemicolon
\KwIn{$\mathcal{S}_2\subset \mathcal{S}_{d-1}$ $\qquad$ \color{blue}{\# \emph{collection} of all 2-cells from $\mathcal{S}_{d-1}$ input in $\E^{d}$}}
\KwOut{$[\partial_2]$ $\qquad$ \color{blue}{\# CSC signed matrix}}
$\widetilde{\mathcal{S}_2} = \emptyset$ $\qquad$ {\color{blue}{\# initialisation of collection of local fragments }}\;
\For($\qquad$ {\color{blue}{\# for each 2-cell $\sigma$ in the input set}}){$\sigma \in \mathcal{S}_2$}{
	$M = \mathit{SubManifoldMap}(\sigma)$ $\qquad$ {\color{blue}{\# affine transform s.t. $\sigma\mapsto x_3=0$ subspace}}\;
    $\Sigma = M\,\mathcal{I}(\sigma)$ $\qquad$ {\color{blue}{\# apply the transformation to (possible) incidencies to $\sigma$}}\;
	$\mathcal{S}_1(\sigma) = \emptyset$ $\qquad${\color{blue}{\# collection  of line segments in $x_3=0$}}\;
    \For($\qquad$ {\color{blue}{\# for each 2-cell $\tau$ in $\Sigma$}}){$\tau \in \Sigma$}{
        $\mathcal{P}(\tau),\mathcal{L}(\tau) = \emptyset,\emptyset$ $\qquad$ {\color{blue}{\# intersection points and int.~segment(s) with $x_3=0$}}\;
        \For($\qquad$ {\color{blue}{\# for each 1-cell $\lambda$ in $X_1(\tau)$}}){$\lambda \in X_1(\tau)$}{
        	\lIf($\qquad$ {\color{blue}{\# append the intersection point of $\lambda$ with $x_3=0$}}){$\lambda\not\subset\{\p{q}\,|\,x_3(\p{q})=0\}$}{
				$\mathcal{P}(\tau)\ $+=$\ \{\p{p}\}$}
		}
		$\mathcal{L}(\tau) = \mathit{Points2Segments}(\mathcal{P}(\tau))$ $\qquad$ {\color{blue}{\# Compute a set of collinear intersection segments}}\;
		$\mathcal{S}_1(\sigma)\ $+=$\ \mathcal{L}(\tau)$ $\qquad$ {\color{blue}{\# accumulate intersection segments with $\sigma$ generated by $\tau$}}\;
	}
	$X_2(\sigma) = \mathcal{A}(\mathcal{S}_1(\sigma))$ $\qquad$ {\color{blue}{\# arrangement of $\sigma$ space induced by a soup of 1-complexes}} \;
	$\widetilde{\mathcal{S}_2}\ $+=$\ M^{-1}\,X_2$  $\qquad$ {\color{blue}{\# accumulate local fragments, back transformed in $\E^d$ }}\;
}
$[\partial_1] = \mathit{QuotientBases}(\widetilde{\mathcal{S}_2})$  $\qquad$ {\color{blue}{\# identification of 0- and 1-cells using $kd$-trees and canonical LAR  }}\;
$[\partial_2] = \mathit{Algorithm\_1}([\partial_1])$ $\qquad$ {\color{blue}{\# output computation}}\;
\Return{$[\partial_2]$}
\caption{ \emph{{Subdivision} of 2-cells }}
\label{alg:two}
\end{algorithm}

\subsection{{Subdivision} of 2-cells}
\label{sec:fragmentation}

Algorithm~\ref{alg:two} was already introduced in Section~\ref{sec:basic-step} for the basic case concerning the arrangement of lines in $\E^2$. We give in this section the pseudocode and a more detailed exposition, relative to the handling of data elements producing a {subdivision} of a $(d-1)$-cell $\sigma$.  Of course, when $d=3$, we have $\mathcal{S}_{d-1}=\mathcal{S}_2$, and only this case is discussed here. {In the multidimensional case}, after {that} the 2-cells have been topologically extracted from the $(d-1)$-cells, Algorithm~\ref{alg:two} still works,  reducing to the arrangement of the 2-cell space $\sigma$ induced by a soup $\mathcal{S}_1$ of 1-complexes, i.e., of line segments, as shown in Figures~\ref{fig:cartoon-2}d, ~\ref{fig:cartoon-3}a, and~\ref{fig:cartoon-3}b, as well as in the more complex  Example~\ref{example3} and Figure~\ref{fig:2D-complex}.

\subsubsection{Complexity of 2-cells {subdivision}}

The time complexity of Algorithm~\ref{alg:two} is given by the number of 2-cells times the worst case cost required by the {subdivision} of one of them. In turn this depends on the size of the actual input, i.e., on the number of 2-cells possibly intersecting each other. It is fair to say that, in all the regular cases we usually meet in computer graphics, CAD meshes and engineering applications, the number of 2-cells incident on (even on the boundaries of) a fixed one is bounded by a constant number $k_1$. If $k_2$ is the maximum number of 1-cells on the boundary of a 2-cell, then the whole computation of  Algorithm~\ref{alg:two} requires time $O(k_1 k_2 n + A)$, where $n$ is the number of the 2-cells in the input, and $A$ is the time needed to glue all the $X_2(\sigma)$ in $\E^d$ space.  When $d=3$, the affine transformations of each set $\Sigma$ (see Section~\ref{sec:facet-arrangment}) are computable in $O(1)$ time; building a static $kd$-tree generated by $m$ points requires $O(m \log^2 m)$; and each query for finding the nearest neighbor in a balanced $kd$-tree requires $O(\log m)$ time on average. The number of occurrences of the same vertex on incident 2-cells is certainly bounded by a small constant $k_3$, approximately equal to $m/v$, where $v=\#X_0$ is the number of 0-cells after the identification processing. The transformation of output LAR in canonical form (sorted 1-array of integers)  is done in $O(1)$ for each edge, so giving $A = O(m \log^2 m) + O(m\log m) + O(1) = O(m \log^2 m)$. In conclusion, {the total running time of Algorithm~\ref{alg:two} is}  $O(k_1 k_2 n + m \log^2 m)$.
 
%\subsection{Section of a 2-cell with $x_3 = 0$ hyperplane}
%\label{sec:intersection}
%
%The robustness of the computation of the 1-cells that are non-empty \emph{section} of a 2-cell $\rho \in \mathcal{I}(\sigma)$ with the subspace $x_3 = 0$ is critical for the correctness of the whole \emph{Merge} algorithm. Hence we compute them trying to avoid all sources of numerical instability, and in particular the introduction of partially superimposed or quasi-coincident 1-cells.
%
%\subsubsection{Complexity of 2-cells section}
%
%Most of sections of a generic 2-cell $\sigma$ with $x_3=0$ are connected 1-cells, i.e.~single line segments. This fact always holds for convex $\rho$ cells that are not located upon the section hyperplane.
%
%% Algorithm
%\begin{algorithm}[h]
%\SetAlgoNoLine
%\DontPrintSemicolon
%
%\caption{ \emph{Fragmentation of 2-cells }}
%\label{alg:three}
%\end{algorithm}
%

\section{Applications and examples}\label{applications-and-examples}

An important application of the \emph{Merge} algorithm is for implementing Boolean operations between solids represented as cellular complexes, by using decompositions  of either the boundary or the interior. For example, the free space for motion planning of robots in 2D or 3D can be obtained by merging the (grown) models of obstacles within a cubical mesh of the workspace, and by computing   the generated arrangement. Other important applications may be found in biomedical applications, {e.g.,} the extraction of solid models of {small-scale biological structures} from high-resolution 3D medical images~\cite{ClementiSSPP-CAD16,doi:10.1080/16864360.2016.1168216}.

\begin{figure}[htbp] %  figure placement: here, top, bottom, or page
   \centering
\includegraphics[width=0.2\textwidth]{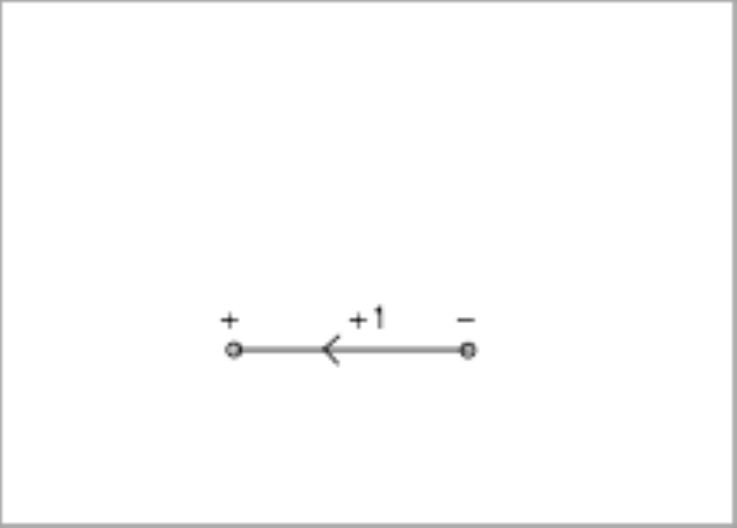}\includegraphics[width=0.2\textwidth]{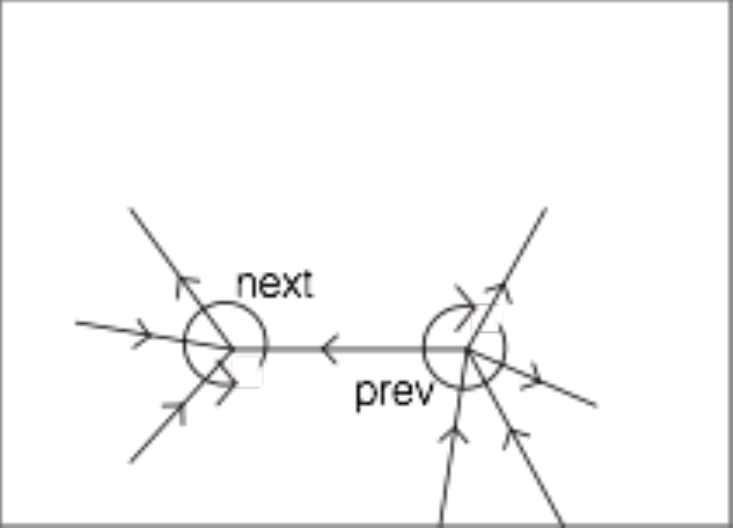}\includegraphics[width=0.2\textwidth]{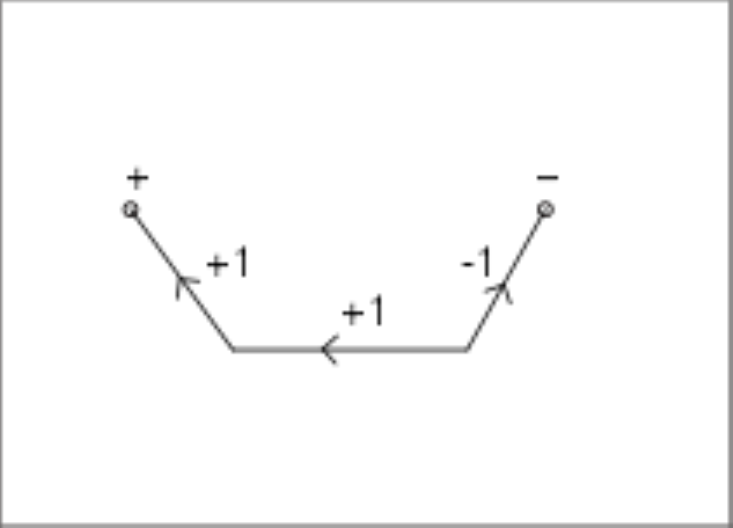}\includegraphics[width=0.2\textwidth]{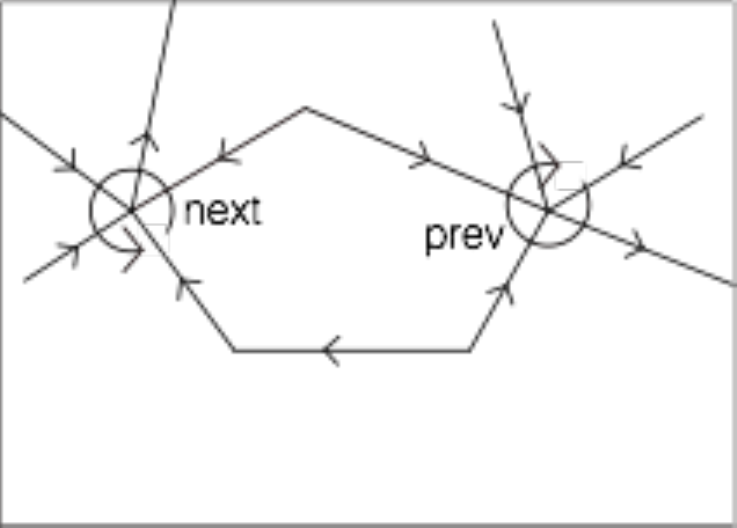}\includegraphics[width=0.2\textwidth]{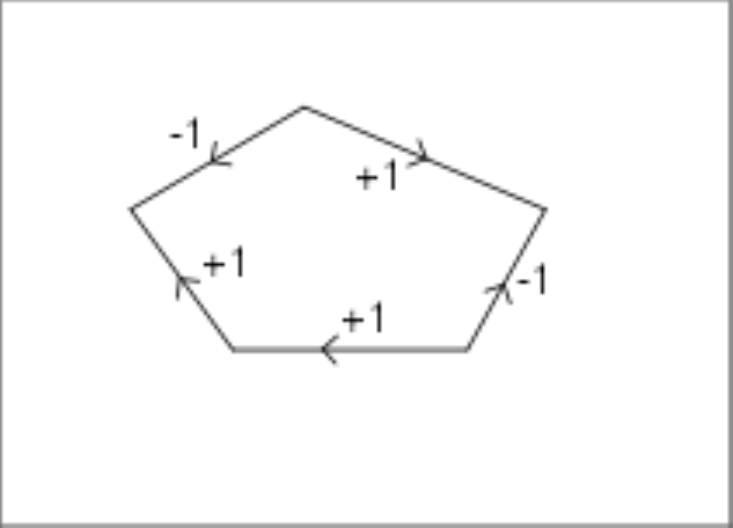}

{\footnotesize\hspace{.09\textwidth}(a)\hfill(b)\hfill(c)\hfill(d)\hfill(e)\hspace{.09\textwidth}}
\vspace{-2mm}   
\caption{Extraction of a minimal 1-cycle from $\mathcal{A}(X_1)$: (a) the initial value for $c\in C_1$ and the signs of its oriented boundary; (b) cyclic subgroups on $\delta\partial c$; (c) new (coherently oriented) value of $c$ and $\partial c$; (d) cyclic subgroups on $\delta\partial c$; (e) final value of $c$, with $\partial c = \emptyset$.}
\label{fig:step-by-step}
\end{figure}

\begin{figure}[htbp] %  figure placement: here, top, bottom, or page
   \centering
   \includegraphics[width=0.33\textwidth]{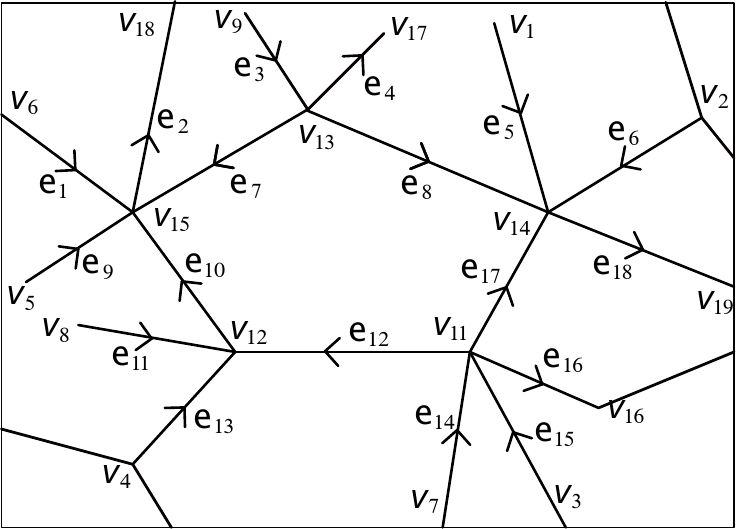} 
   \caption{A portion of the 1-complex used by Example~\ref{example1}.}
   \label{fig:2D-full}
\end{figure}

\begin{example}\label{example1}

In Figure~\ref{fig:2D-full} we show a fragment of a 1-complex $X=X_1$ in $\E_2$, with cells $v_k\in X_0$ and $e_h\in X_1$. Here we compute stepwise the 1-chain representation $c\in C_1$ of a 2-cell of the complex $X_2 = \mathcal{A}(X_1)$. Look also at Figures~\ref{fig:step-by-step}a-e to follow stepwise the extraction of the 2-cell.
\begin{enumerate}
\item[(a)]
Set $c = e_{12}$. Then $\partial c=v_{12}-v_{11}$;
\item[(b)]
then $\delta\partial c = \delta v_{12}-\delta v_{11}$ by linearity. Hence, $\delta\partial c = (e_{10}+e_{11}+e_{12}+e_{13})-(+e_{12}+e_{14}+e_{15}+e_{16}+e_{17})$.
\item[(c)]
Actually, by computing $\mbox{\emph{corolla}}(c)$ we get
\begin{align*}
\mbox{\emph{corolla}}(c) 
&= c+\mbox{\emph{next}}(c \cap \delta \partial c)\\
&= c+\mbox{\emph{next}}(e_{12})(\delta v_{12}) - \mbox{\emph{next}}(e_{12})(\delta v_{11})\\
&= e_{12}+\mbox{\emph{next}}(e_{12})(\delta v_{12}) + \mbox{\emph{prev}}(e_{12})(\delta v_{11})\\
&= e_{12}+e_{10}+e_{17}
\end{align*}
If we orient  $c$ coherently, we get $c = e_{10}+e_{12}-e_{17}$, and $\partial c=v_{15}-v_{12}+v_{12}-v_{11}+v_{11}-v_{14} = v_{15}-v_{14}$.
\item[(d)]
As before, we repeat and orient coherently the computed 1-chain:
\begin{align*}
\mbox{\emph{corolla}}(c) 
&= c+\mbox{\emph{next}}(c \cap \delta \partial c)\\
&= c+\mbox{\emph{next}}(e_{10})(\delta v_{15}) - \mbox{\emph{next}}(e_{17})(\delta v_{14})\\
&= e_{10}+e_{12}-e_{17}+\mbox{\emph{next}}(e_{10})(\delta v_{15}) + \mbox{\emph{prev}}(e_{17})(\delta v_{14})\\
&= e_{10}+e_{12}-e_{17}-e_{7}+e_{8}
\end{align*}
\item[(e)]
Finally, $\partial\, \mbox{\emph{corolla}}(c) = \emptyset $,
and the extraction algorithm terminates, giving 
$e_{10}+e_{12}-e_{17}-e_{7}+e_{8}$ as the $C_1(X)$ representation for a basis element of $C_2(X)$, with $X = \mathcal{A}(X_1)$,  and hence as a column for the oriented matrix of the unknown $\partial_2: C_2\to C_1$. 
\end{enumerate}

\end{example}

\begin{example}\label{example2}

An example of extraction of a minimal 2-cycle $c$, representation in $C_2$ of a basis element of $C_3$, is shown in Figure~\ref{fig:3D}. The coefficients of the linear combination provide a matrix column of the coordinate representation of the operator $\partial_3: C_3\to C_2$. 
%For the sake of simplicity, we number the used cell incrementally. This is not the concrete case, of course.  

Let us preliminary recall some notions about $p$-chains as linear combinations of oriented $(p-1)$-chains, for $0\leq p\leq d$ (see Section~\ref{sec:cells-chains}).
It is possible to see that the ordering (numbering) of vertices, edges, and faces (i.e.~of 0-, 1-, and 2-cells) completely determines the pattern of signs in the matrices of boundary/coboundary operators.

\end{example}

\begin{example}[{Arrangement of line segments}]\label{example3}
In Figure~\ref{fig:2D-complex} we show the whole algorithm pipeline for the computation of the plane arrangement $X_2 = \mathcal{A}(\mathcal{L})$ generated by a set $\mathcal{L}$ of random line segments in $\E^2$. {The merge algorithm introduced in this paper directly produces the regularized arrangement. However,  the  non-regularized solution could readily be recovered by  \emph{a posteriori} addition of the dangling edges and trees, previously removed.}
\end{example}

\begin{figure}[htbp] %  figure placement: here, top, bottom, or page
   \centering
   \includegraphics[height=0.25\textwidth,width=0.2\textwidth]{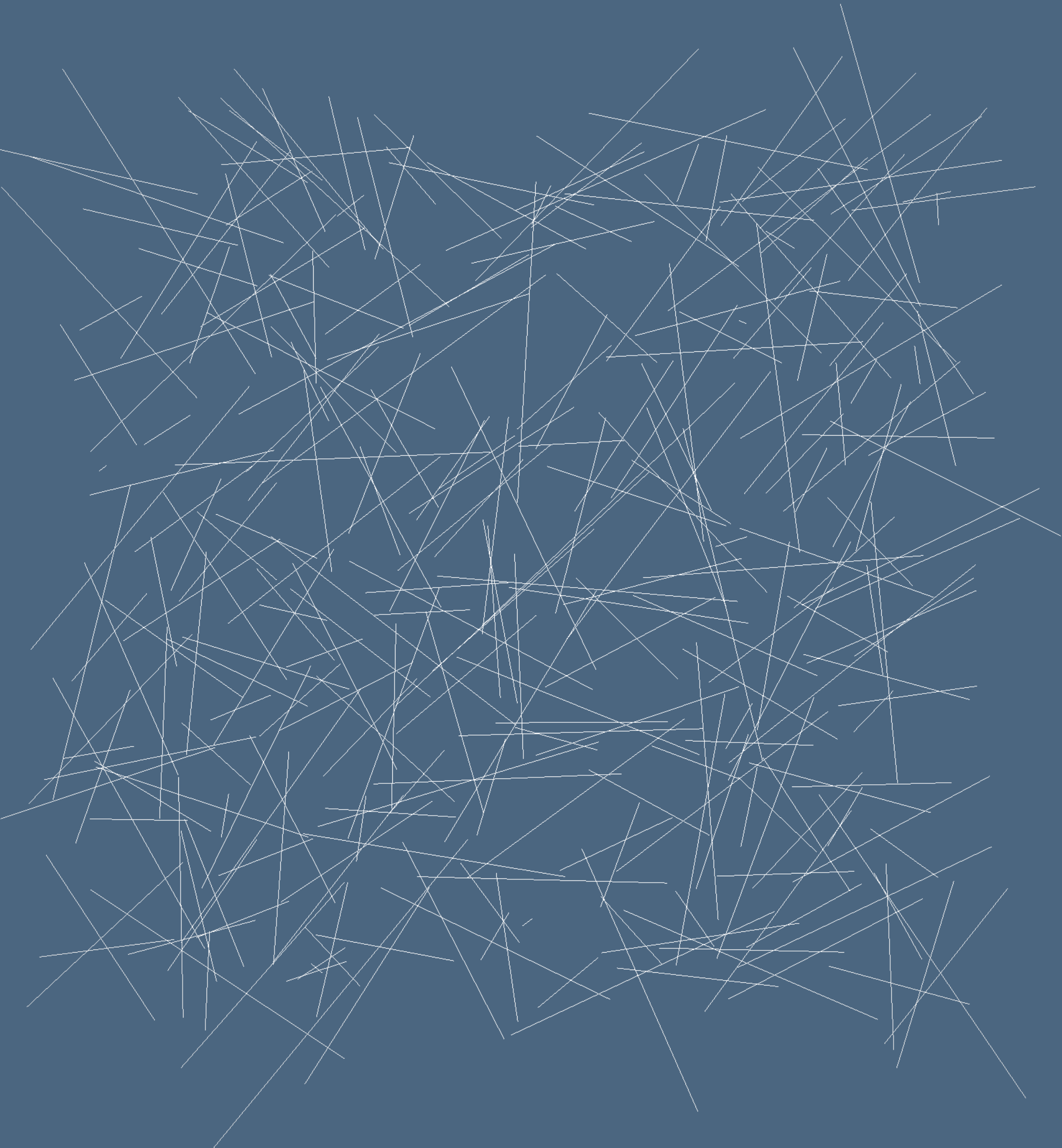}\includegraphics[height=0.25\textwidth,width=0.2\textwidth]{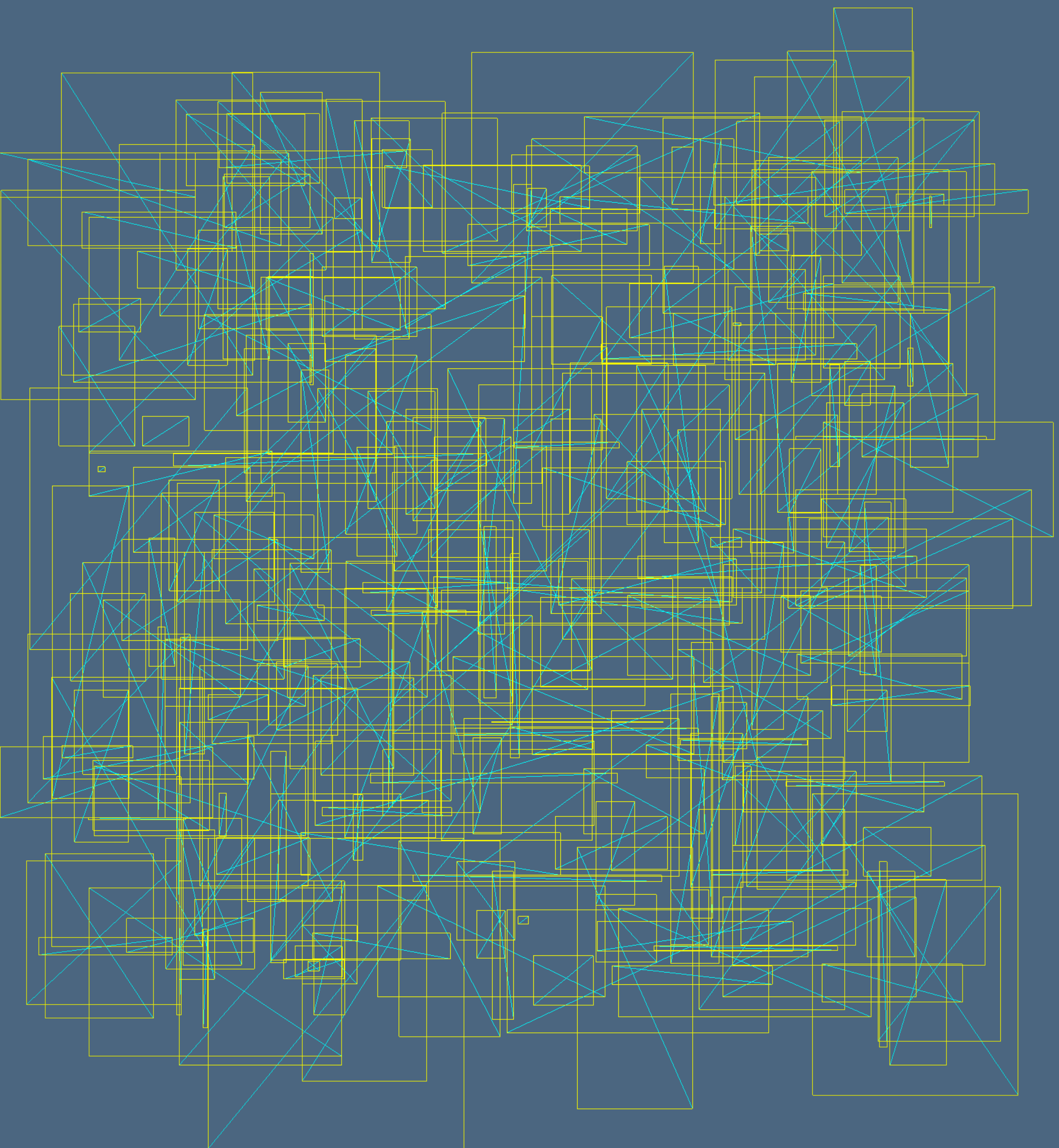}\includegraphics[height=0.25\textwidth,width=0.2\textwidth]{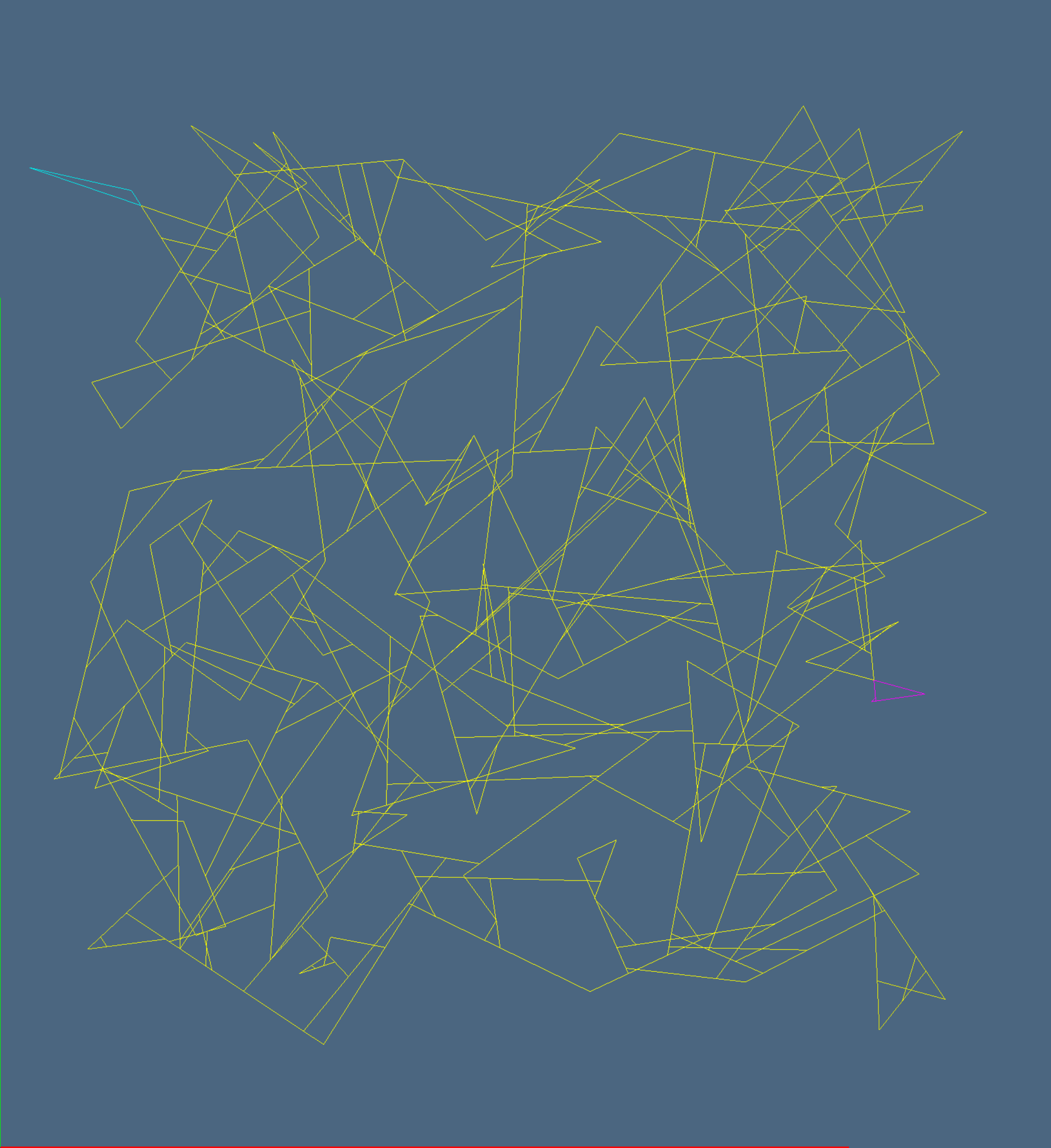}\includegraphics[height=0.25\textwidth,width=0.2\textwidth]{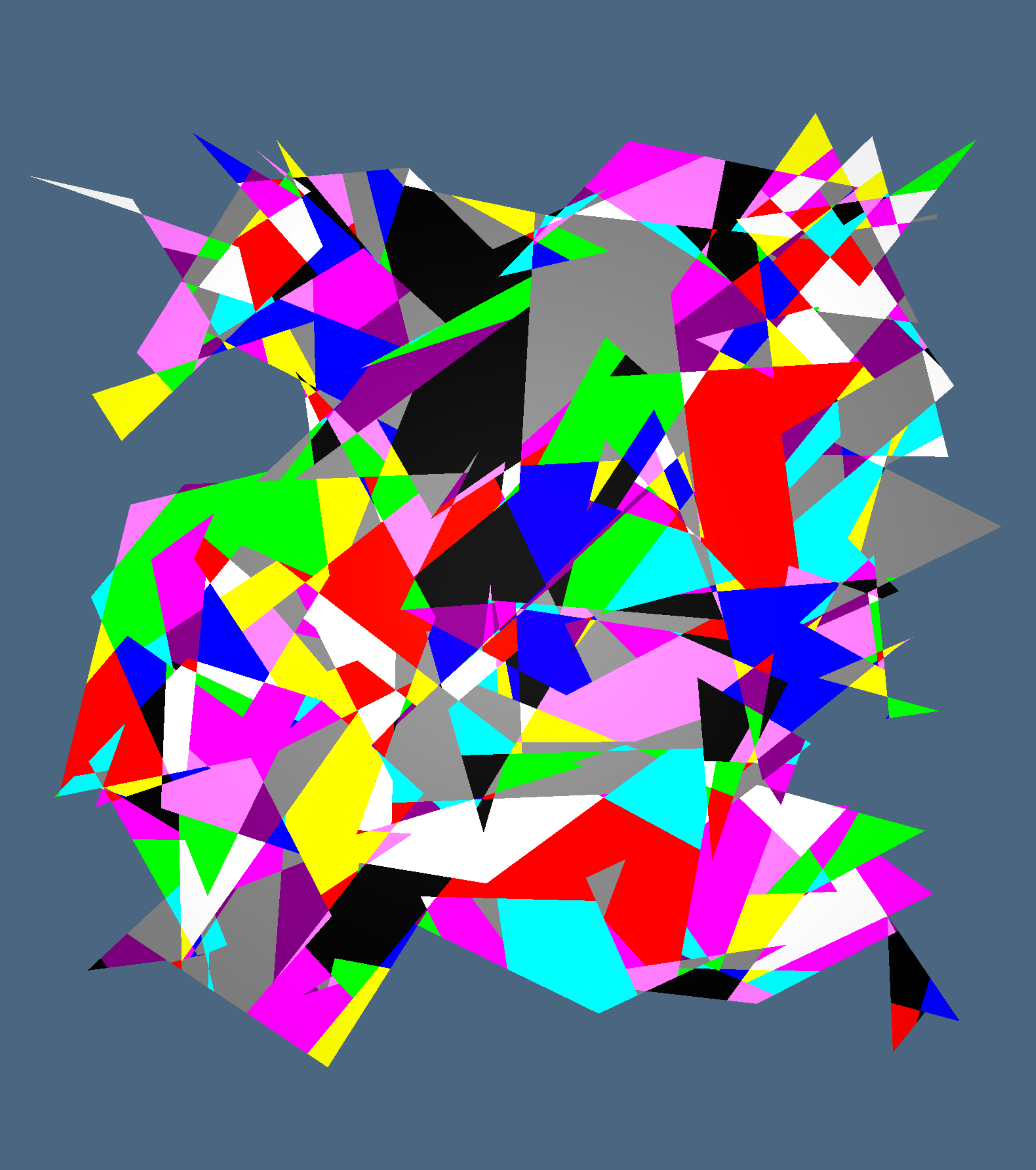}\includegraphics[height=0.25\textwidth,width=0.2\textwidth]{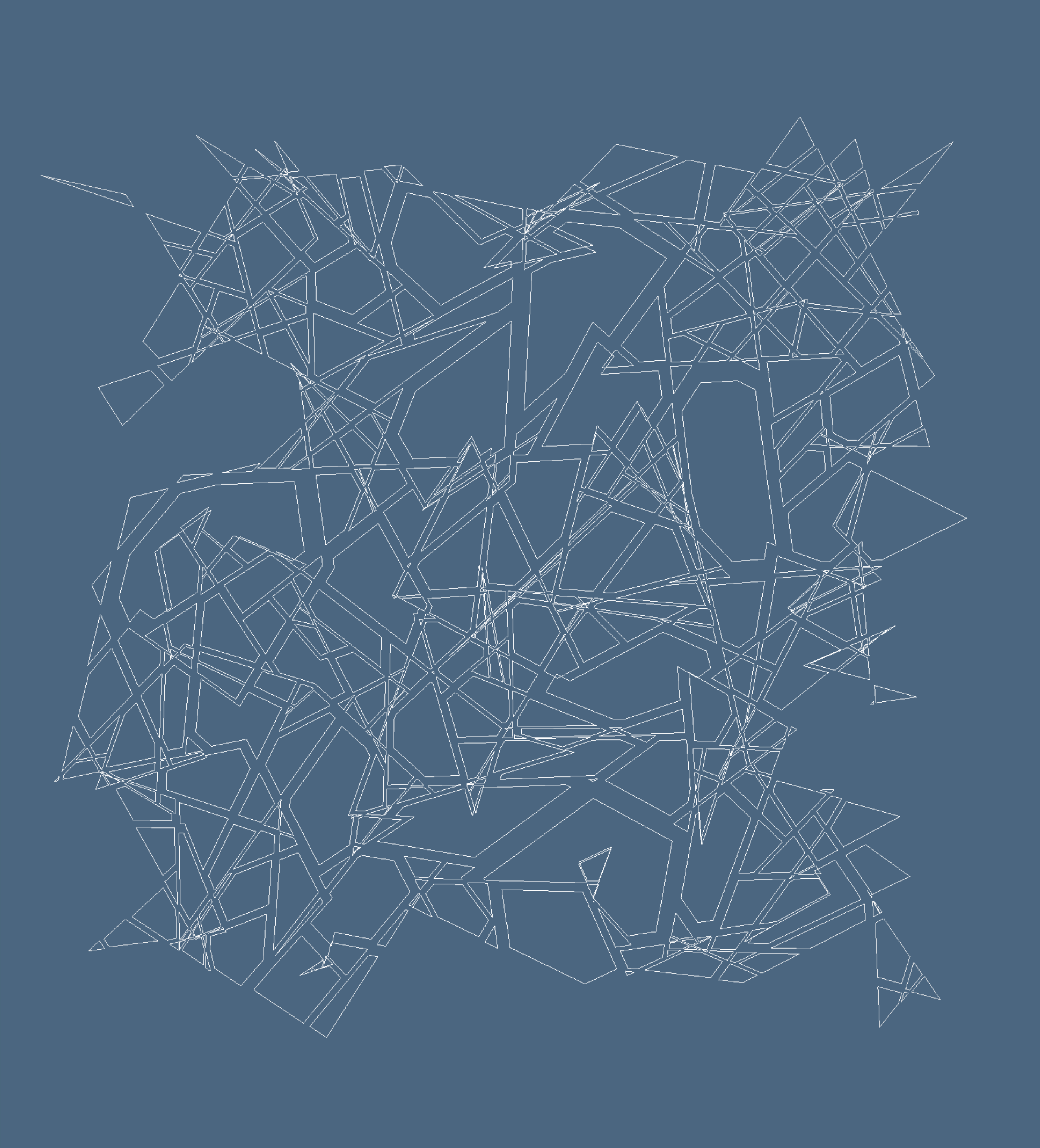}
   
{\footnotesize\hspace{.09\textwidth}(a)\hfill(b)\hfill(c)\hfill(d)\hfill(e)\hspace{.09\textwidth}}
   \vspace{-3mm}\caption{Computation of the plane arrangement $\mathcal{A}(\mathcal{L})$ generated by a set of line segments: (a) set of random lines in $\E^2$; (b) spatial index based on containment boxes; (c) plane partition $\mathcal{A}(X_1)$, where the 1-complex $X_1$ is the union of maximal 2-node-connected subgraphs in $\mathcal{A}(\mathcal{L})$, drawn here in yellow and cyan; (d) cells of $X_2$ in $X=\mathcal{A}(X_1)$, randomly colored; (e) exploded boundaries of the 2-cells in $X_2$.}
   \label{fig:2D-complex}
\end{figure}

\section{Conclusion}\label{conclusion}

In this paper we have introduced a computational topology method for constructing the space arrangement induced by a collection of cellular complexes. The most attractive feature of this approach consists in its minimal {subdivision} of the output cells:  no part of the boundary of any output cell  was not present in the boundary of input cells. 

The \emph{Merge} algorithm applies to any kind of  decompositions of either the interior or the boundary of the merged spaces, including polytopal (e.g.~Voronoi) complexes or complexes with non convex cells, even punctured, i.e., non contractible to a point {(with holes)}. 

The output of the \emph{Merge} algorithm produces not only a complete generation of the cellular $d$-complex induced by a collection of $d$-complexes, but also the associated \emph{chain complex}, hence including all the signed boundary/coboundary operators, a consistent orientation of all cells, and hence the complete knowledge of the topology. 

The introduced method differs significantly from other approaches, not only because of the language, based on chain complexes and their operators, but also for the actual computer implementation, that uses only two classical data structures ($kd$-tree of vertices and 1D interval trees) as computation accelerators, without introducing %fancy 
representations typical of non-manifold solid modeling, that  might seem mandatory in this kind of algorithms and applications. 

The actual implementation, based on LAR~\cite{Dicarlo:2014:TNL:2543138.2543294} and  {implemented} in Julia language~\cite{BEKS14} in a parallel computational environment using CUDA, efficiently describes cells and chains with either dense or sparse arrays of signed integers, and their linear operators with sparse matrices. It is currently being used for deconstruction modeling of buildings~\cite{visigrapp17:cvdlab}, as well as for the extraction of solid models of biomedical structures at the cellular scale~\cite{doi:10.1080/16864360.2016.1168216,ClementiSSPP-CAD16}.

\section*{Acknowledgements}
This work is supported in part by a grant 2016/17 from SOGEI, the ICT company of Italian Ministry of Economy and Finance, and by the  EU project \href{https://www.medtrain3dmodsim.eu}{medtrain3dmodsim}. {Vadim Shapiro is supported in part by National Science Foundation grant CMMI-1344205 and National Institute of Standards and Technology. }{The authors would like to acknowledge their gratitude to Giorgio Scorzelli, implementor and maintainer of the dimension-independent library \texttt{pyplasm}, without which this work could have not been possible. }

\appendix
\label{appendix}

\section{APPENDIX: Examples of LAR}
\label{sec:lar-examples}

In this Appendix we give some examples of LAR use, to {show} the reader how a cellular complex is defined, how to compute a signed boundary operator, and how some topological adjacency relations may be computed in 2D and 3D, respectively, by multiplication of sparse matrices.

\begin{example}[2D cellular complex] % example 1

In Figure~\ref{fig:ex1} we show an example of 2D cellular complex, with $\#\texttt{V} = 22$ vertices (0-cells), $\#\texttt{EV} = 34$ edges (1-cells), $\#\texttt{FV} = 13$ faces (2-cells). The full user-readable representation is given below. Note that \texttt{EV} and \texttt{FV} respectively provide the canonical (sorted) representations  of edges and faces as lists of lists of vertex indices, that can be interpreted as the user-readable CSR sparse characteristic matrices $M_1$ and $M_2$ of the bases of 1-chains and 2-chains, respectively.

\begin{verbatim}
V = [[0.5,0.2475],[0.5,0.0],[0.5,0.7525],[0.7525,0.0],[0.0,0.0],[0.7525,0.7475],
[0.8787,0.5],[0.0,0.5],[0.2475,0.7525],[0.5,0.5],[0.2475,0.0],[0.8787,0.2475],
[0.2475,0.5],[0.2475,0.2475],[0.7525,0.2475],[1.0,0.5],[0.0,1.0],[0.7525,0.5],
[0.5,1.0],[1.0,0.0],[1.0,0.2475],[0.2475,1.0]]

EV = [[5,15],[5,17],[5,18],[6,15],[15,20],[6,17],[11,20],[11,14],[6,11],[3,19], 
[19,20],[3,14],[1,3],[14,17],[0,14],[9,17],[2,18],[18,21],[2,9],[8,21],[8,12], 
[2,8],[16,21],[7,16],[0,1],[1,10],[0,9],[12,13],[10,13],[0,13],[7,12],[4,10], 
[4,7],[9,12]]

FV = [[5,6,15,17],[2,5,9,17,18],[6,11,15,20],[6,11,14,17],[3,11,14,19,20], 
[0,1,3,14],[0,9,14,17],[2,8,18,21],[2,8,9,12],[7,8,12,16,21],[0,1,10,13], 
[0,9,12,13],[4,7,10,12,13]]
\end{verbatim}

We would like to note that, by means of the canonical representation, we can execute efficiently, via string syntax, both equality tests and/or set operations on the textual representation of chains and cells. This approach is used, e.g., in order to perform the identification of coincident cells and their reduction to quotient sets.

\begin{figure}[htbp] 
   \centering
   \includegraphics[height=.327\textwidth,width=.327\textwidth]{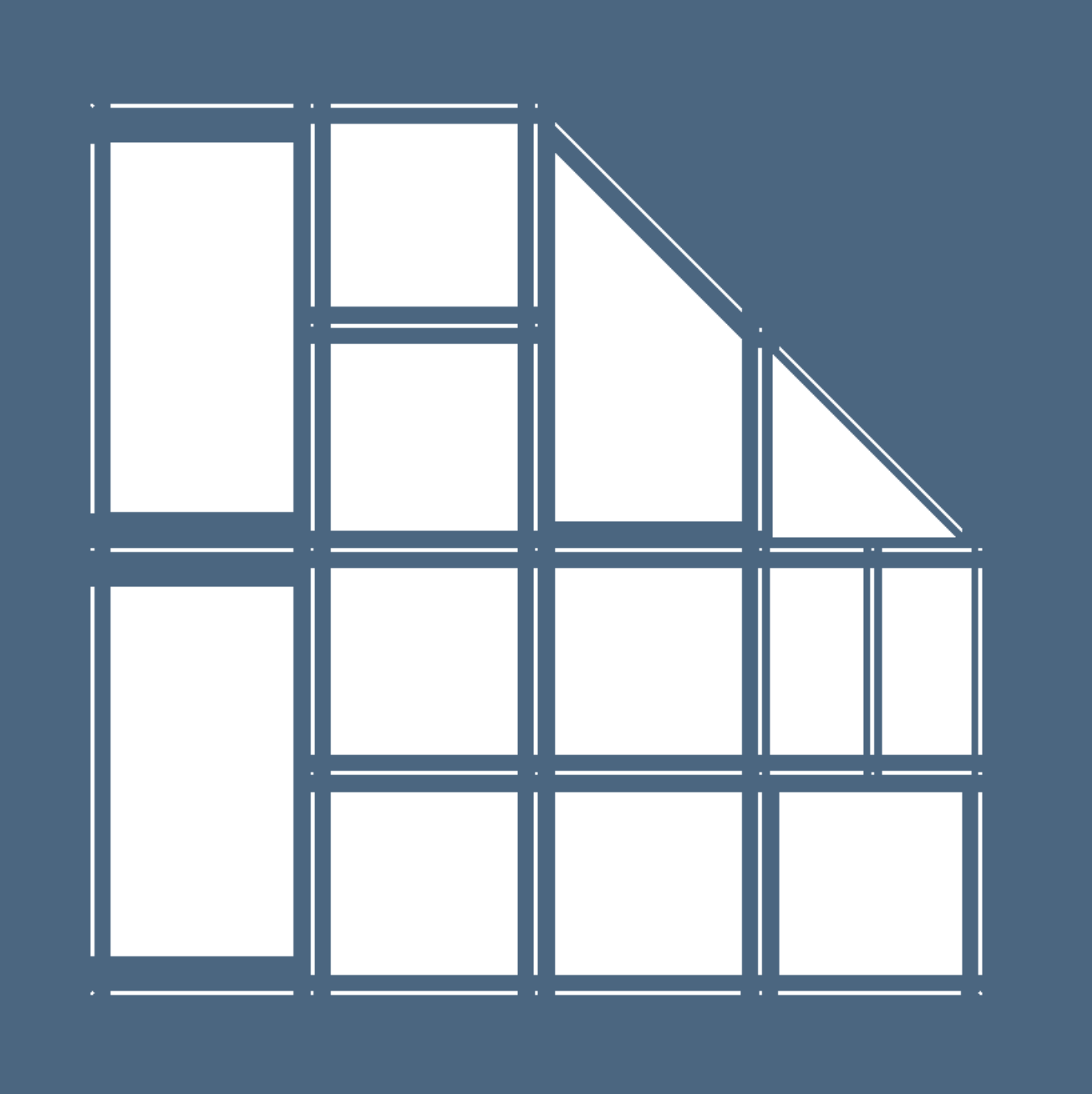} 
   \includegraphics[height=.327\textwidth,width=.327\textwidth]{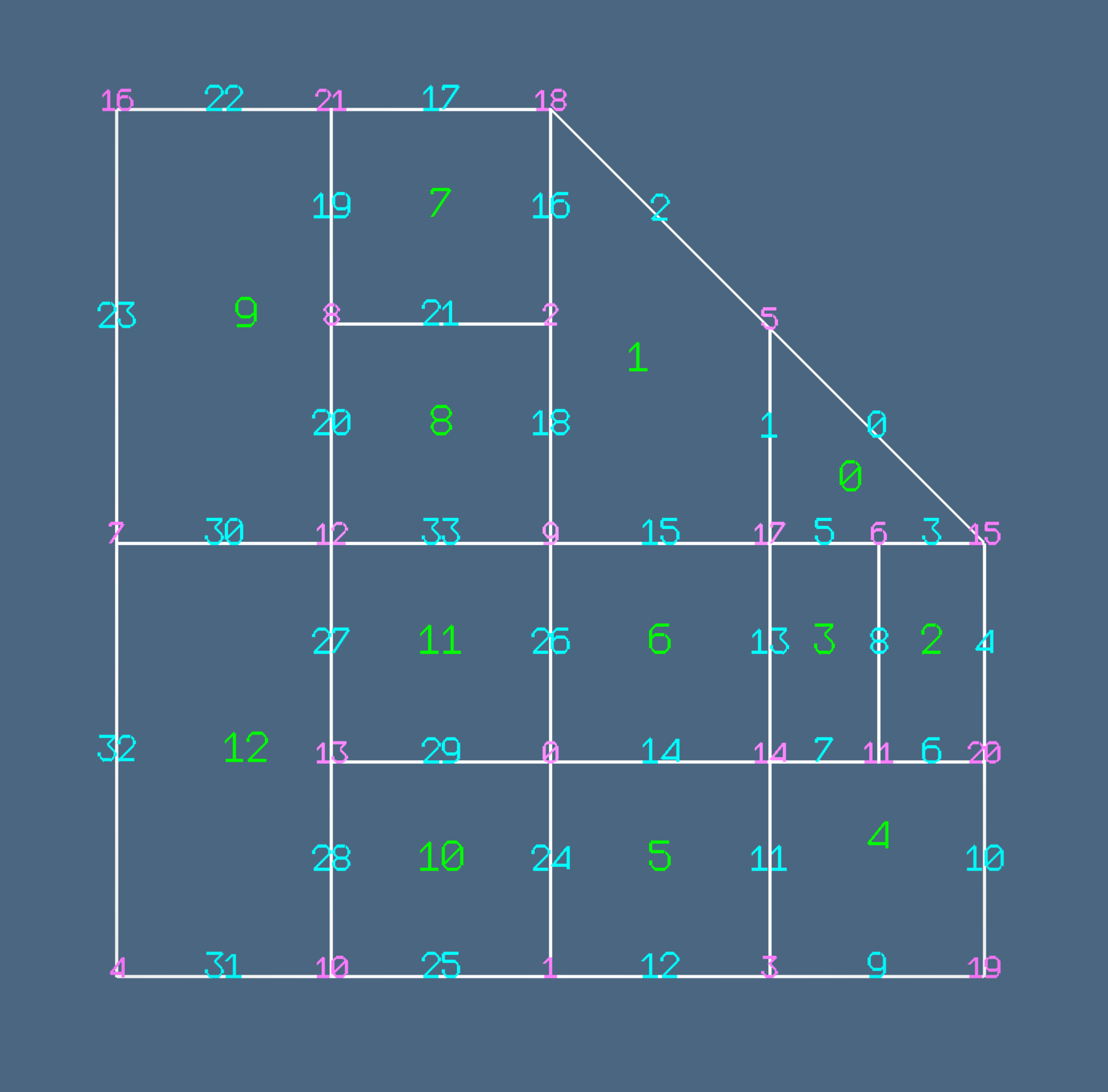} 
   \includegraphics[height=.327\textwidth,width=.327\textwidth]{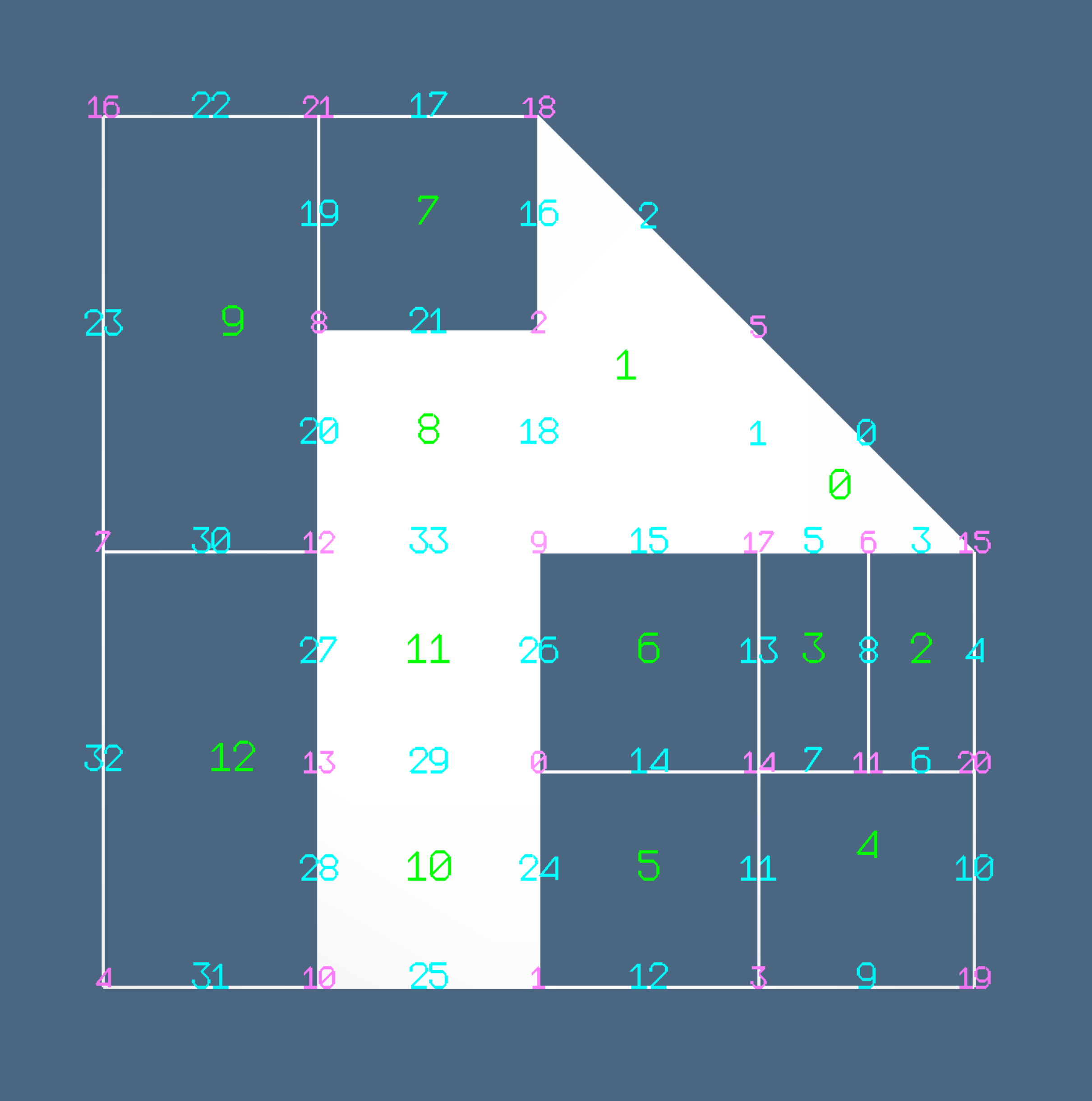} 
   \caption{LAR representation of a computation on  a cellular complex: (a) 2-complex exploded; (b) numbering of basis elements of $C_0$ (magenta), $C_1$ (cyan), and $C_2$ (green); (c) the 2-chain $\texttt{LAR}([g]) = [0,1,8,11,12]$ (white) in $C_2$, and its boundary  $\texttt{LAR}([\partial_2][g]) = [-0, 2, 3, -5,15,-16,20,21,-24,-25,26,27,-28]$, 1-cycle in $C_1$.}
   \label{fig:ex1}
\end{figure}

\end{example} % example 1

\begin{example}[Boundary computation] % example 2 
\label{ex:ex1}
In order to compute the boundary of a 2-chain $g = c_0 + c_1 + c_8 + c_{11} +c_{12}$, simply represented in LAR as \texttt{[0,1,8,11,12]}, i.e.~the 1-cycle bounding its cells, we must multiply the coordinate matrix representation  $[\partial_2]$ of the boundary operator $\partial_2: C_2\to C_1$, times the coordinate representation of $g$. Once fixed an ordering of the $C_2$ basis, the coordinate representation of any cell is unique. In this case we have
\[
[g] = [1,1,0,0,0,0,0,0,1,0,0,1,1]^t,
\]
and
\[
[\partial_2]^t = \mat{
{\tiny\arraycolsep=2pt\def\arraystretch{0.8}
\begin{array}
{rrrrrrrrrrrrrrrrrrrrrrrrrrrrrrrrrrrrrrr}
-1 & 1 & 0 & 1 & 0 &-1 & 0 & 0 & 0 & 0 & 0 & 0 & 0 & 0 & 0 & 0 & 0 & 0 & 0 & 0 & 0 & 0 & 0 & 0 & 0 & 0 & 0 & 0 & 0 & 0 & 0 & 0 & 0 & 0\\
 0 &-1 & 1 & 0 & 0 & 0 & 0 & 0 & 0 & 0 & 0 & 0 & 0 & 0 & 0 & 1 &-1 & 0 & 1 & 0 & 0 & 0 & 0 & 0 & 0 & 0 & 0 & 0 & 0 & 0 & 0 & 0 & 0 & 0\\
 0 & 0 & 0 & 1 & 1 & 0 &-1 & 0 &-1 & 0 & 0 & 0 & 0 & 0 & 0 & 0 & 0 & 0 & 0 & 0 & 0 & 0 & 0 & 0 & 0 & 0 & 0 & 0 & 0 & 0 & 0 & 0 & 0 & 0\\
 0 & 0 & 0 & 0 & 0 &-1 & 0 & 1 & 1 & 0 & 0 & 0 & 0 & 1 & 0 & 0 & 0 & 0 & 0 & 0 & 0 & 0 & 0 & 0 & 0 & 0 & 0 & 0 & 0 & 0 & 0 & 0 & 0 & 0\\
 0 & 0 & 0 & 0 & 0 & 0 & 1 &-1 & 0 &-1 &-1 & 1 & 0 & 0 & 0 & 0 & 0 & 0 & 0 & 0 & 0 & 0 & 0 & 0 & 0 & 0 & 0 & 0 & 0 & 0 & 0 & 0 & 0 & 0\\
 0 & 0 & 0 & 0 & 0 & 0 & 0 & 0 & 0 & 0 & 0 & 1 & 1 & 0 &-1 & 0 & 0 & 0 & 0 & 0 & 0 & 0 & 0 & 0 & 1 & 0 & 0 & 0 & 0 & 0 & 0 & 0 & 0 & 0\\
 0 & 0 & 0 & 0 & 0 & 0 & 0 & 0 & 0 & 0 & 0 & 0 & 0 & 1 & 1 &-1 & 0 & 0 & 0 & 0 & 0 & 0 & 0 & 0 & 0 & 0 &-1 & 0 & 0 & 0 & 0 & 0 & 0 & 0\\
 0 & 0 & 0 & 0 & 0 & 0 & 0 & 0 & 0 & 0 & 0 & 0 & 0 & 0 & 0 & 0 & 1 & 1 & 0 &-1 & 0 &-1 & 0 & 0 & 0 & 0 & 0 & 0 & 0 & 0 & 0 & 0 & 0 & 0\\
 0 & 0 & 0 & 0 & 0 & 0 & 0 & 0 & 0 & 0 & 0 & 0 & 0 & 0 & 0 & 0 & 0 & 0 &-1 & 0 & 1 & 1 & 0 & 0 & 0 & 0 & 0 & 0 & 0 & 0 & 0 & 0 & 0 &-1\\
 0 & 0 & 0 & 0 & 0 & 0 & 0 & 0 & 0 & 0 & 0 & 0 & 0 & 0 & 0 & 0 & 0 & 0 & 0 & 1 &-1 & 0 &-1 &-1 & 0 & 0 & 0 & 0 & 0 & 0 & 1 & 0 & 0 & 0\\
 0 & 0 & 0 & 0 & 0 & 0 & 0 & 0 & 0 & 0 & 0 & 0 & 0 & 0 & 0 & 0 & 0 & 0 & 0 & 0 & 0 & 0 & 0 & 0 &-1 &-1 & 0 & 0 &-1 & 1 & 0 & 0 & 0 & 0\\
 0 & 0 & 0 & 0 & 0 & 0 & 0 & 0 & 0 & 0 & 0 & 0 & 0 & 0 & 0 & 0 & 0 & 0 & 0 & 0 & 0 & 0 & 0 & 0 & 0 & 0 & 1 & 1 & 0 &-1 & 0 & 0 & 0 & 1\\
 0 & 0 & 0 & 0 & 0 & 0 & 0 & 0 & 0 & 0 & 0 & 0 & 0 & 0 & 0 & 0 & 0 & 0 & 0 & 0 & 0 & 0 & 0 & 0 & 0 & 0 & 0 &-1 & 1 & 0 &-1 & 1 &-1 & 0\\
\end{array}
}}
\]

By multiplication of $[\partial_2]: C_2\to C_1$ times $[g]\in C_2$ we get the boundary 1-cycle in coordinate notation:
\[
[\partial_2][g] = 
[-1,0,1,1,0,-1,0,0,0,0,0,0,0,0,0,1,-1,0,0,0,1,1,0,0,-1,-1,1,1,-1,0,0,0,0,0]^t
\]
that, in the LAR formalism of signed integers for cells and chains, becomes the 1-array:
\begin{verbatim}
[-0, 2, 3, -5, 15, -16, 20, 21, -24, -25, 26, 27, -28]
\end{verbatim}
i.e.\footnote{Of course, $-0$ is just a symbolic notation here. In Python, where array indexing is 0-based, the indices of cells and their sign are maintained separated within sparse arrays. Conversely, the notation of $d$-chains as arrays of signed (nonzero) integers can be used directly  in Julia, where arrays are 1-based. }, in mathematical notation for chains:
\[
\partial_2 g = (-c_0+c_2+c_3-c_5+c_{15}-c_{16}+c_{20}+c_{21}-c_{24}-c_{25}+c_{26}+c_{27}-c_{28})\in C_1
\]
that the reader may readily check as  counterclockwise coherently oriented.
\end{example} % example 2

\begin{example}[Adjacency of vertices] % example 3

Here we compute the adjacency relation of vertices of Example~\ref{ex:ex1}, simply represented by a sparse matrix $\texttt{VV}=M_1^t M_1$, defined by this product of the characteristic matrix $M_1$ of edges (given in readable form by the array \texttt{EV} of Example~\ref{ex:ex1}). The reader is sent to~\cite{Dicarlo:2014:TNL:2543138.2543294} for details and other topological relationships.

\[
\texttt{VV}=M_1^t M_1 = \mat{
{\tiny\arraycolsep=1.9pt%\def\arraystretch{2.2}
\begin{array}
{cccccccccccccccccccccc}
4 & 1 & 0 & 0 & 0 & 0 & 0 & 0 & 0 & 1 & 0 & 0 & 0 & 1 & 1 & 0 & 0 & 0 & 0 & 0 & 0 & 0\\
1 & 3 & 0 & 1 & 0 & 0 & 0 & 0 & 0 & 0 & 1 & 0 & 0 & 0 & 0 & 0 & 0 & 0 & 0 & 0 & 0 & 0\\
0 & 0 & 3 & 0 & 0 & 0 & 0 & 0 & 1 & 1 & 0 & 0 & 0 & 0 & 0 & 0 & 0 & 0 & 1 & 0 & 0 & 0\\
0 & 1 & 0 & 3 & 0 & 0 & 0 & 0 & 0 & 0 & 0 & 0 & 0 & 0 & 1 & 0 & 0 & 0 & 0 & 1 & 0 & 0\\
0 & 0 & 0 & 0 & 2 & 0 & 0 & 1 & 0 & 0 & 1 & 0 & 0 & 0 & 0 & 0 & 0 & 0 & 0 & 0 & 0 & 0\\
0 & 0 & 0 & 0 & 0 & 3 & 0 & 0 & 0 & 0 & 0 & 0 & 0 & 0 & 0 & 1 & 0 & 1 & 1 & 0 & 0 & 0\\
0 & 0 & 0 & 0 & 0 & 0 & 3 & 0 & 0 & 0 & 0 & 1 & 0 & 0 & 0 & 1 & 0 & 1 & 0 & 0 & 0 & 0\\
0 & 0 & 0 & 0 & 1 & 0 & 0 & 3 & 0 & 0 & 0 & 0 & 1 & 0 & 0 & 0 & 1 & 0 & 0 & 0 & 0 & 0\\
0 & 0 & 1 & 0 & 0 & 0 & 0 & 0 & 3 & 0 & 0 & 0 & 1 & 0 & 0 & 0 & 0 & 0 & 0 & 0 & 0 & 1\\
1 & 0 & 1 & 0 & 0 & 0 & 0 & 0 & 0 & 4 & 0 & 0 & 1 & 0 & 0 & 0 & 0 & 1 & 0 & 0 & 0 & 0\\
0 & 1 & 0 & 0 & 1 & 0 & 0 & 0 & 0 & 0 & 3 & 0 & 0 & 1 & 0 & 0 & 0 & 0 & 0 & 0 & 0 & 0\\
0 & 0 & 0 & 0 & 0 & 0 & 1 & 0 & 0 & 0 & 0 & 3 & 0 & 0 & 1 & 0 & 0 & 0 & 0 & 0 & 1 & 0\\
0 & 0 & 0 & 0 & 0 & 0 & 0 & 1 & 1 & 1 & 0 & 0 & 4 & 1 & 0 & 0 & 0 & 0 & 0 & 0 & 0 & 0\\
1 & 0 & 0 & 0 & 0 & 0 & 0 & 0 & 0 & 0 & 1 & 0 & 1 & 3 & 0 & 0 & 0 & 0 & 0 & 0 & 0 & 0\\
1 & 0 & 0 & 1 & 0 & 0 & 0 & 0 & 0 & 0 & 0 & 1 & 0 & 0 & 4 & 0 & 0 & 1 & 0 & 0 & 0 & 0\\
0 & 0 & 0 & 0 & 0 & 1 & 1 & 0 & 0 & 0 & 0 & 0 & 0 & 0 & 0 & 3 & 0 & 0 & 0 & 0 & 1 & 0\\
0 & 0 & 0 & 0 & 0 & 0 & 0 & 1 & 0 & 0 & 0 & 0 & 0 & 0 & 0 & 0 & 2 & 0 & 0 & 0 & 0 & 1\\
0 & 0 & 0 & 0 & 0 & 1 & 1 & 0 & 0 & 1 & 0 & 0 & 0 & 0 & 1 & 0 & 0 & 4 & 0 & 0 & 0 & 0\\
0 & 0 & 1 & 0 & 0 & 1 & 0 & 0 & 0 & 0 & 0 & 0 & 0 & 0 & 0 & 0 & 0 & 0 & 3 & 0 & 0 & 1\\
0 & 0 & 0 & 1 & 0 & 0 & 0 & 0 & 0 & 0 & 0 & 0 & 0 & 0 & 0 & 0 & 0 & 0 & 0 & 2 & 1 & 0\\
0 & 0 & 0 & 0 & 0 & 0 & 0 & 0 & 0 & 0 & 0 & 1 & 0 & 0 & 0 & 1 & 0 & 0 & 0 & 1 & 3 & 0\\
0 & 0 & 0 & 0 & 0 & 0 & 0 & 0 & 1 & 0 & 0 & 0 & 0 & 0 & 0 & 0 & 1 & 0 & 1 & 0 & 0 & 3\\
\end{array}
}}
\]

Of course, the incidence relation \texttt{VV} in form of array of arrays of integers is immediately derivable, 
\begin{verbatim}
VV = [[1,9,13,14],[0,3,10],[8,9,18],[1,14,19],[7,10],[15,17,18],[11,15,17],
[4,12,16],[2,12,21],[0,2,12,17],[1,4,13],[6,14,20],[7,8,9,13],[0,10,12],
[0,3,11,17],[5,6,20],[7,21],[5,6,9,14],[2,5,21],[3,20],[11,15,19],[8,16,18]]
\end{verbatim}
but it is more useful and efficient to solve several queries at the same time, on modern computational architectures, by multiplication of sparse matrices~\cite{Davis:2006:DMS:1196434,Kepner:2011:GAL:2039367}.
\end{example} % example 3

\begin{example}[3D cellular complex] % example 4
The LAR scheme is, of course, dimension-independent. Here we give the LAR description of a the simplest simplicial decomposition of a mesh $3\times 2\times 1$ of 3-cubes into tetrahedra, shown in Figure~\ref{fig:ex4}.

\begin{verbatim}
V = [[0,0,0],[1,0,0],[2,0,0],[3,0,0],[0,1,0],[1,1,0],[2,1,0],[3,1,0],[0,2,0],
[1,2,0],[2,2,0],[3,2,0],[0,0,1],[1,0,1],[2,0,1],[3,0,1],[0,1,1],[1,1,1],
[2,1,1],[3,1,1],[0,2,1],[1,2,1],[2,2,1],[3,2,1]]

TV = [0,1,4,12],[1,4,12,13],[4,12,13,16],[1,4,5,13],[4,5,13,16],[5,13,16,17],[1,2,
5,13],[2,5,13,14],[5,13,14,17],[2,5,6,14],[5,6,14,17],[6,14,17,18],[2,3,6,14],[3,6,
14,15],[6,14,15,18],[3,6,7,15],[6,7,15,18],[7,15,18,19],[4,5,8,16],[5,8,16,17],[8,
16,17,20],[5,8,9,17],[8,9,17,20],[9,17,20,21],[5,6,9,17],[6,9,17,18],[9,17,18,21],
[6,9,10,18],[9,10,18,21],[10,18,21,22],[6,7,10,18],[7,10,18,19],[10,18,19,22],[7,
10,11,19],[10,11,19,22],[11,19,22,23]]
\end{verbatim}

\begin{figure}[htbp] 
   \centering
   \includegraphics[height=.245\textwidth,width=.245\textwidth]{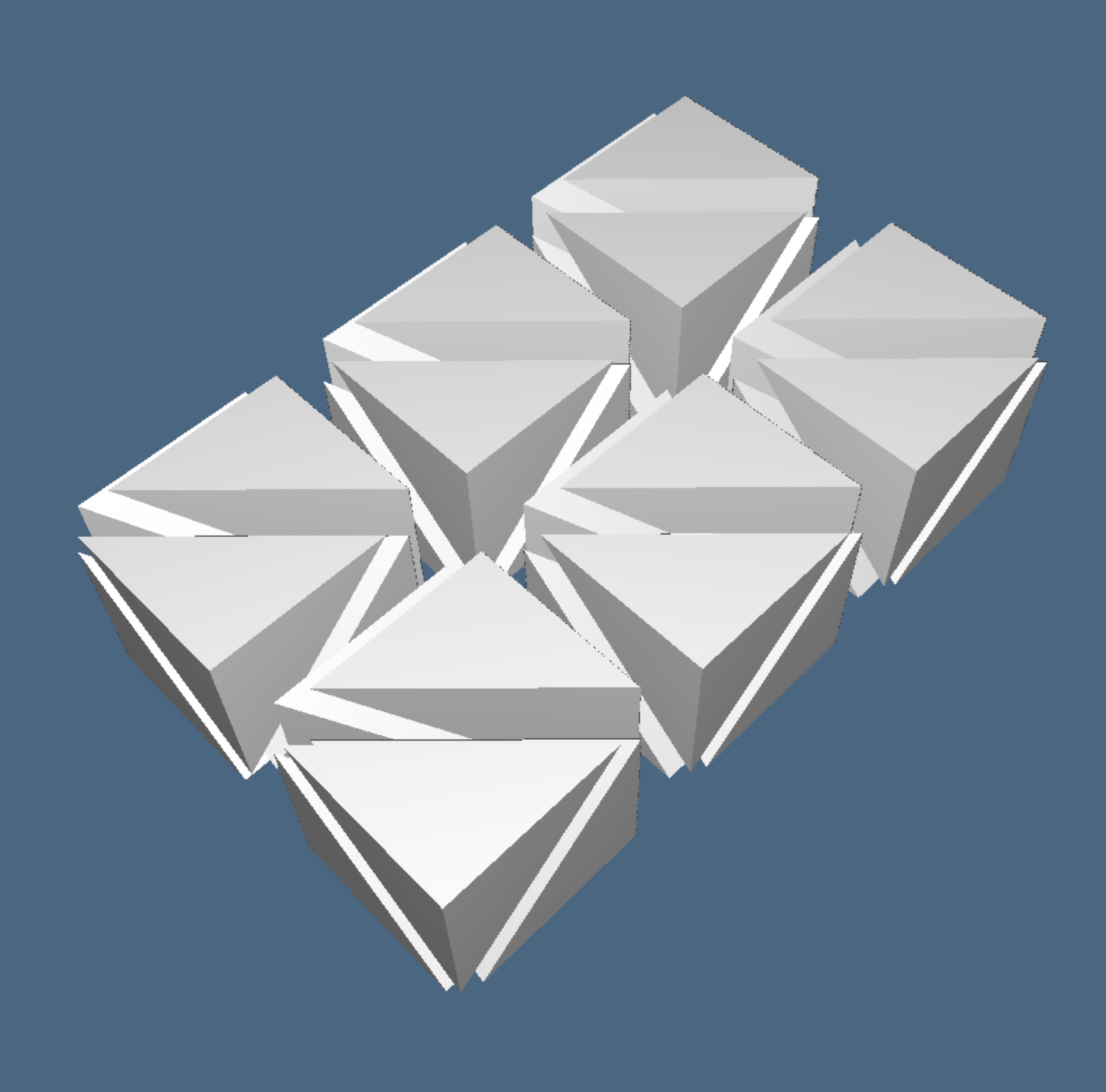} 
   \includegraphics[height=.245\textwidth,width=.245\textwidth]{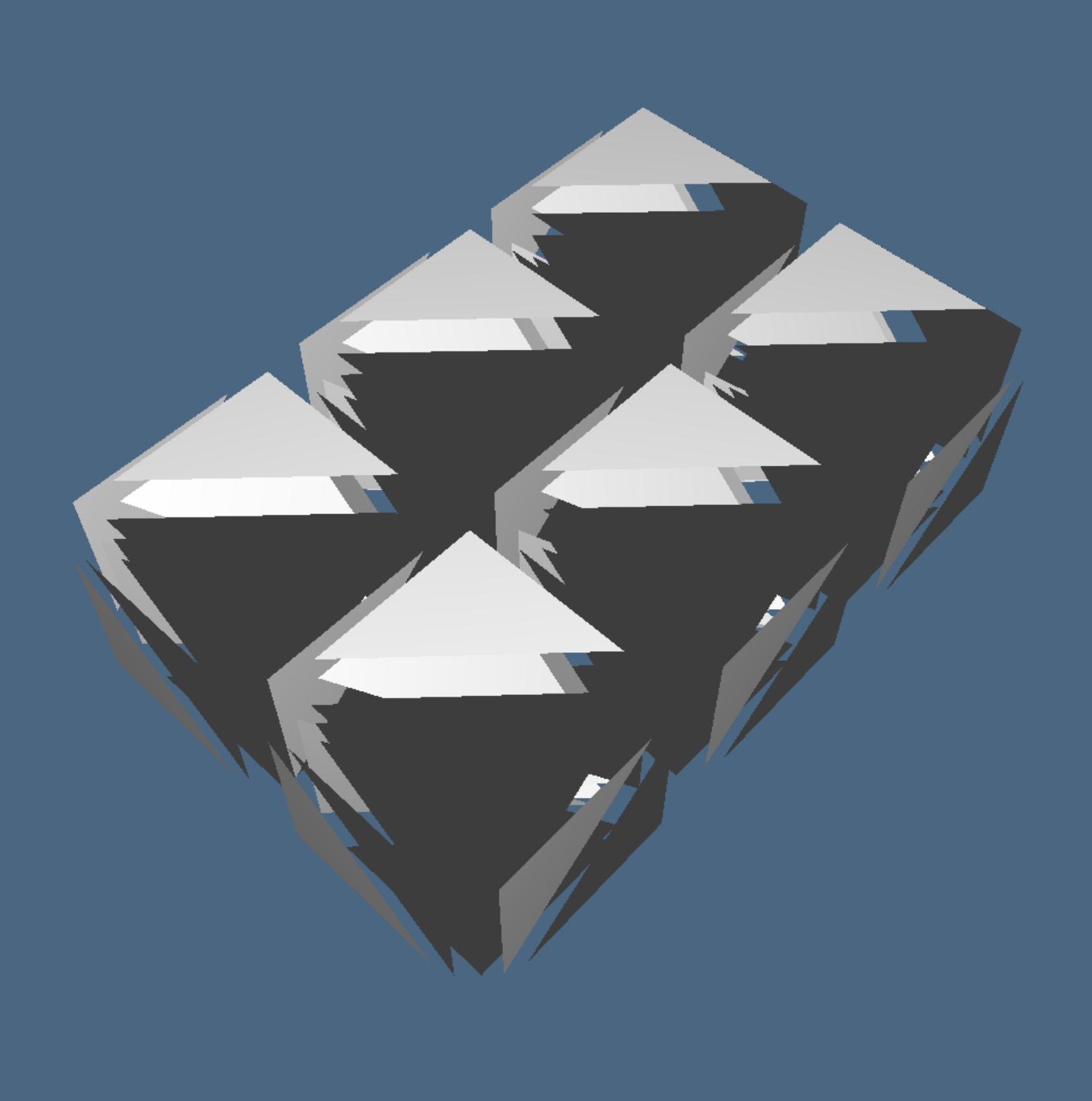} 
   \includegraphics[height=.245\textwidth,width=.245\textwidth]{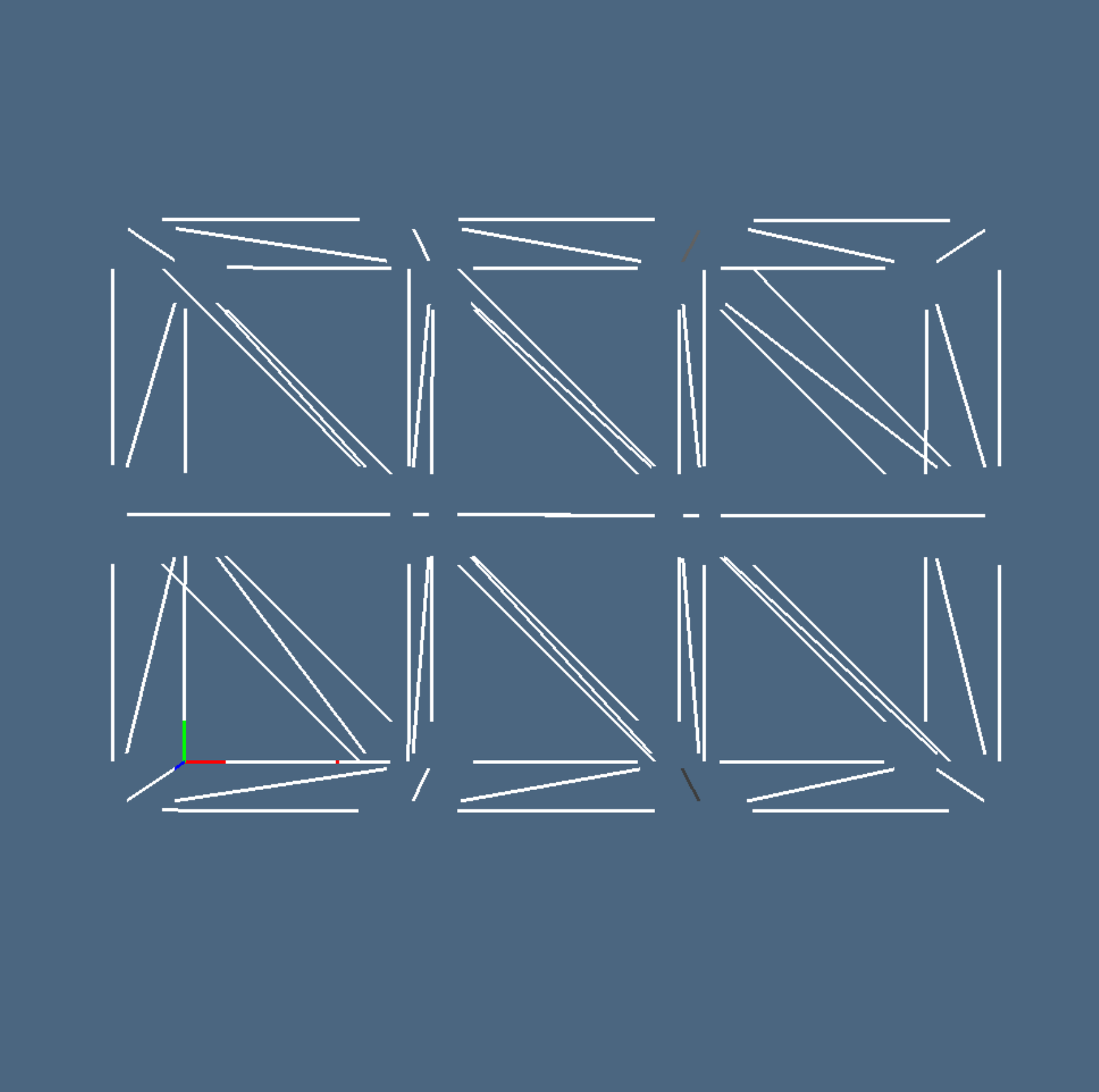} 
   \includegraphics[height=.245\textwidth,width=.245\textwidth]{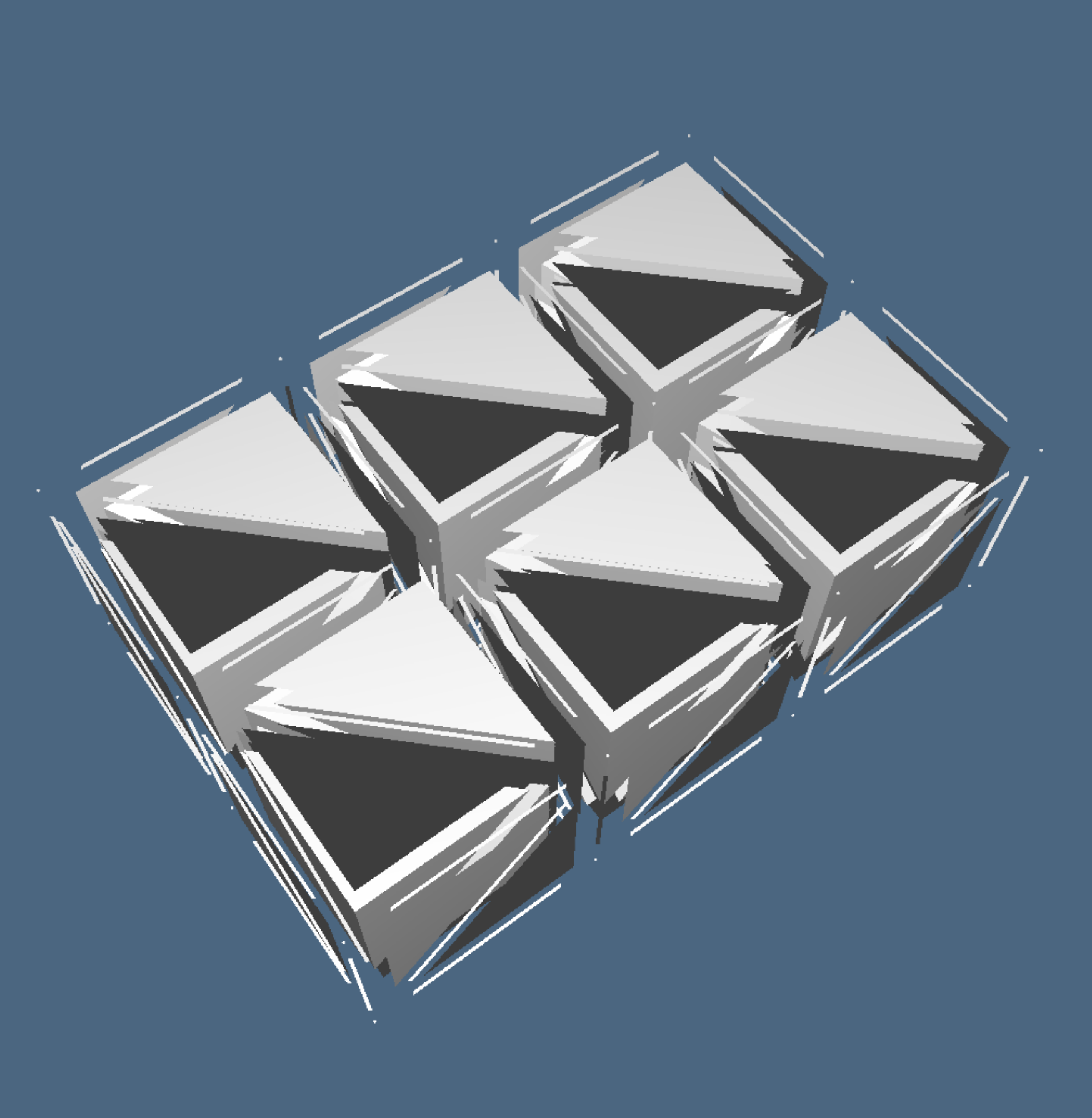} 
   \caption{A small ($3\times 2\times 1$) mesh (exploded) of tetrahedra: (a) 3-cells; (b) 2-cells; (c) 1-cells; (d) the whole 3-complex. Note that to $(d-1)$-cells in $\E^d$ space cannot be given a preferred orientation.}
   \label{fig:ex4}
\end{figure}

The above \texttt{TV} array of ``tetrahedra by vertices" is the user-readable format of the sparse characteristic matrix $M_3$ of 3-cells, and provides a \emph{complete} LAR representation of the solid model, allowing for efficient queries on topology. For example the adjacency relation \texttt{TT} between tetrahedra is computed by filtering the elements with value 3 in the (sparse) matrix $M_3M_3^t$. In this case we have:
\[
\texttt{TT} = \mathrm{filter}(M_3 M_3^t,3) = \mat{
{\tiny\arraycolsep=1.9pt%\def\arraystretch{2.2}
\begin{array}
{cccccccccccccccccccccccccccccccccccc}
- & {1} & - & - & - & - & - & - & - & - & - & - & - & - & - & - & - & - & - & - & - & - & - & - & - & - & - & - & - & - & - & - & - & - & - & -\\
{1} & - & {1} & {1} & - & - & - & - & - & - & - & - & - & - & - & - & - & - & - & - & - & - & - & - & - & - & - & - & - & - & - & - & - & - & - & -\\
- & {1} & - & - & {1} & - & - & - & - & - & - & - & - & - & - & - & - & - & - & - & - & - & - & - & - & - & - & - & - & - & - & - & - & - & - & -\\
- & {1} & - & - & {1} & - & {1} & - & - & - & - & - & - & - & - & - & - & - & - & - & - & - & - & - & - & - & - & - & - & - & - & - & - & - & - & -\\
- & - & {1} & {1} & - & {1} & - & - & - & - & - & - & - & - & - & - & - & - & {1} & - & - & - & - & - & - & - & - & - & - & - & - & - & - & - & - & -\\
- & - & - & - & {1} & - & - & - & {1} & - & - & - & - & - & - & - & - & - & - & {1} & - & - & - & - & - & - & - & - & - & - & - & - & - & - & - & -\\
- & - & - & {1} & - & - & - & {1} & - & - & - & - & - & - & - & - & - & - & - & - & - & - & - & - & - & - & - & - & - & - & - & - & - & - & - & -\\
- & - & - & - & - & - & {1} & - & {1} & {1} & - & - & - & - & - & - & - & - & - & - & - & - & - & - & - & - & - & - & - & - & - & - & - & - & - & -\\
- & - & - & - & - & {1} & - & {1} & - & - & {1} & - & - & - & - & - & - & - & - & - & - & - & - & - & - & - & - & - & - & - & - & - & - & - & - & -\\
- & - & - & - & - & - & - & {1} & - & - & {1} & - & {1} & - & - & - & - & - & - & - & - & - & - & - & - & - & - & - & - & - & - & - & - & - & - & -\\
- & - & - & - & - & - & - & - & {1} & {1} & - & {1} & - & - & - & - & - & - & - & - & - & - & - & - & {1} & - & - & - & - & - & - & - & - & - & - & -\\
- & - & - & - & - & - & - & - & - & - & {1} & - & - & - & {1} & - & - & - & - & - & - & - & - & - & - & {1} & - & - & - & - & - & - & - & - & - & -\\
- & - & - & - & - & - & - & - & - & {1} & - & - & - & {1} & - & - & - & - & - & - & - & - & - & - & - & - & - & - & - & - & - & - & - & - & - & -\\
- & - & - & - & - & - & - & - & - & - & - & - & {1} & - & {1} & {1} & - & - & - & - & - & - & - & - & - & - & - & - & - & - & - & - & - & - & - & -\\
- & - & - & - & - & - & - & - & - & - & - & {1} & - & {1} & - & - & {1} & - & - & - & - & - & - & - & - & - & - & - & - & - & - & - & - & - & - & -\\
- & - & - & - & - & - & - & - & - & - & - & - & - & {1} & - & - & {1} & - & - & - & - & - & - & - & - & - & - & - & - & - & - & - & - & - & - & -\\
- & - & - & - & - & - & - & - & - & - & - & - & - & - & {1} & {1} & - & {1} & - & - & - & - & - & - & - & - & - & - & - & - & {1} & - & - & - & - & -\\
- & - & - & - & - & - & - & - & - & - & - & - & - & - & - & - & {1} & - & - & - & - & - & - & - & - & - & - & - & - & - & - & {1} & - & - & - & -\\
- & - & - & - & {1} & - & - & - & - & - & - & - & - & - & - & - & - & - & - & {1} & - & - & - & - & - & - & - & - & - & - & - & - & - & - & - & -\\
- & - & - & - & - & {1} & - & - & - & - & - & - & - & - & - & - & - & - & {1} & - & {1} & {1} & - & - & - & - & - & - & - & - & - & - & - & - & - & -\\
- & - & - & - & - & - & - & - & - & - & - & - & - & - & - & - & - & - & - & {1} & - & - & {1} & - & - & - & - & - & - & - & - & - & - & - & - & -\\
- & - & - & - & - & - & - & - & - & - & - & - & - & - & - & - & - & - & - & {1} & - & - & {1} & - & {1} & - & - & - & - & - & - & - & - & - & - & -\\
- & - & - & - & - & - & - & - & - & - & - & - & - & - & - & - & - & - & - & - & {1} & {1} & - & {1} & - & - & - & - & - & - & - & - & - & - & - & -\\
- & - & - & - & - & - & - & - & - & - & - & - & - & - & - & - & - & - & - & - & - & - & {1} & - & - & - & {1} & - & - & - & - & - & - & - & - & -\\
- & - & - & - & - & - & - & - & - & - & {1} & - & - & - & - & - & - & - & - & - & - & {1} & - & - & - & {1} & - & - & - & - & - & - & - & - & - & -\\
- & - & - & - & - & - & - & - & - & - & - & {1} & - & - & - & - & - & - & - & - & - & - & - & - & {1} & - & {1} & {1} & - & - & - & - & - & - & - & -\\
- & - & - & - & - & - & - & - & - & - & - & - & - & - & - & - & - & - & - & - & - & - & - & {1} & - & {1} & - & - & {1} & - & - & - & - & - & - & -\\
- & - & - & - & - & - & - & - & - & - & - & - & - & - & - & - & - & - & - & - & - & - & - & - & - & {1} & - & - & {1} & - & {1} & - & - & - & - & -\\
- & - & - & - & - & - & - & - & - & - & - & - & - & - & - & - & - & - & - & - & - & - & - & - & - & - & {1} & {1} & - & {1} & - & - & - & - & - & -\\
- & - & - & - & - & - & - & - & - & - & - & - & - & - & - & - & - & - & - & - & - & - & - & - & - & - & - & - & {1} & - & - & - & {1} & - & - & -\\
- & - & - & - & - & - & - & - & - & - & - & - & - & - & - & - & {1} & - & - & - & - & - & - & - & - & - & - & {1} & - & - & - & {1} & - & - & - & -\\
- & - & - & - & - & - & - & - & - & - & - & - & - & - & - & - & - & {1} & - & - & - & - & - & - & - & - & - & - & - & - & {1} & - & {1} & {1} & - & -\\
- & - & - & - & - & - & - & - & - & - & - & - & - & - & - & - & - & - & - & - & - & - & - & - & - & - & - & - & - & {1} & - & {1} & - & - & {1} & -\\
- & - & - & - & - & - & - & - & - & - & - & - & - & - & - & - & - & - & - & - & - & - & - & - & - & - & - & - & - & - & - & {1} & - & - & {1} & -\\
- & - & - & - & - & - & - & - & - & - & - & - & - & - & - & - & - & - & - & - & - & - & - & - & - & - & - & - & - & - & - & - & {1} & {1} & - & {1}\\
- & - & - & - & - & - & - & - & - & - & - & - & - & - & - & - & - & - & - & - & - & - & - & - & - & - & - & - & - & - & - & - & - & - & {1} & -\\
\end{array}
}}
\]
The symmetric \texttt{TT} matrix can be immediately transformed into the array of arrays of tetrahedra indices giving all pairs $(k\times \texttt{TT}[k])$ ($0\leq k\leq \#\texttt{TV}$) of tetrahedra that share a 2-cell, i.e.~a triangle. Clearly such product matrix returns in element $(i,j)$ the number of vertices shared by tetrahedron $\texttt{TV}[i]$ and  tetrahedron $\texttt{TV}[j]$.
Of course, several queries on tetrahedra adjacency can be answered in parallel by a computational kernel \emph{SpMSpM} (sparse matrix times sparse matrix) working on \texttt{TT} and a compatible matrix with unit columns.

\begin{verbatim}
TT = [[1],[0,2,3],[1,4],[1,4,6],[2,3,5,18],[4,8,19],[3,7],[6,8,9],[5,7,10],[7,10,
12],[8,9,11,24],[10,14,25],[9,13],[12,14,15],[11,13,16],[13,16],[14,15,17,30],
[16,31],[4,19],[5,18,20,21],[19,22],[19,22,24],[20,21,23],[22,26],[10,21,25],
[11,24,26,27],[23,25,28],[25,28,30],[26,27,29],[28,32],[16,27,31],[17,30,32,33],
[29,31,34],[31,34],[32,33,35],[34]]
\end{verbatim}

\end{example} % example 4